\documentclass{aa}

\usepackage[varg]{txfonts}

\usepackage{graphicx}
\usepackage{natbib}
\usepackage{tikz}
\usepackage{here}
\usepackage{bm}
\usepackage{url}
\usepackage{verbatim}
\usepackage{hyperref}

\hypersetup{
    colorlinks=true,
    linkcolor=blue,
    filecolor=blue,      
    urlcolor=blue,
    pdfpagemode=FullScreen,
    citecolor = blue,
    }

\newcommand{\sparkling}{\textsc{Sparkling }}
\newcommand{\revolver}{\textsc{Revolver }}

\begin{document}

\title{Evidence for a sign change of the ISW effect in the very recent Universe? Hot voids and cold overdensities at $z<0.03$
}

\author{Frode K. Hansen\inst{1},  Diego Garcia Lambas\inst{2,3}, Andrés N. Ruiz\inst{2,3}, Facundo Toscano\inst{2} \and Luis A. Pereyra\inst{2,3}}

\institute{Institute of Theoretical Astrophysics, University of Oslo, PO Box 1029 Blindern, 0315 Oslo, Norway, \email{frodekh@astro.uio.no}
\and
Instituto de Astronomía Teórica y Experimental, CONICET-UNC, Laprida 854, X5000BGR, Córdoba, Argentina
\and
Observatorio Astronómico de Córdoba, UNC, Laprida 854, X5000BGR, Córdoba, Argentina}

\abstract
    {

      We find a significant CMB temperature excess in the direction of local underdensities within $z<0.03$. By contrast, less than $0.2\%$ of simulated CMB maps show a similarly significant temperature excess in nearby voids. Combined with earlier findings showing a $>5\sigma$ cooling of CMB photons in galactic filaments in the same redshift range, we now may have possible evidence for a negative Integrated Sachs-Wolfe (ISW) effect in the very recent Universe. In addition to having opposite sign, the observed amplitude is an order of magnitude larger than the predicted Rees-Sciama and ISW effects, pointing to an unknown physical process, possibly related with a non-standard model of dark energy. We discuss the results in light of the latest Data Release 2 results of the Dark Energy Spectroscopic Instrument (DESI) showing hints for dynamical dark energy. Removing the quadrupole, we find the CMB temperatures measured in nearby voids to a large degree uncorrelated with the temperature measured around nearby galaxies and the observed mean difference between these temperatures is almost $6.5\sigma$ larger than found in simulations.

}

\authorrunning{Frode K. Hansen et al.} 
\titlerunning{Evidence for a sign change of the ISW effect}


\maketitle
\nolinenumbers

\section{Introduction}
\label{sect:intro}

In \cite{luparello, hansen2023, garcialambas2023, santander, parameters, hansen2025}, substantial evidence for an unexpected cooling of CMB photons passing through galactic halos in nearby ($z<0.02$) massive galactic filaments was shown. Looking particularly for late type spiral galaxies, the temperature of the CMB fluctuations was shown to be $27\mu$K lower than the mean, more than $5\sigma$ below expected temperature fluctuations in simulations. In \citet[hereafter, DF2025]{nacho}, we showed that a similar cooling of CMB photons also occur in galactic filaments in the higher redshift shell $0.02<z<0.04$. In \citet[hereafter, H2025]{hansen2025}, we speculated on a possible explanation in terms of an interaction of CMB photons with dark matter. Here, based on the observed achromatic nature of the cooling, we look at an alternative explanation in terms of changing gravitational potentials.

In the well-known Rees-Sciama effect \citep[RS,][]{reessciama}, photons passing through the changing gravitational well of an overdensity during non-linear growth of matter perturbations, gain energy in entering the well, but loose more energy exiting due to the well growing as the photons pass. A cooling of these photons is therefore expected, but the predicted RS-effect is more than an order of magnitude smaller than what we observe in our previous works \citep{iswlocal,iswfull}. The Integrated Sachs-Wolfe effect \citep[ISW,][]{sw} is similar and takes place when dark energy causes accelerated expansion of the Universe and therefore causing an opposite change in the potentials, giving rise to heating of CMB photons in overdensities and cooling in underdense regions. The latter effect has been detected at higher redshifts $z\gtrsim0.1$ through cross-correlating CMB maps with tracers of large scale structure (see \cite{iswref2} and \cite{iswref1} for recent estimates and an extensive list of references). Note that both these effects give rise to a frequency independent change of CMB temperature which is consistent with the properties of the observed cooling of the CMB around galaxies.

At the redshifts where we have observed cooling of CMB photons in galactic halos at the order of $20-30\mu$K, the expected amplitude of the RS-effect (about $1\mu$K) is far too small to explain the observations. Furthermore, as shown in \cite{iswfull}, the ISW effect is expected to be dominating over the RS effect for structures in the nearby Universe (at the $2-4\mu$K level). One would therefore expect a heating and not a cooling of photons passing through nearby potential wells. The latest results of the Dark Energy Spectroscopic Instrument (DESI) \citep{desi1,desi2} have shown evidence for dynamical dark energy and an unexpected behaviour of the dark energy equation of state during the recent history of the Universe. Therefore, the equation of state and behaviour of dark energy at redshifts lower than those considered by DESI are unknown.

Extrapolating DESI results for the dark energy equation of state, we here speculate if this may open the possibility for a negative ISW-effect where the growth of the gravitational potentials are amplified during the recent ($z\lesssim0.1$) Universe, giving rise to cooling instead of heating of CMB photons passing through these potentials. As we discuss later, some theories for dark energy predict ISW effects which change sign \citep{ula,galilean2}. Evidence for an unexpected behaviour of the ISW effect at higher redshifts has already been reported in several studies using different data sets (see i.e. \cite{bossquasars, iswpuzzle} and references therein). The hypothesis of a possible negative ISW effect in the very recent Universe could be supported by the fact that the observed CMB cooling appears weaker around elliptical galaxies residing in clusters of galaxies \citet[hereafter, L2023]{luparello} which are virialized at an earlier stage.

If the hypothesis of a negative dark energy equation of state in the recent Universe is true, we expect to see heating of CMB photons passing through local voids. In this paper, we test this hypothesis by identifying all low redshift voids ($z<0.03$) and comparing the mean Planck CMB temperature in the areas of these voids with the corresponding temperature in simulations.

\section{Data and Method}
\label{sect:method}

\subsection{CMB and galaxy data}
Our main aim in this paper is to measure the mean CMB temperature in the areas of the sky corresponding to the projected voids of the nearby Universe. We use CMB intensity maps taken by the Planck satellite\footnote{\url{http://pla.esac.esa.int/pla}}. Both data from the Public Release 3 (PR3) \citep{pr3} as well as the recent Planck Public Release 4 (PR4) \citep{pr4} is used. The Planck data were cleaned for galactic foregrounds by 4 different component separation methods, SMICA, SEVEM, NILC and Commander, as described in \cite{compsep2018}. In general the temperature maps for all 4 foreground methods agree very well over all parts of the sky outside the Planck common mask. We here focus on two methods, SMICA for PR3 as 10 000 simulated maps for this method were made available, and SEVEM for PR4 since this was the only method providing a substantial number of simulations (600) for that release. We also use the SEVEM frequency maps where component separations was applied to individual Planck frequency maps and the NILC and Commander maps for consistency tests. We use the Planck common mask created for PR3 \citep{compsep2018} to mask possible galactic residuals as well as resolved extragalactic point sources, leaving $78\%$ of the sky available for analysis.

In addition to Planck data, we also use data from the WMAP satellite \citep{wmap}. In order to test frequency dependence, we use the lowest foreground cleaned channel of the WMAP experiment. We present results using the WMAP Q-band centred at 41GHz as well as simulated maps including the WMAP beam and noise realizations for this channel. All data and instrumental properties used to generate the simulations were taken from the publicly available WMAP Data Release 5 \footnote{\url{https://lambda.gsfc.nasa.gov/product/wmap/dr5/m_products.html}}

For identifying voids, we use the 2MASS Redshift Survey (2MRS)\footnote{\url{http://tdc-www.harvard.edu/2mrs/}} catalogue \citep{2mrs}. We aim to study the voids within a similar redshift range as for the galaxies in the previous papers. We identify voids in the range $z=[0,0.03]$ as the density of tracers strongly limit reliable identification of voids above $z>0.03$.

\subsection{Measurements of CMB temperature in voids}

In order to calculate the mean CMB temperature in voids, we define a set of void weight maps $M^\mathrm{void}_i$. In these maps, each pixel $i$ has a weight in the range $[0.0,1.0]$ depending on the number of times the index of the pixel was identified as being inside a projected void in the different void finding procedures. The weight maps thereby take into account the uncertainty in position and size of the voids. The higher the weight, the higher the probability that the pixel is indeed located in a void. Below we describe in detail how these void weight maps are constructed from the different void finding algorithms. When calculating the mean CMB temperature in voids, we take a weighted mean temperature using these void weight maps. We define the void temperature as
\begin{equation}
  \label{eq:voidtemp}
T_\mathrm{void} = \frac{\sum_i T^\mathrm{CMB}_iM^\mathrm{void}_iM^\mathrm{Planck}_i}{\sum_i M^\mathrm{void}_iM^\mathrm{Planck}_i}
\end{equation}

where $T^\mathrm{CMB}_i$ is the CMB temperature in HEALPix\footnote{\url{https://healpix.sourceforge.io/}} pixel $i$ (as discussed below, this is often replaced by the temperature of the wavelet transformed CMB map), $M^\mathrm{void}_i$ is the void weight map in pixel $i$ and $M^\mathrm{Planck}_i$ is the Planck common mask for pixel $i$, excluding the parts of the sky where the CMB temperature is uncertain.

The angular radius of the nearby voids which we use in our analysis varies between $\sim6^\circ$ to about $\sim30^\circ$ with a median of about $\sim10^\circ$. Using the full angular extension of these voids to create the void weight maps, most projected void areas overlap with neighbouring voids and a substantial fraction of the full sky would be used for calculating the mean temperature. We need to reduce the fraction of the void radii used to create the void maps in order to focus the void temperature measurement on their central parts, seeking a balance between not covering too much area and maintaining sufficient statistics at the same time. We do not know the expected temperature profile in the voids, but still expect a peak towards the centre where the photons have spent a larger time in an underdense environment. We therefore use 1/4, 1/2 and 3/4 of the void radius and create a corresponding void weight map and thereby a void temperature for each of these fractions of the radius.

The typical angular extension of the voids in our redshift range is $\sim10^\circ$. We therefore expect a possible ISW imprint of these voids to give rise to fluctuations in the CMB at a similar angular scale. CMB fluctuations at much smaller or lager angular scales,  then act as noise when trying to identify these void imprints in the CMB fluctuation field. In order to remove these fluctuations at scales outside of the expected angular size of a possible ISW effect, we apply a wavelet transformation to the CMB map before measuring the void temperature. We use the Spherical Mexican Hat Wavelets \citep[SMHW]{smhw} which have strong localisation properties on the sphere (and less so in harmonic space) which is important for identifying focused void imprints in their correct positions. The median angular scale of $\sim10^\circ$ of our void sample corresponds to a SMHW wavelet scale of $4^\circ$ \citep{vielva}.  All the void temperatures are based on this wavelet scale, unless otherwise stated. We also test the robustness of our results with slightly smaller or larger scales as well as use the original temperature map for profile calculations.

Another reason for choosing the SMHW is their symmetric shape. Depending on the identification criteria used, the voids may have different shapes \citep{colberg08,paillas18}, but on average they are well described as spherical volumes \citep{sheth04,demchenko16,wojtak16}. Furthermore, one of the two algorithms which we use for identifying voids (\sparkling , see details below) assumes voids as regions with spherical shape. A radially symmetric wavelet transform like the SMHW amplifies such fluctuations.

We further compare the void temperatures with the temperatures previously measured in galactic halos. For CMB temperatures around galactic halos, unless otherwise stated, we use the sample described in H2025 based on large, late type spiral galaxies in the redshift range $0.004<z<0.02$ in the three areas with the most massive nearby cosmic filaments. From here on, we use the term ``galaxy temperature'' to denote the mean temperature of CMB fluctuations taken over these galactic halos. Note that for all galaxy temperature estimates, we always remove the first five multipoles of the CMB map to increase the signal-to-noise ratio, as detailed in H2025.

\subsection{Void finding algorithms and void weight maps}

As described above, we estimate the mean CMB temperature in voids by calculating the weighted mean of the temperature over the CMB map (or wavelet transformed map) using the void weight maps to upweight the pixels which have been frequently flagged as belonging to a void. We now explain in detail how voids are identified and how the void weight maps are constructed from the two different void finding algorithms, \sparkling and \revolver.

\subsubsection{\sparkling}

The voids were identified with the public void finder \sparkling\footnote{\url{https://gitlab.com/andresruiz/Sparkling}} \citep{ruiz15,ruiz19} using as tracers (1) all 2MRS galaxies up to $z=0.03$ and (2) those with absolute $K$-band magnitudes $M_K < -23.93$, which constitute a volume limited sample at $z=0.03$ assuming $h=0.674$. We have included the volume limited sample since this assures a constant density of tracers at different redshifts, but at the cost of a smaller density of tracers at lower redshifts. Briefly, the void identification procedure starts with an estimation of the galaxy density field via a Voronoi tessellation, which is used to quickly find the underdense regions of the catalogue. Centred in those underdense regions, void candidates are identified as spherical regions with an integrated density contrast satisfying $\Delta(R_{\rm void})<-0.9$, $R_{\rm void}$ being the void radius, which means that voids contains only $10\%$ of the density of tracers. Once a void candidate is identified, the computation of $\Delta$ is repeated starting from a randomly displaced centre, sampling in this way the volume of the corresponding sphere associated with the initial Voronoi cell and updating the void centre only if the new radius is larger than the previous one. This iterative re-centring procedure guarantee that void centres are located as close as possible to the local minima of the density field. The final step in the identification is a cleaning procedure retaining only the largest non-overlapping voids while removing the smallest ones.

\begin{figure}[htbp]
  \includegraphics[width=0.495\linewidth]{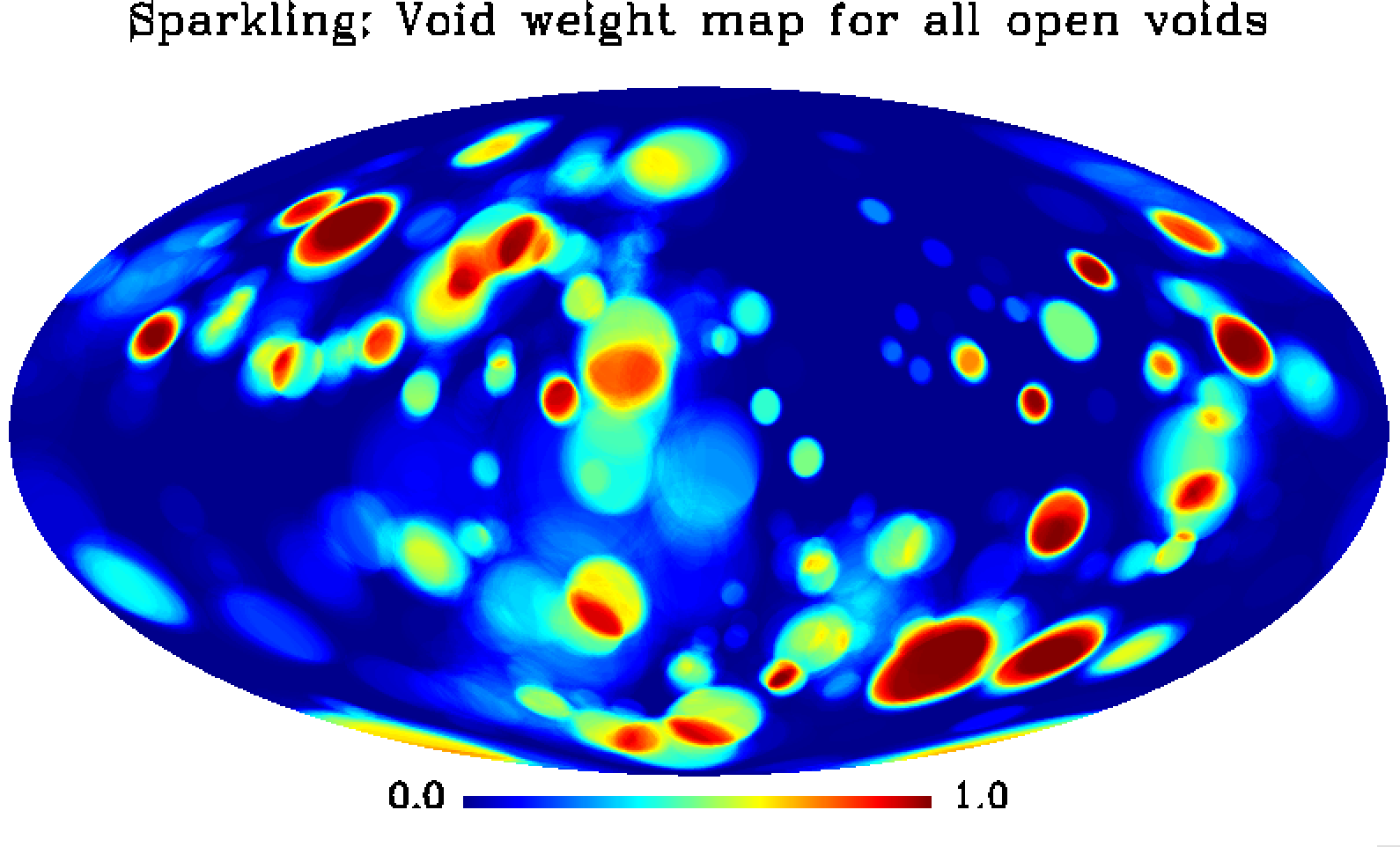}
  \includegraphics[width=0.495\linewidth]{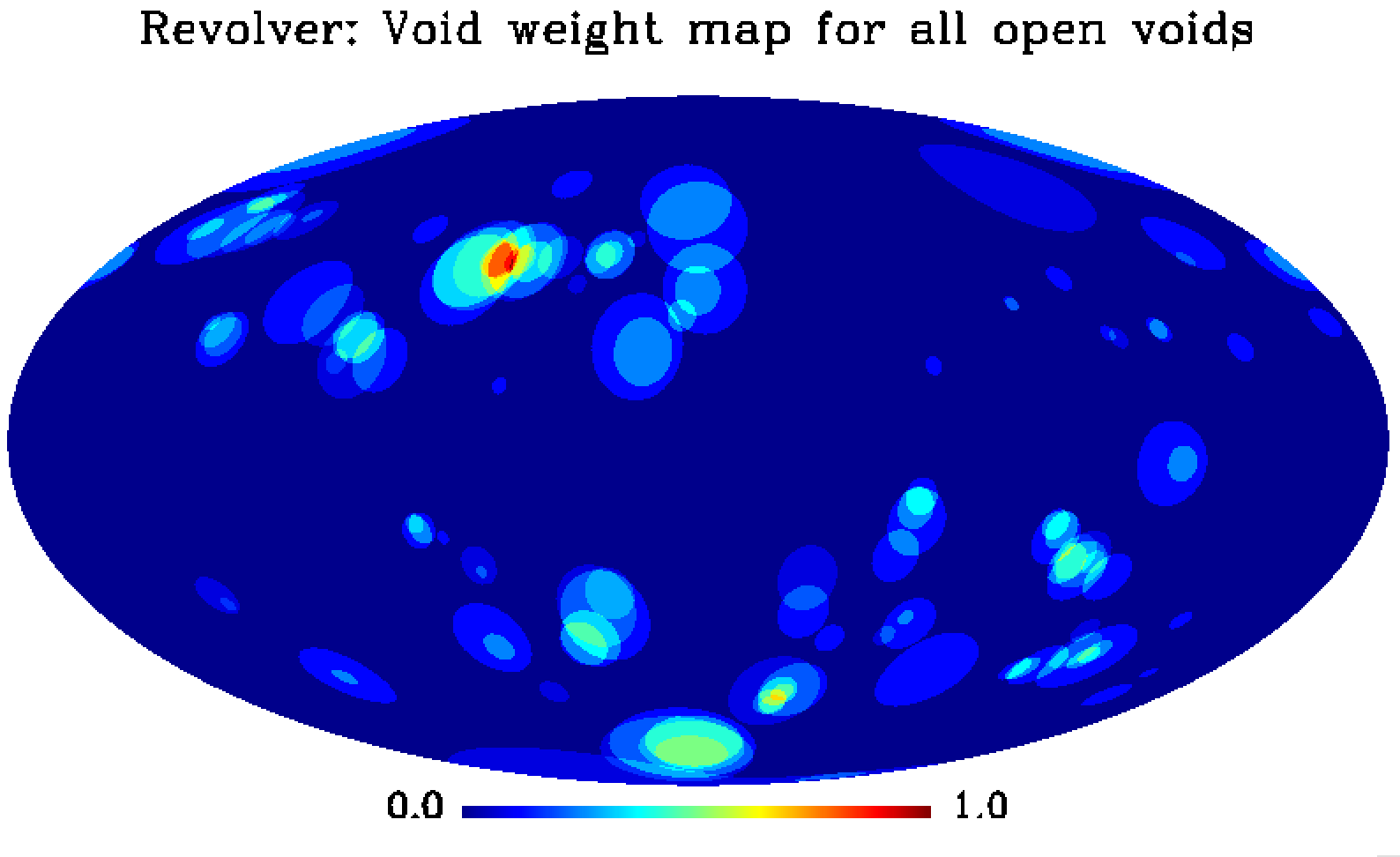}
  \includegraphics[width=0.495\linewidth]{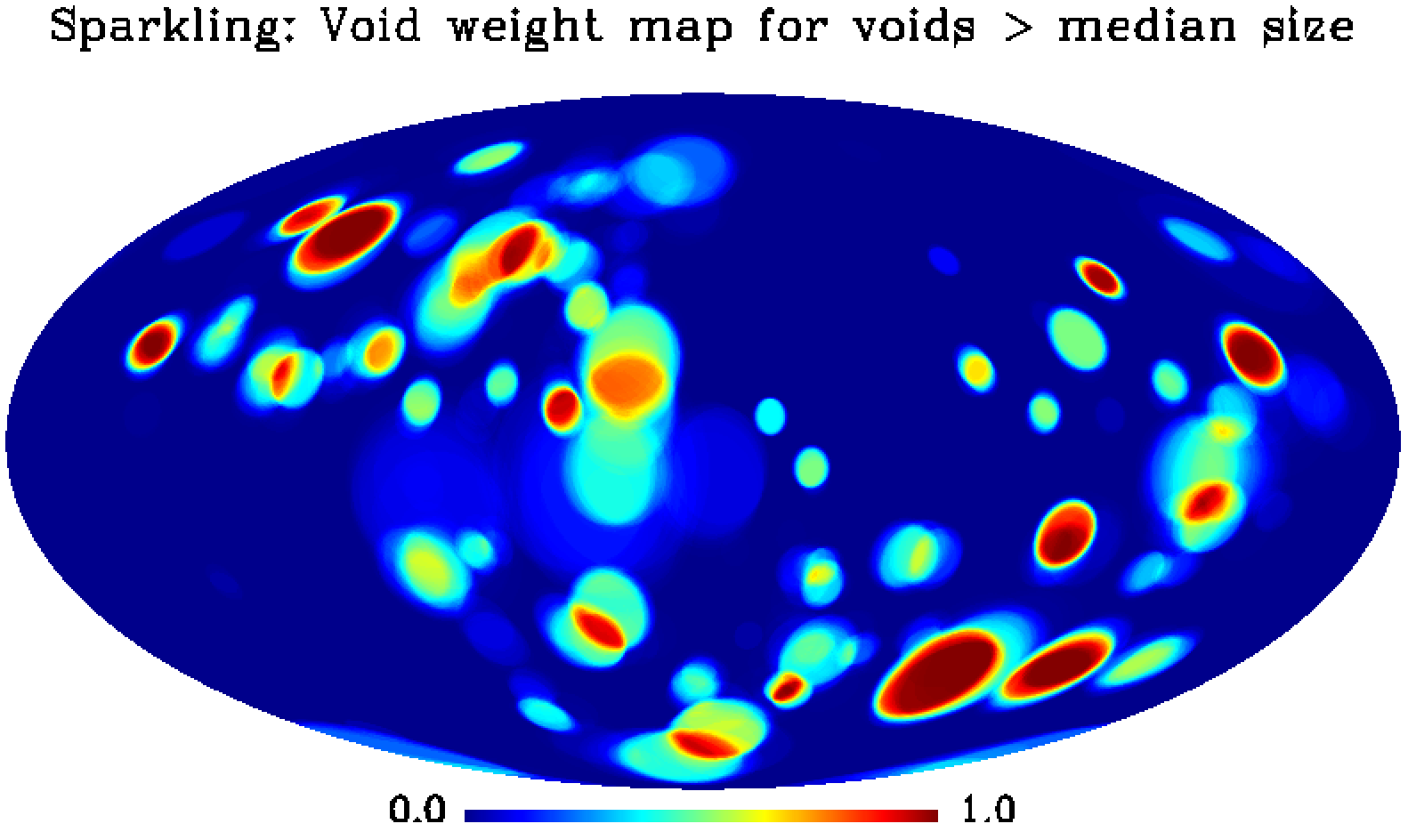}
  \includegraphics[width=0.495\linewidth]{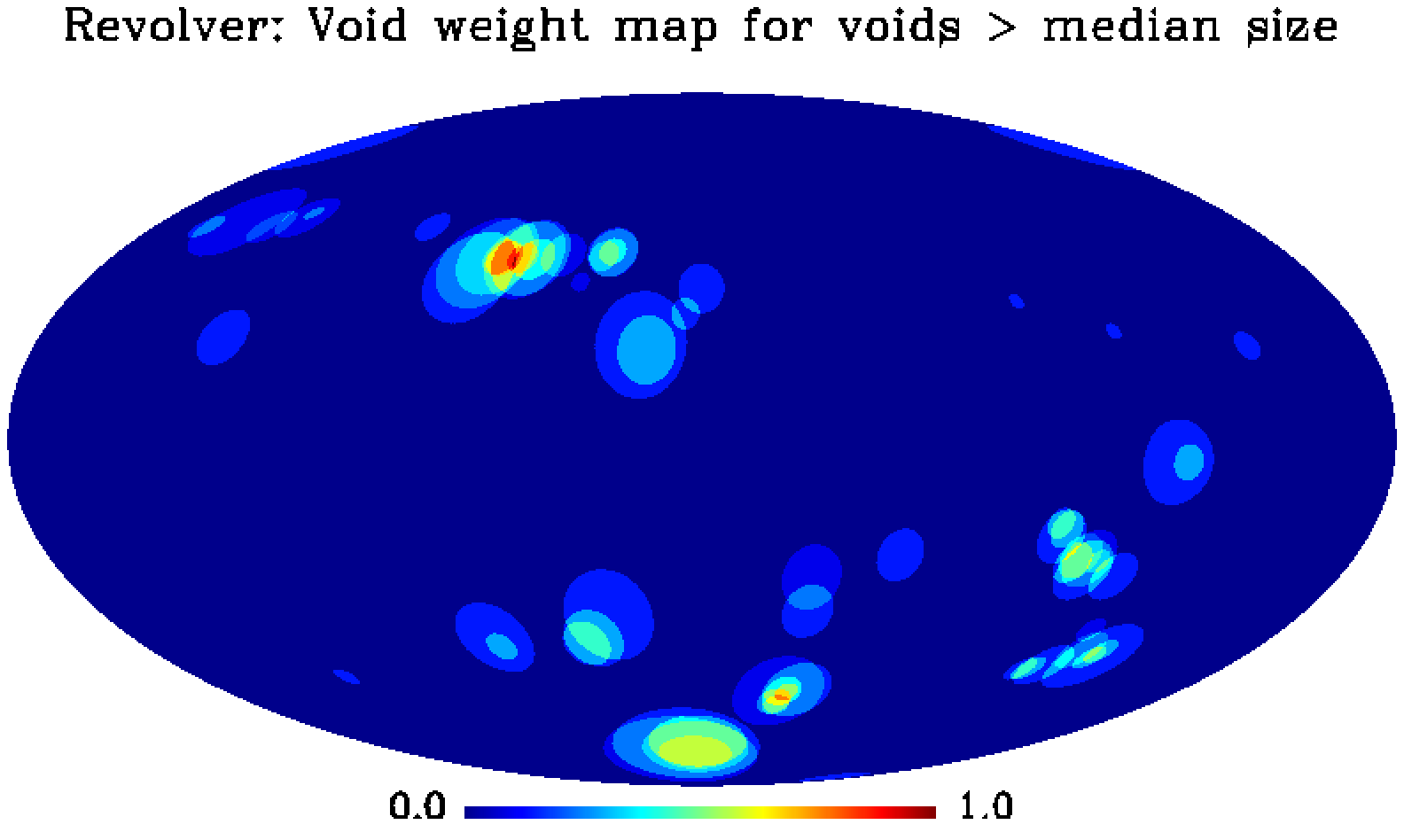}
  \includegraphics[width=0.495\linewidth]{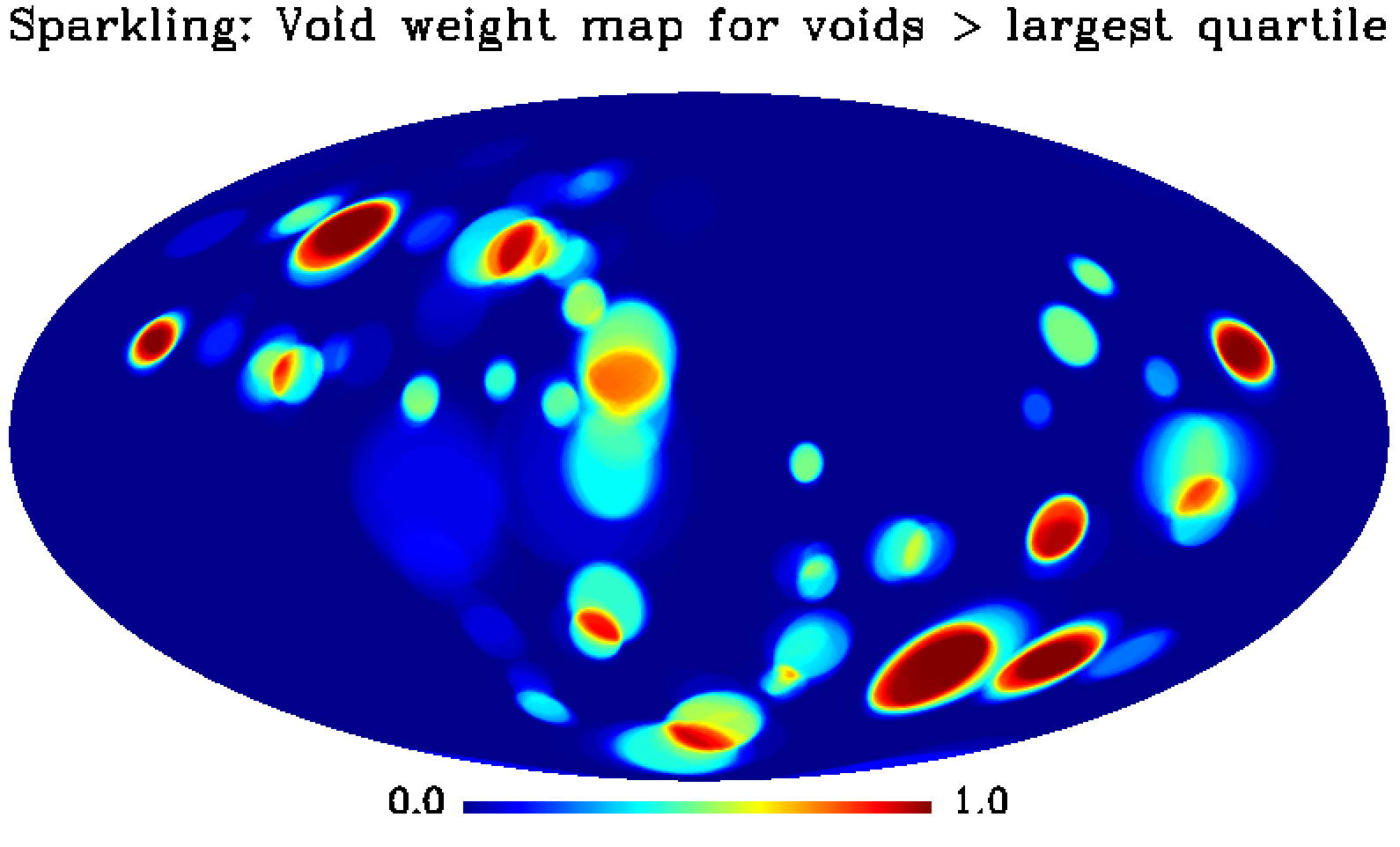}
  \includegraphics[width=0.495\linewidth]{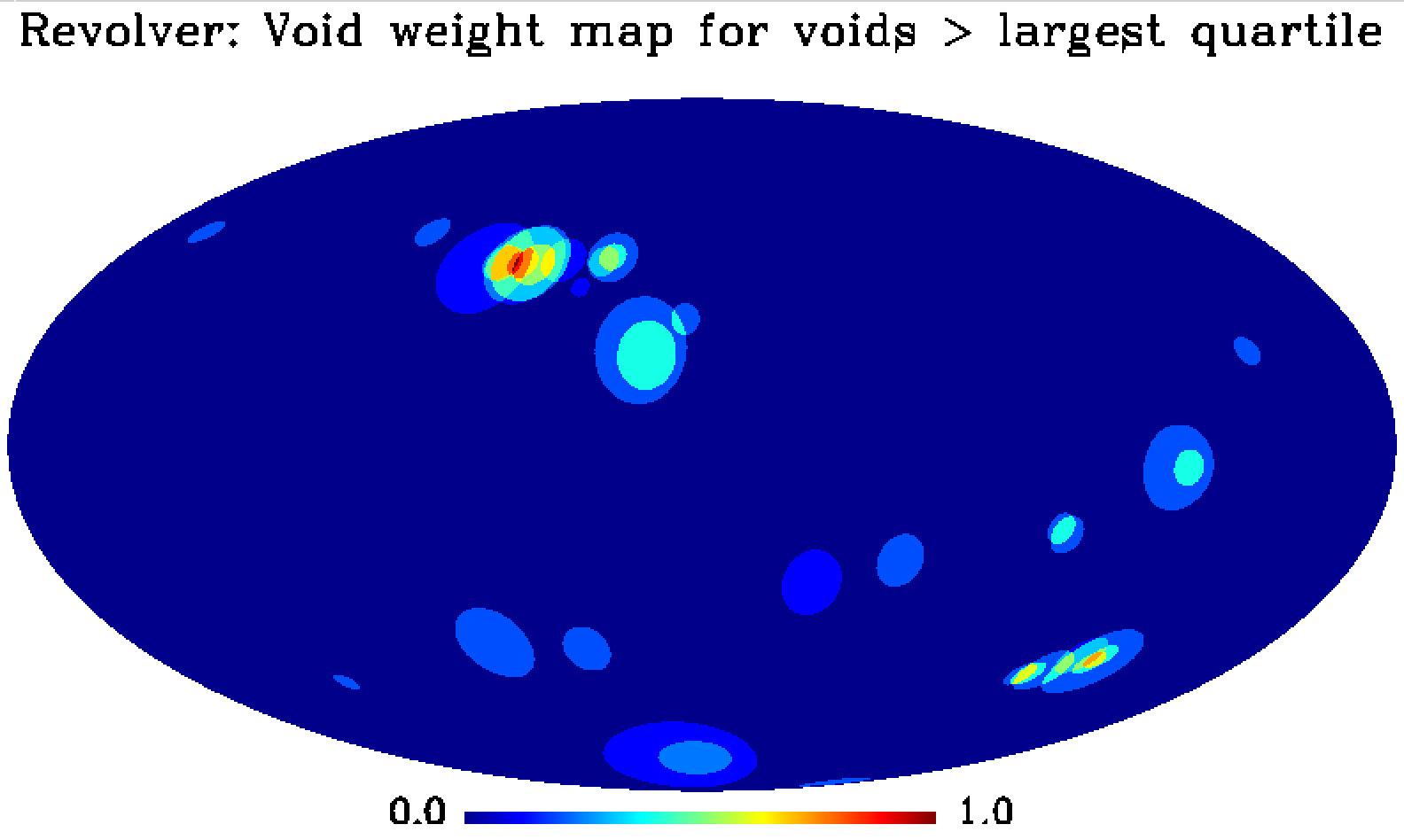}
  \caption{ \label{fig:voidmaps} Void weight maps. Left panels: based on 200 void maps created from \sparkling combining 100 realizations for the volume limited sample of galaxies and 100 realizations from the sample using all galaxies. Right panels: based on Revolver voids identified in the sample using all tracers as well as in the volume limited sample. Each pixel represents a weight proportional to how many times that pixel was flagged as belonging to a void among those 200 void maps (\sparkling) or among overlapping voids (\revolver). In these maps, a pixel was defined as belonging to a void when it was found within $r<0.5R_\mathrm{void}$ for voids identified in the redshift range $z=[0,0.03]$ based on the 2MRS redshift catalogue. The upper panels show the result when all open voids were included, the middle panels for all voids larger than the median size of a given void set and the lower panels for the largest quartile of voids. Similar void maps (not shown) were created using only pixels within $r<0.25R_\mathrm{void}$ and $r<0.75R_\mathrm{void}$.}
\end{figure}

The stochastic nature of the re-centring process means that different runs with different random seeds produce slightly different results in void centres and radii, and, also depending on whether all galaxies, or only the volume limited sample were used as tracers. In general, the volume limited sample shows much larger variation with random seed and is therefore slightly less reliable. In order to construct a map of stable void regions, we created 100 realizations of both sets of tracers, resulting in 200 sets of voids. Among those 200 sets, the void centres and void radii were similar, but some variation was observed. Each pixel in our final void map was then weighted by the number of times the given pixel was flagged as part of a void in those 200 sets. In this way, the void weight map takes into account the uncertainty in void position and radius. The weighted void maps are shown in the left panels of Figure \ref{fig:voidmaps}.

We identified a total number of $192\pm5$ voids ($110\pm3$ voids in the volume limited sample) in this redshift range among each of the 100 sets. The number of voids is clearly much smaller than the number of galaxies, giving considerably poorer statistics in this case. We further expect an RS-like effect to be much stronger for larger voids (in the same way as we found that larger galaxies give rise to stronger cooling \citep{luparello,hansen2023}). We therefore make two samples where we look at the voids with a void radius larger than the median size of the sample (varying between the 200 sets, but typically $R_\mathrm{void}>13$\;Mpc) and the voids of the largest quartile of the sample (typically $R_\mathrm{void}>17$\;Mpc). We only consider voids where the mean integrated density contrast out to 2-3 void radii, $\Delta_{23}<0$ (defined as the maximum value of the integrated density contrast between 2 and 3 void radii), is negative, meaning that the voids are surrounded by an underdense region (in the following, we call these ``open voids'') and not by a shell of matter.  We know that the latter voids may be shrinking in contrast to the open voids which are normally expanding during non-linear growth of perturbations \citep{sheth04,ceccarelli13,paz13}. The total number of open voids is $69\pm4$ ($40\pm3$ in the volume limited sample). The weighted void maps limited to these larger voids (median and quartile) are also seen in Figure \ref{fig:voidmaps}. In the following, we calculate the void temperatures (Equation \ref{eq:voidtemp}) using the void weight maps for all the combinations of fractions of void radius (1/4, 1/2 or 3/4 void radius) and for the sample of all the open voids, all the voids larger than the median void radius or all the voids belonging to the largest quartile of void radii.

\begin{figure}[htbp]
  \includegraphics[width=0.9\linewidth]{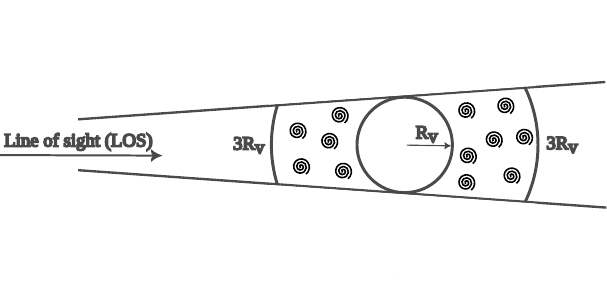}
    \caption{ \label{fig:losvoid} Definition of $\Delta_\mathrm{LOS}$ considering the density within the solid angle subtended by the void between $1$ and $3R_\mathrm{void}$ }
\end{figure}

As a further test of consistency, we also identify open voids where the low density region outside the void is directed along the line of sight, making that the CMB photons travelling through these voids spend more time in a low density area. If our working hypothesis involving an anomalous ISW or RS-effect holds, we would expect an enhanced positive temperature from these voids. On the other hand, with a standard ISW effect based on a cosmological constant, we would expect these voids to be colder. In the same way that we calculate $\Delta_{23}$, we now define a line of sight (LOS) integrated overdensity, $\Delta_\mathrm{LOS}$, by considering only the galaxies that are between $1$ and $3R_\mathrm{void}$ in front of and behind the void, confined to the solid angle defined by the line of sight and the radius of the void, as we show schematically in Fig. \ref{fig:losvoid}.

\subsubsection{\revolver}

As different void finders and algorithm use different definitions of voids and therefore commonly may identify quite different regions of the sky as voids \citep{colberg08,paillas18}, we also use \revolver\footnote{\url{https://github.com/seshnadathur/Revolver}} \citep{revolver} as an alternative method. This void finder starts with density field computed using a cloud-in-cell algorithm to assign galaxy positions to a regular 3D grid. Once the density field is estimated, a watershed algorithm is used to identify the underdense regions, selecting the density minima as potential void centres. As the void's boundaries are defined by the crests of the density field, this algorithm identify voids with irregular shapes. 

Applying \revolver to the same two samples of galaxies, we find 115 voids when using all galaxies and 65 voids when using the volume limited sample. We define void radii in a similar way as for \sparkling, requiring $\Delta(R_{\rm void})<-0.9$ around all void centres identified by \revolver, ensuring a very low density region within the void. We find a median void radius of $15\,$Mpc, similar to the voids identified by \sparkling. In the right panels of Figure \ref{fig:voidmaps}, we show the \revolver void weight maps side by side with the corresponding \sparkling maps. For \revolver, the voids overlap to a much larger degree than for \sparkling and high weights indicate pixels where several voids overlap (note that for \revolver, there is no random process in the identification of the voids and the weights are therefore a result only of overlapping voids from one single void map and not overlapping voids from different sets of voids). A pixel with a high weight means that this pixel was part of several voids identified by \revolver in both galaxy samples (volume limited and all galaxies). Comparing to the \sparkling void maps in the left panels, we can see that some regions of the sky are identified as voids by both methods whereas others are identified by only one method depending on the morphology of the density distribution in the given region. Note also that since a large part of the \revolver voids are overlapping, the total sky area occupied by \revolver voids is substantially less than for \sparkling. Therefore, results based only on \revolver voids alone (see Appendix) have a larger uncertainty.

\subsubsection{Combining algorithms}

\begin{figure}[htbp]
  \begin{center}
  \includegraphics[width=0.6\linewidth]{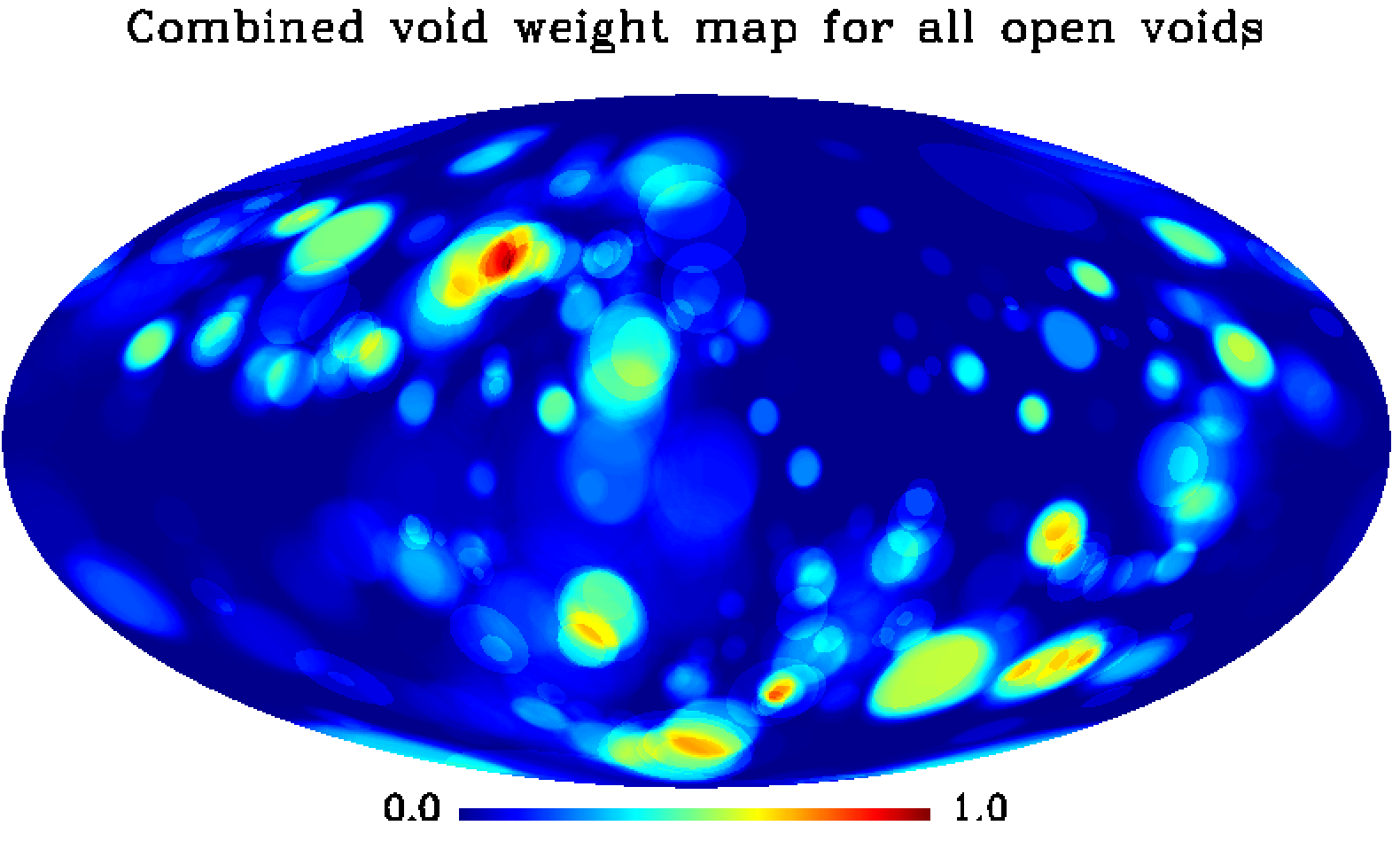}
  \includegraphics[width=0.6\linewidth]{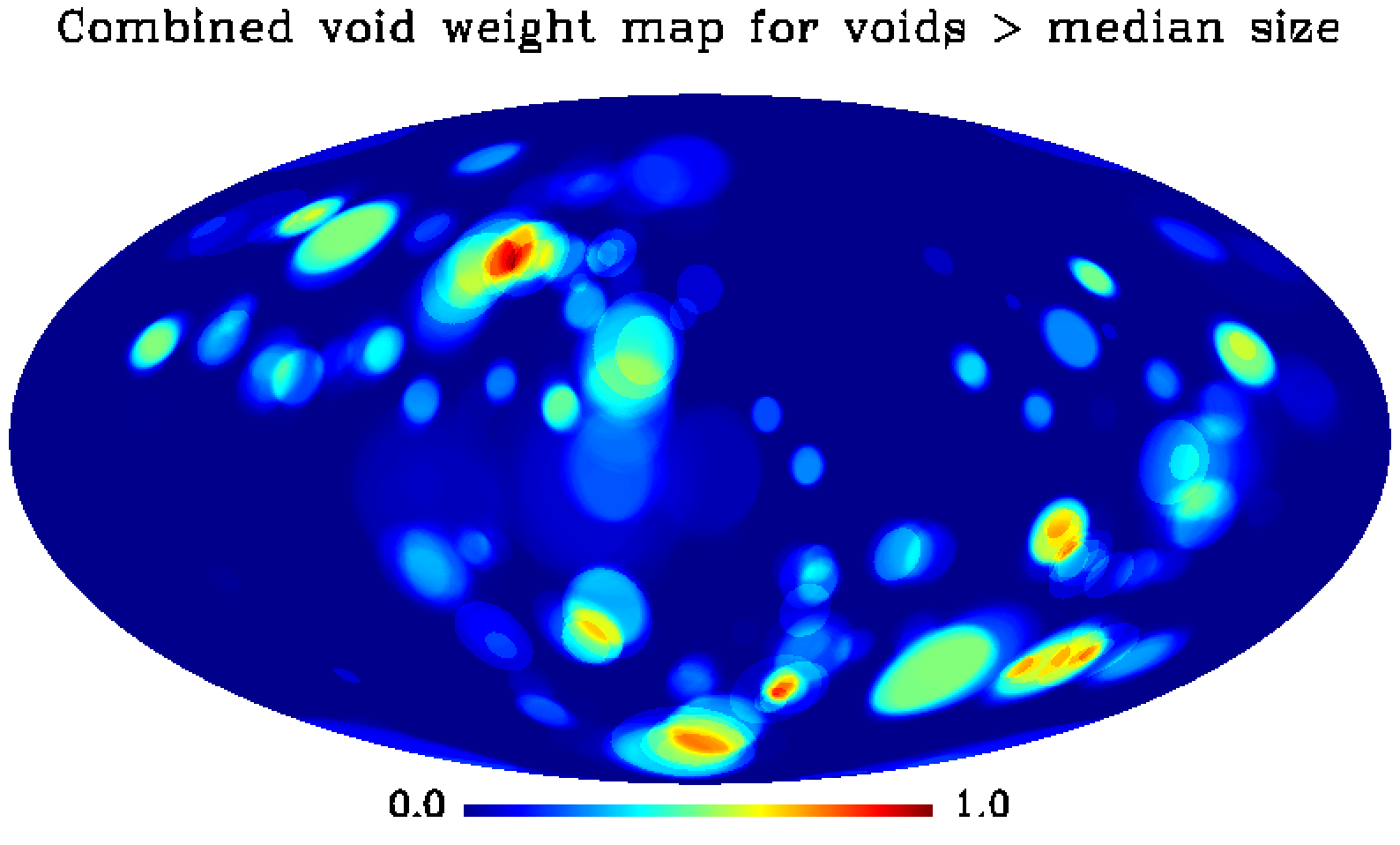}
  \includegraphics[width=0.6\linewidth]{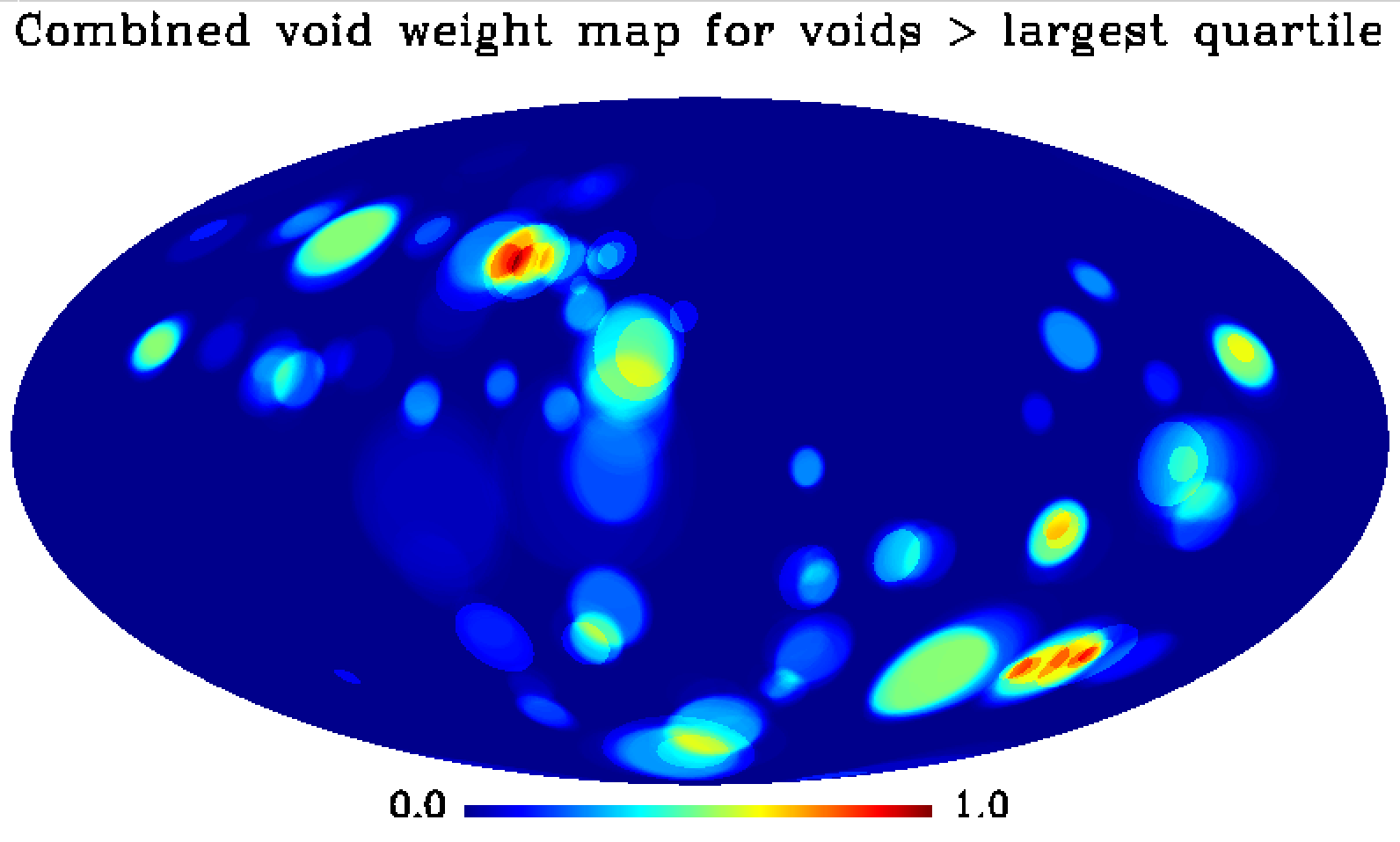}
  \end{center}
  \caption{ \label{fig:combinedmaps} The combined weight maps created from the mean between the \revolver void maps and the \sparkling void maps in Figure \ref{fig:voidmaps}. The larger number of times a pixel has been flagged as belonging to a void in \sparkling and \revolver, the higher the pixel weight. A weight larger than 0.5 indicates that the pixel has been flagged as a void several times in both methods. From upper to lower panel, void map based on: all open voids, the largest half of the voids and the larger quartile of voids.}
\end{figure}

While we test the robustness of our results using the \sparkling void map and \revolver void map separately (see Appendix), the main results are presented by the mean of these two voids maps: The pixels of these void maps represent the number of times the given pixel was flagged as being part of a void in \sparkling and in \revolver. A pixel weight larger than 0.5 indicates that the pixel has been part of a void a considerable number of times in both \revolver and \sparkling. A weight of 1 means that the pixel was identified as belonging to a void for all random realization of \sparkling and that \revolver identified several voids overlapping the given pixel. This pixel thus has a very high probability of being in a low density region. These combined void maps are shown in Figure \ref{fig:combinedmaps}. In addition to give weight to the mean CMB temperature calculations in voids, the void weight maps are used to create temperature profiles around voids where only voids with high ($>0.5$) central weights are used for profile calculations.

\subsection{Mock galaxy catalogues}

In addition to measuring the void weighted CMB temperature in the Planck data and comparing it with the mean temperature in simulated CMB maps using the same void weightings, we also test our results with an opposite approach: We apply \sparkling to 100 mock catalogues of galaxies, create the corresponding 100 void weight maps from each of these mock catalogues, and measure the mean temperature in the Planck CMB map using these 100 void weight maps, comparing it with the temperature obtained using the void weight map from the actual 2MRS survey. Mock galaxy catalogues are constructed by extracting random spherical volumes of radius $\sim 120$ Mpc from the MultiDark Planck 2 (MDPL2) catalogue \citep{md2}, chosen to match the observational volume of the 2MRS survey. We use the $z=0$ snapshot, publicly available through the CosmoSim database\footnote{\url{https://www.cosmosim.org}}. We increase the radius of the voids in the mocks by a factor 1.15 such that the median radius of the mock void sample coincides with the median size of the real 2MRS voids to take into account a small difference in the mean density of tracers in the mock and the data.

Note that we have not generated random realizations of the void distribution for each of these 100 mocks, such that the void weight maps for these 100 \sparkling sets of voids are similar to the ones obtained for \revolver: A high pixel weight means that there are many voids overlapping at the given pixel. For the actual 2MRS data however, we do have 100 random realizations, and we create one void weight map for each of these in exactly the same manner as we create the void weight maps for the mock catalogues. In this way, we can compare the 100 mean temperatures obtain on Planck data for each of the 100 mocks as well as for the 100 random realizations of \sparkling voids for the actual 2MRS data.

\subsection{Correlations between galaxy and void temperatures}

A main worry when measuring the void temperature in Planck data where we already know that the CMB temperature around galaxies is much colder than expected, is a possible anticorrelation between the temperature measured in voids and the temperature measured around galaxies. Voids are generally in a part of the sky where there are few galaxies and cold galaxy temperatures could therefore increase the possibility of hot void temperatures. In the results, we show that such an anticorrelation does indeed exist in simulated maps with a quadrupole present. Removing the quadrupole, the anticorrelation is reduced to $2-5\%$ depending on the void sample used. We therefore always remove the quadrupole estimated outside the Planck common mask from Planck data and all simulated CMB maps. The quadrupole is estimated inverting the coupling matrix of the $a_{\ell m}$ coefficients as described in detail in H2025.

\section{Results}
\label{sect:results}

\begin{figure}[htbp]
  \begin{center}
  \includegraphics[width=0.6\linewidth]{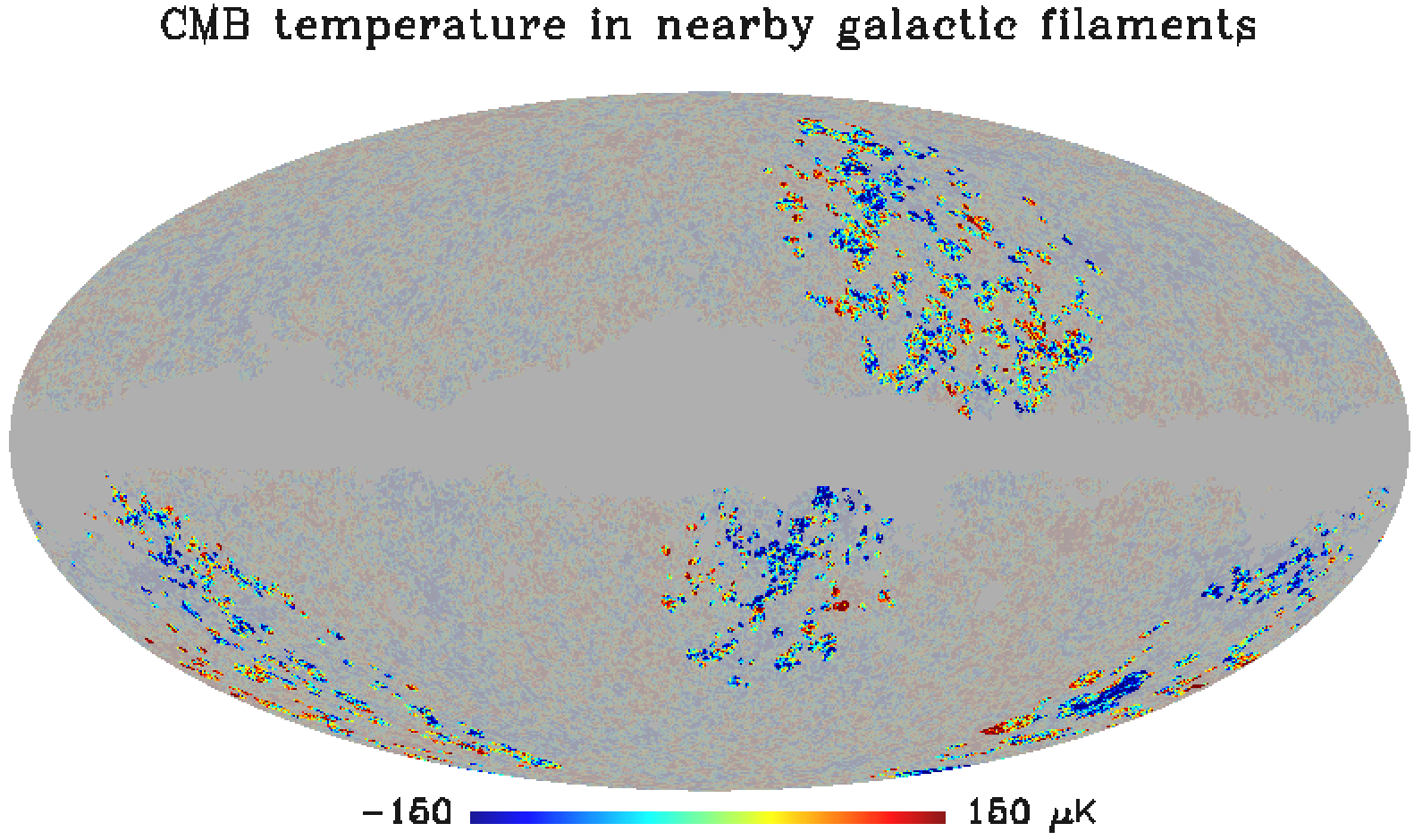}
  \includegraphics[width=0.6\linewidth]{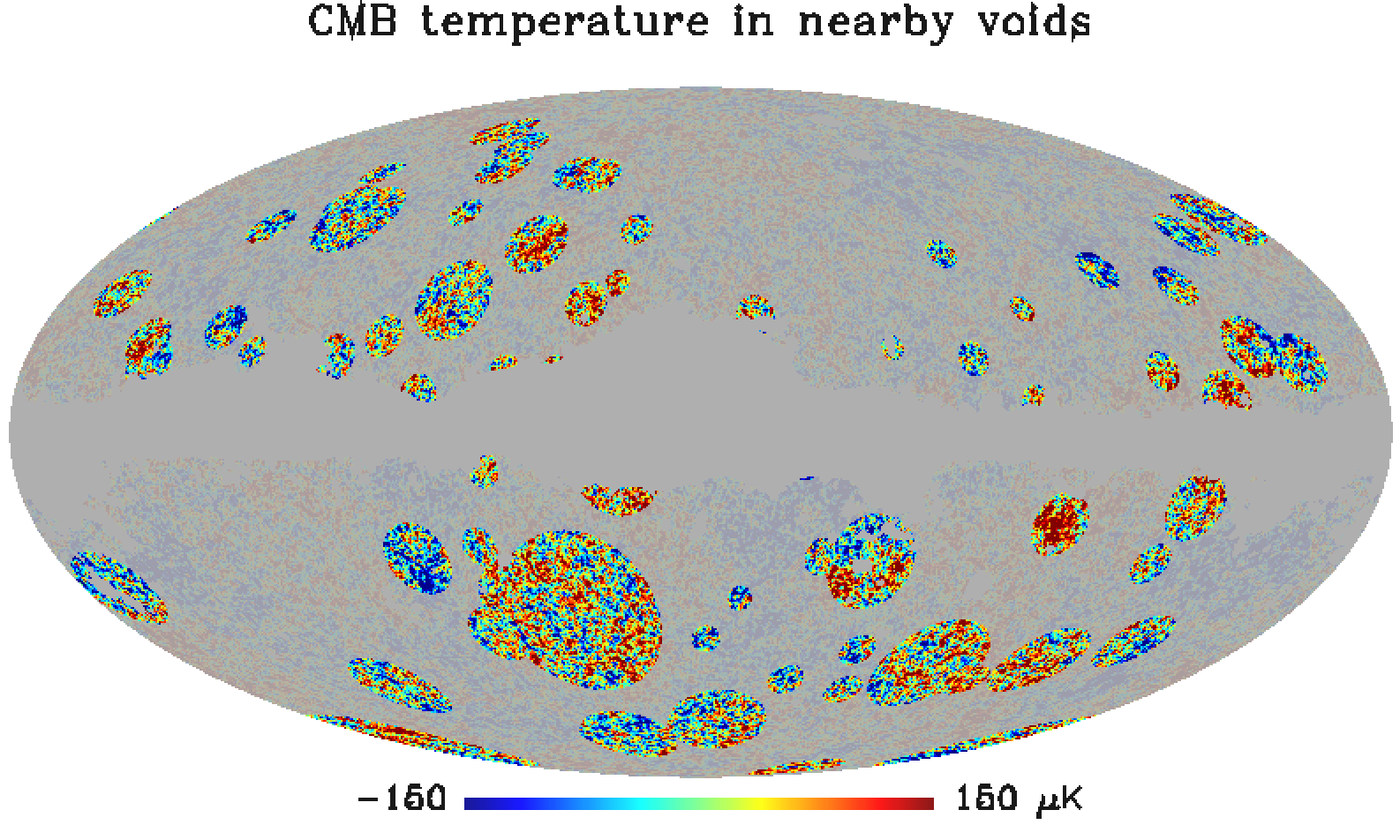}
  \end{center}
  \caption{The Planck SMICA CMB temperature fluctuation map, masked such that only the areas around nearby galaxies (upper panel) and voids (lower panel) are clearly visible. The mask consists of holes of $1$\,Mpc radius around all large nearby spiral galaxies in the most massive galactic filaments within $z<0.02$ (see H2025 for details). The void holes includes one realization of \sparkling voids, including all voids with $\Delta_{23}<0$ out to $z<0.03$ and extend to 1/2 void radius. The full CMB temperature map can be seen in the background. It can be clearly seen that the CMB temperature is mostly negative around galaxies and mostly positive in voids. \label{fig:tempmap}}
\end{figure}

\begin{table}[htbp]
\caption{Significances of void temperatures on wavelet transformed SMICA maps (with quadrupole removed). }
  \begin{center}
    \label{tab:significances_comb}
    \begin{tabular}{c|c|c|c|c}
wav. sc. & ph. sc. & $\Delta_{23}<0$ & $R>R_\mathrm{median}$ & $R>R_\mathrm{quartile}$\\
\hline
$4^\circ$ & $10^\circ$ & $2.2\sigma$ & $2.4\sigma$ &  $3.4\sigma$\\
$4^\circ$ (0.75$R$)& $10^\circ$ & $2.0\sigma$  & $2.4\sigma$ & $3.5\sigma$\\
$4^\circ$ (0.25$R$) & $10^\circ$ & $2.2\sigma$  & $2.3\sigma$ & $3.0\sigma$\\
$5^\circ$ & $12.5^\circ$ & $2.3\sigma$ & $2.6\sigma$ & $3.4\sigma$ \\
$3^\circ$ & $7.5^\circ$ &  $2.0\sigma$ & $2.3\sigma$ & $3.2\sigma$ \\
\hline
    \end{tabular}
  \end{center}
\tablefoot{Standard deviations are calibrated using 1000 simulated SMICA CMB maps. Results are given for different samples, wavelet scales and fractions of void radii (results given for 1/2 void radius when not otherwise specified).}
\end{table}

\begin{table}[htbp]
\caption{Significances of void temperatures on wavelet transformed SMICA maps (with quadrupole removed) using only voids with negative density along the line of sight, $\Delta_{LOS}<0$. }
  \begin{center}
    \label{tab:significances_los}
    \begin{tabular}{c|c|c|c|c}
wav. sc. & ph. sc. & $\Delta_{23}<0$ & $R>R_\mathrm{median}$ & $R>R_\mathrm{quartile}$\\
\hline
$4^\circ$ & $10^\circ$ & $2.3\sigma$ & $2.7\sigma$ &  $3.7\sigma$\\
$4^\circ$ (0.75$R$)& $10^\circ$ & $2.2\sigma$  & $2.8\sigma$ & $4.0\sigma$\\
$4^\circ$ (0.25$R$) & $10^\circ$ & $2.1\sigma$  & $2.3\sigma$ & $2.9\sigma$\\
$5^\circ$ & $12.5^\circ$ & $2.4\sigma$ & $2.7\sigma$ & $3.6\sigma$ \\
$3^\circ$ & $7.5^\circ$ &  $2.1\sigma$ & $2.6\sigma$ & $3.7\sigma$ \\
\hline
    \end{tabular}
  \end{center}
\tablefoot{Standard deviations are calibrated using 1000 simulated SMICA CMB maps. Results are given for different samples, wavelet scales and fractions of void radii (results given for 1/2 void radius when not otherwise specified).}
\end{table}

In Figure \ref{fig:tempmap} we show the SMICA Planck CMB temperature map masked in such a way that only the areas with galaxies and voids are clearly shown. For the galaxies, discs of 1Mpc around each galaxy is shown whereas for the voids, the discs correspond to half the void radius. Note that we have taken one realization of the \sparkling voids here as an illustration. It is clearly visibly by eye that the areas around galaxies are cold whereas the voids on average are warm. Note in particular the prominent hotspot at the right hand side, just below the galactic mask. This spot was noted already in \cite{spots} where the five most prominent peaks in the CMB were identified. Among these, there are 3 cold spots, all coinciding with areas of high galactic density seen in the upper panel of Figure \ref{fig:tempmap} and 2 hot spots where one is the prominent hot spot seen in the lower panel. The void corresponding to this hot spot is one of the largest voids in the nearby Universe with a void radius of $22$\,Mpc situated at $z=0.021$. Its position is $(l=263.0^\circ,b=-19.6^\circ)$ while the position of the CMB hot spot in wavelet space is $(l=264.1^\circ,b=-19.7^\circ)$, a distance of $1.1^\circ$ apart. As a counterpart, in \cite{garcialambas2023}, we found a significant coincidence between the CMB Cold Spot \citep{vielva} and the largest local supergroup, the Eridanus group. This large region shows signs of infalling dynamics as revealed by the large fraction of HI-deficient late-type galaxies. 

\begin{figure}[htbp]
  \begin{center}
    \includegraphics[width=0.8\linewidth]{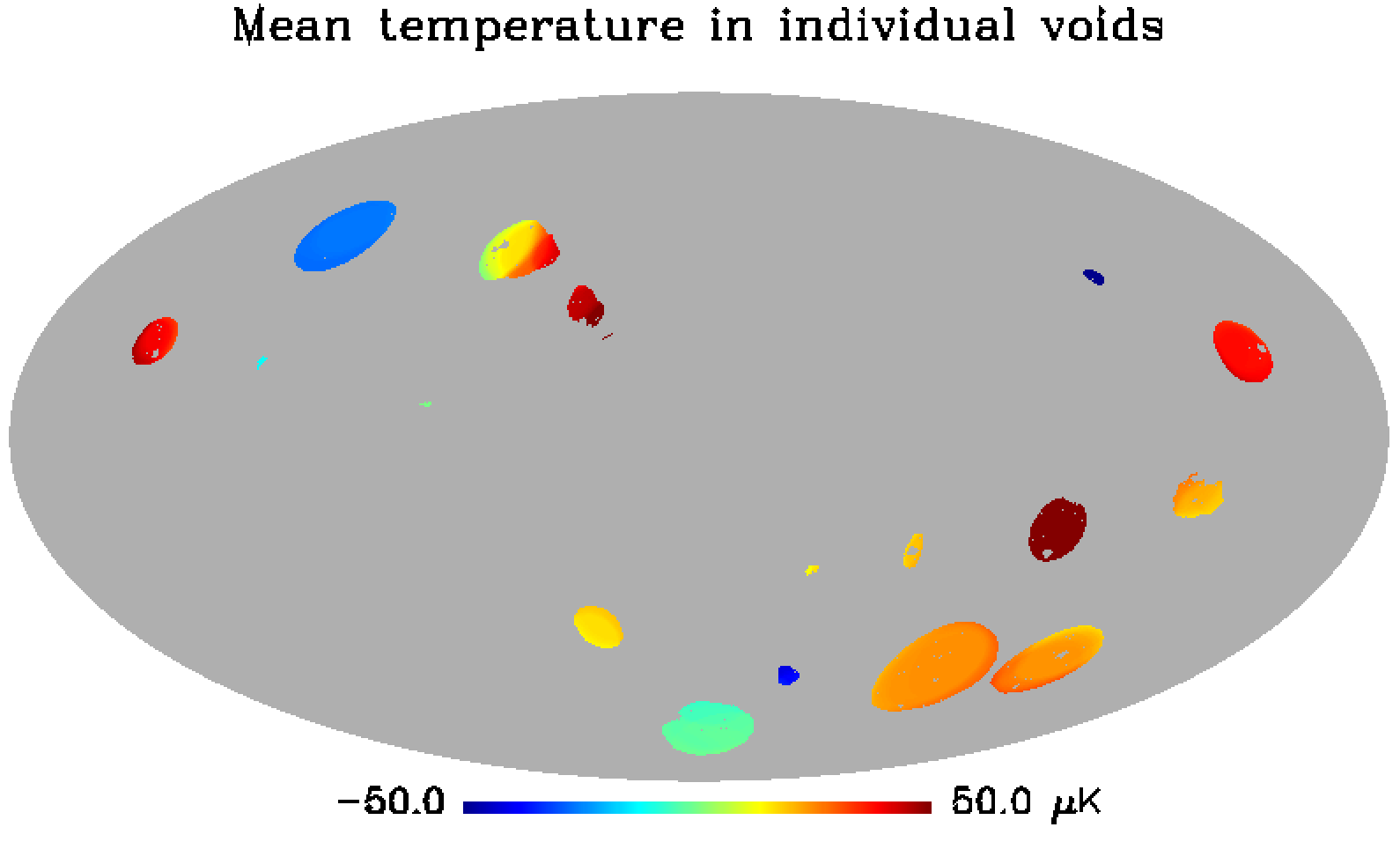}
  \end{center}
  \caption{The mean CMB temperature within $0.5R_\textrm{void}$ of each individual void for voids within the larges quartile. Voids are taken from the combined Sparkling sample where only voids appearing in $>50\%$ of the random realizations are included, corresponding to the voids having the largest weights in the void weight maps of Figure \ref{fig:voidmaps}. The fact that the void position and radius differ slightly for different realizations is reflected in the varying temperature within one single void for some of the voids. \label{fig:individual_void_temps} } 
\end{figure}

We measured the weighted mean temperature over all the voids in the data as well as in 10 000 SMICA simulations. In Table \ref{tab:significances_comb} we show the mean temperature for the data measured in terms of standard deviations estimated from simulations. We see that the CMB temperature in voids are significantly warmer than expected in simulations and the mean temperature increases with the void size, reaching $3.5\sigma$ for the largest voids. We see that already using all open voids, regardless of size, the mean temperature is always $\geq 2\sigma$. The significances are then gradually increasing with void size as expected if an ISW/RS like effect is at play. The signal is not a result of a few very hot voids, but is rather driven by a trend where the larger voids generally are warmer. This can be seen in Figure \ref{fig:individual_void_temps} where we show the mean temperature within $0.5R_\textrm{void}$ of each void, including only voids with weights larger than 0.5 in Figure \ref{fig:combinedmaps} and therefore having larger probability of belonging to an actual void. In the figure, which is limited to the voids in the largest quartile, we can see that almost all voids have a positive temperature.

To test consistency, in Table \ref{tab:significances_comb}, we also show the corresponding significances for other wavelet scales and radii showing generally consistent results. In Table \ref{tab:significances_los} we show the same results using only the voids with $\Delta_\mathrm{LOS}<0$ having an extended underdense region along the line of sight around the void. We see clearly that these voids show higher significances, reaching $4\sigma$ for the largest voids. Note further that all significance values here are taken with respect to a zero expectation value. In the $\Lambda$CDM model, the expected void temperatures given the standard ISW effect, is negative and about $-4\mu$K (judging from Figure 1 in \cite{iswfull} taken from a simulated slab of $100h^{-1}$Mpc at $z=0$) while we, on the contrary, have detect a large positive temperature. Therefore, comparing to the standard cosmological constant scenario, the actual significance of the deviation is even larger than the significances quoted here.

\begin{figure}[htbp]
  \includegraphics[width=0.9\linewidth]{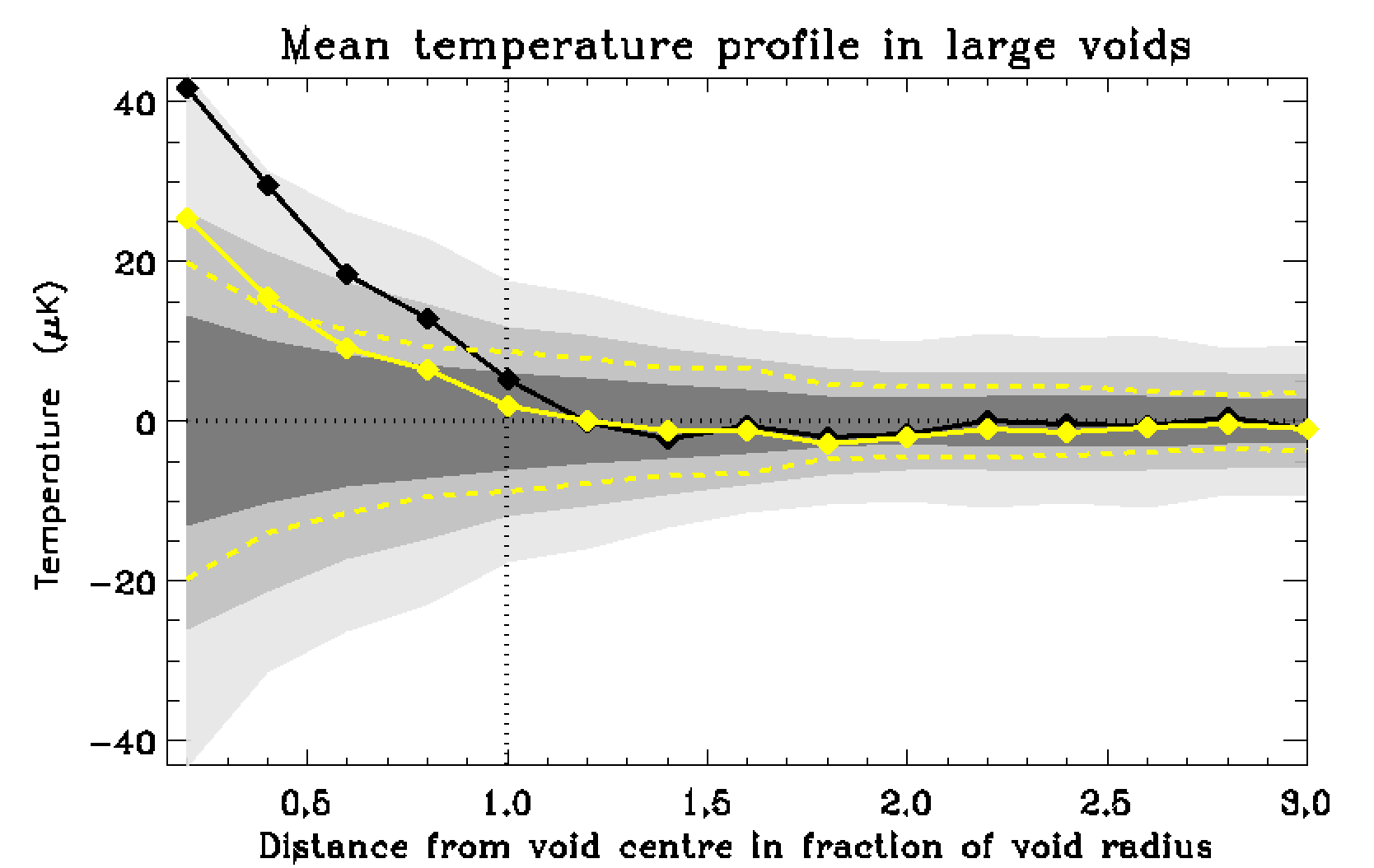}
  \caption{The mean CMB temperature in radial bins around voids: black line shows the temperature taken from the largest quartile and yellow line from all open nearby voids in the Planck SMICA CMB temperature fluctuation map with the quadrupole removed. Each bin is taken as 1/5 of the void radius. The grey bands show the $67$\%, $95$\% and $99.7$\% confidence intervals based on 10 000 simulated maps for the largest quartile of voids and the yellow dashed line show the corresponding $95\%$ interval for the sample of all voids. Note that the profile are only taken around voids with a central weight of $>0.5$ in the void weight map (Figure \ref{fig:combinedmaps}), ensuring that the voids used for measuring the profile is flagged as a void in most random realizations of \sparkling as well as in \revolver. \label{fig:voidprof}} 
\end{figure}

\begin{figure}[htbp]
  \includegraphics[width=0.9\linewidth]{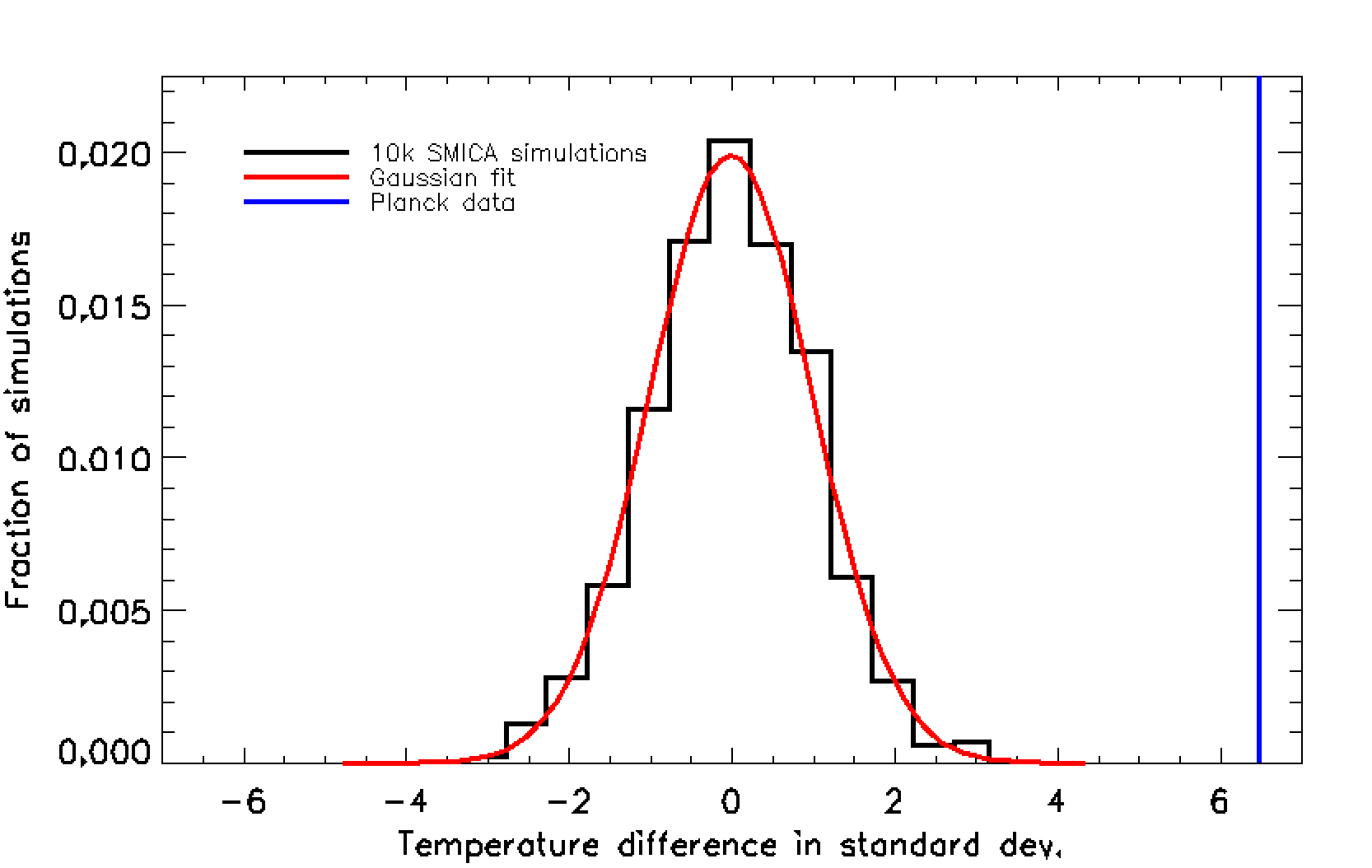}
  \caption{The variance weighted temperature difference between the mean CMB temperature in voids (inner temperature profile bin of the largest quartile of voids from Fig. \ref{fig:voidprof}) and the mean CMB temperature around galaxies (from H2025). The histogram shows the temperature difference in 10 000 SMICA simulations in terms of standard deviations calibrated on simulations. The red curve shows a Gaussian fit to the histogram. The blue line shows the corresponding difference in the data. \label{fig:tempdiff}} 
\end{figure}

Figure \ref{fig:voidprof} shows a detailed mean temperature profile. Note that this profile is obtained from the original CMB temperature map and not from the wavelet transformed map. We can see that the temperature goes to zero at a distance of about one void radius. The temperature profile in large voids in Figure \ref{fig:voidprof} shows that the inner parts of these voids have a mean temperature of about $\sim30-40\mu$K, similar to the $\sim-27\mu$K galaxy temperatures reported in H2025. In Figure \ref{fig:tempdiff}, we show the variance weighted temperature difference between the first bin of this temperature profile of the voids (for the largest quartile of voids) and the mean galaxy temperature for large spiral galaxies in massive filaments. The observed temperature difference of $69\mu $K is not found in any of the 10 000 simulated maps. Indeed the temperature difference observed in the data is more than $40\%$ larger than the largest difference found in simulations. A Gaussian fit to the temperature differences is drawn as a red line in the figure and shows that the distribution is close to Gaussian. As an approximate measure of significance, assuming a nearly Gaussian distribution, the temperature difference in the data is $6.5\sigma$ larger than expected. Note again that we have compared this difference of $6.5\sigma$ to a zero expectation value whereas we should have compared to the expectation given by the standard $\Lambda$CDM model giving a negative difference of about $-8\mu $K (judging again from Figure 1 in \cite{iswfull} where voids appear to have a temperature of $\sim-4\mu$K and overdensities about $\sim4\mu$K giving $8\mu$K difference) which gives an even larger tension with the standard model based on a cosmological constant.

\section{Discussion}
\label{sect:concl}

Based on the findings of H2025 where CMB temperatures are significantly colder around nearby galactic halos than expected in the standard $\Lambda$CDM model, we put forward the hypothesis that recent changes in the gravitational potential evolution give rise to a negative Integrated Sachs-Wolfe effect. Such a hypothesis predicts that not only are the CMB photons passing through overdensities cooled, but CMB photons passing through underdensities would be heated. We tested the hypothesis by measuring the mean temperature of the CMB temperature as measured by the Planck satellite in all nearby voids. We have shown that the mean CMB temperature in these voids is indeed larger than expected at the $2-4\sigma$ level depending on void size, fraction of void and line of sight densities around the void. If the ISW hypothesis is correct, one would expect an increasing signal for (1) increasing void sizes and (2) for voids with underdense global environment, particularly when these low density surroundings are elongated along the line of sight making the photon path inside an underdense environment longer. We have shown that both of these properties are seen in the data. We further estimate the temperature difference between the mean temperature around galactic halos and the mean temperature in voids and find a temperature difference which is $\sim6.5\sigma$ larger in the data as compared to 10 000 simulated maps, assuming a Gaussian distribution of temperature differences in simulations.

We now discuss several aspects of the statistics used and the reliability of the results.

\subsection{Look-elsewhere effect}

The importance of correcting for the impact of the number of choices made in the analysis is particularly important for serendipitous discoveries such as the cooling of CMB photons around galaxies described in L2023 and followed up in H2025. In the latter paper, a thorough analysis and correction for the look-elsewhere effect was performed and it was shown that even when taking all choices into account, the significance was still at the $p<0.0001$ level.

The discovery of warmer CMB temperature in nearby voids described in this paper was not serendipitous. Based on our previous findings, we made the conjecture that the observed cooling of CMB photons in nearby galaxies could have a possible origin in an anomalous ISW or RS effect. This conjecture was based on the fact that the frequency independence of the effect is difficult to explain in terms of other known mechanisms. In the lack of alternative explanations, we here test if an altered ISW/RS effect may cause the observed cooling.

Assuming this is indeed the case, we would expect a priori, based on what we know about ISW/RS effects,
\begin{itemize}
\item that voids in the same redshift range as the galaxies for which cooling is observed, and particularly open voids which we expect to be expanding and hence having an opposite change of potential than the contracting galactic filaments, would be warmer than expected for a standard model CMB sky.
\item that the central parts of the voids are warmer and gradually decreasing towards zero mean temperature outside of the void radius.
\item that the larger voids would be warmer than the smaller voids.
\item that voids with a surrounding underdense region elongated along the line of sight ($\Delta_\mathrm{LOS}<0$) would give a stronger effect 
\end{itemize}

These expectations strongly limits the number of choices which are possible to make when analysing the voids. Many of the choices made in this analysis could not have been very different due to these expectations. These are not parameters which can been fine-tuned to make a significant detection as we assume a possible physical mechanism, limiting these choices.

We nevertheless correct for the look-elsewhere-effect and for the different choices made for the case where there is indeed a possible range of parameters to adjust. The choices made in the analysis can be listed as
\begin{enumerate}
\item {The redshift range:} The cooling in galactic filaments have been observed for $z<0.04$ (H2025, DF2025) within which the 2MRS galaxy catalogue allows a reliable identification of filaments. The identification of voids needs a sufficient number density of tracers in order to obtain void detections not highly affected by Poisson uncertainties. In this context, the 2MRS catalogue has a fair completeness up to $z=0.03$, making the choice of redshift range strongly limited by available data and is not a parameter which we can change. We simply apply our analysis to the largest available volume for each tracer. Nevertheless, in the Appendix, we show that the results are consistent when we divide this redshift range in two sub ranges and analyse void temperatures in each range separately.

\item {The choice of Spherical Mexican Hat Wavelets:}  These wavelets have been the most applied type of symmetric wavelets used in CMB analysis. Their strong localization properties on the sphere \citep{smhw,sandro} is an advantage when searching for localized features on the sky. The wavelet coefficients of the SMHW are equivalent to the curvature of the field smoothed with a Gaussian kernel \citep{spots} such that the peaks of the SMHW are equal to the peaks in the local curvature of the smoothed CMB. For these reasons the SMHW was the first choice in our work. Given that they were used to discover the CMB cold spot \citep{vielva} and was used by the Planck team when testing for general deviations from isotropy and Gaussianity in the CMB \citep{iands15,iands18}, we expect the SMHW to be an optimal choice for identifying possible ISW void induced fluctuations in the temperature field. Furthermore, due to their strong ability to identify symmetrically shaped features, they were also shown optimal for identifying extragalactic point sources in CMB data \citep{mhw2} and was the basis for the Planck point source catalogue \citep{planck_ps}. Note that in the latter work, the MHW2 family of wavelets was used. We repeated our calculations with the version of MHW2 adapted to the sphere, the SMHW2 wavelets used for instance for point source detection in the CBASS survey \citep{smhw2}, and found slightly stronger void temperature significances than with the standard SMHW. Another class of wavelets used extensively in CMB analysis is the needlet class of wavelets \citep{needlet1, needlet2}, used for instance for the NILC component separation method for Planck data \citep{nilc}. The standard needlets have excellent localization properties in harmonic space, but less so in real space which is needed for our work \citep{sandro}. However, the Mexican needlets \citep{mexneed}, also used for the MC-NILC component separation \citep{mcnilc}, have stronger localization properties in real space, but the wavelet kernel for Mexican needlets is almost identical to the SMHW kernel \citep{sandro} and we therefore expect almost identical results.
\item {Using open voids:} This was already explained above. Open voids are expected to be expanding. Based on our conjecture of a non-standard ISW/RS effect, we would expect to see the effect in expanding voids due to the potential change. The same potential change is therefore not expected in closed voids \citep{sheth04,ceccarelli13,paz13}, and we therefore only analyse open voids in this work.
\item {The choice of a wavelet scale of 4 degrees:} As explained in detail above, the median radius of the voids in our sample is $10^\circ$, corresponding to a SMHW scale of $4^\circ$. Most voids therefore have an angular extension around this scale and is therefore a natural first choice. But in order to test the robustness of this wavelet choice, also for smaller and larger voids, we chose to look at the nearest scale below and above the median size. We therefore also list results for $3^\circ$ and $5^\circ$ scales and we account for the 3 possible choices of scales when correcting for the look-elsewhere-effect.
\item {The choice of void size samples}. We have chosen to look at (1) all open voids, (2) only the largest half of the voids, and (3) only the largest quartile of voids. Based again on our assumption about an ISW/RS-like effect, we expect a larger signal in larger voids. The first natural choice is therefore to look at the largest half. Finally, in order to study even larger voids, we divide the largest half again in two, looking therefor also at the largest quartile of void sizes. We correct for the look-elsewhere-effect by looking at these 3 choices of samples also in simulations.
\item {The fraction of the void radius used to measure temperature:} Assuming an ISW effect, we would expect the CMB temperature to be hotter in the centre of the void and gradually decrease towards the edges. As a first natural choice, we use half of the void radius to measure the temperature. A smaller fraction of the void radius gives less statistics and a larger fraction would dilute the signal including parts of the void were we expect less heating of the CMB photons. To test the stability of the results however, we also divide these ranges in two and measure the inner 1/4 of the voids as well as including 3/4 of the void radius. We correct for the look-elsewhere-effect for these three choices.
\end{enumerate}

\begin{figure}[htbp]
  \includegraphics[width=0.495\linewidth]{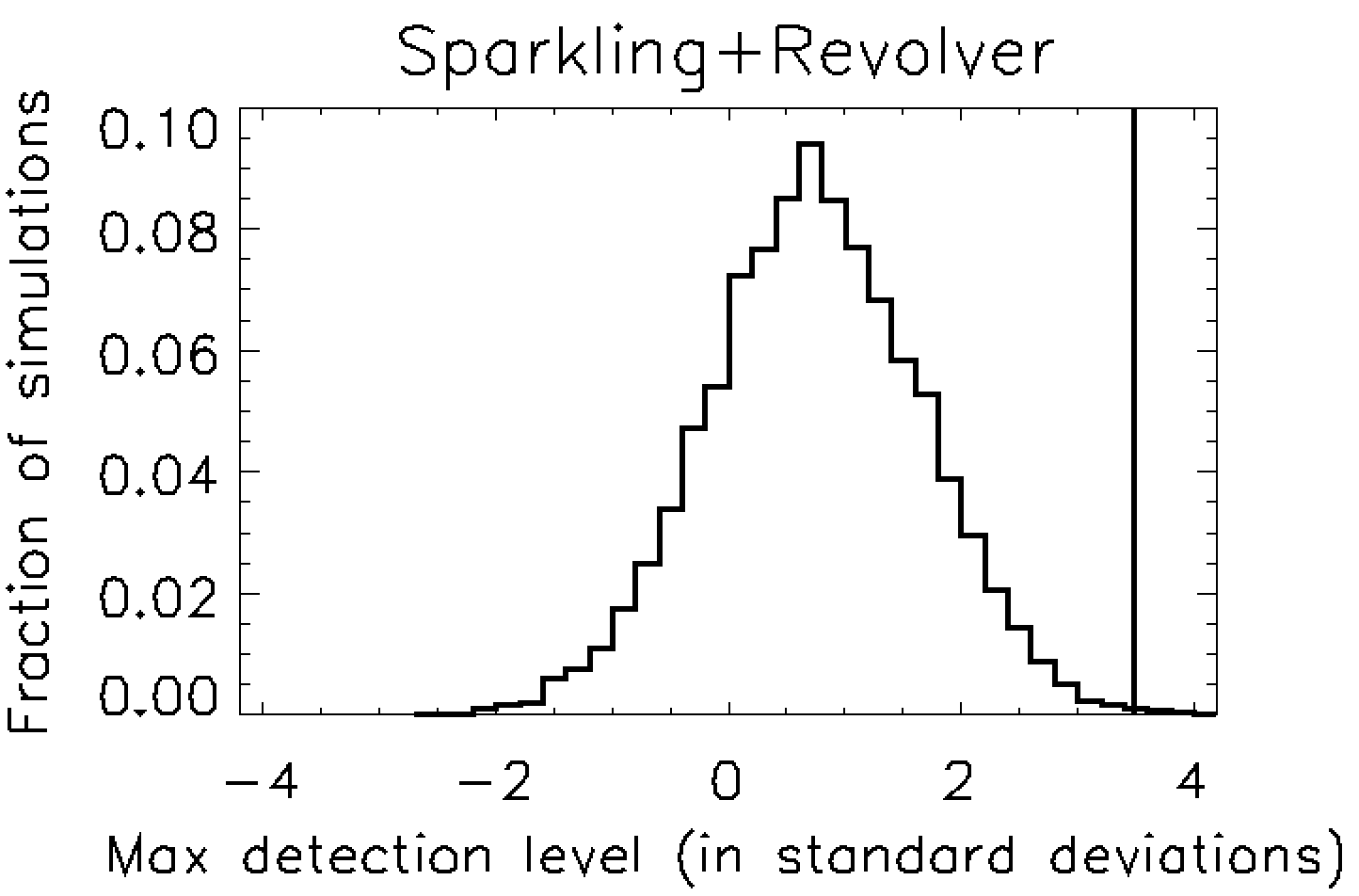}
  \includegraphics[width=0.495\linewidth]{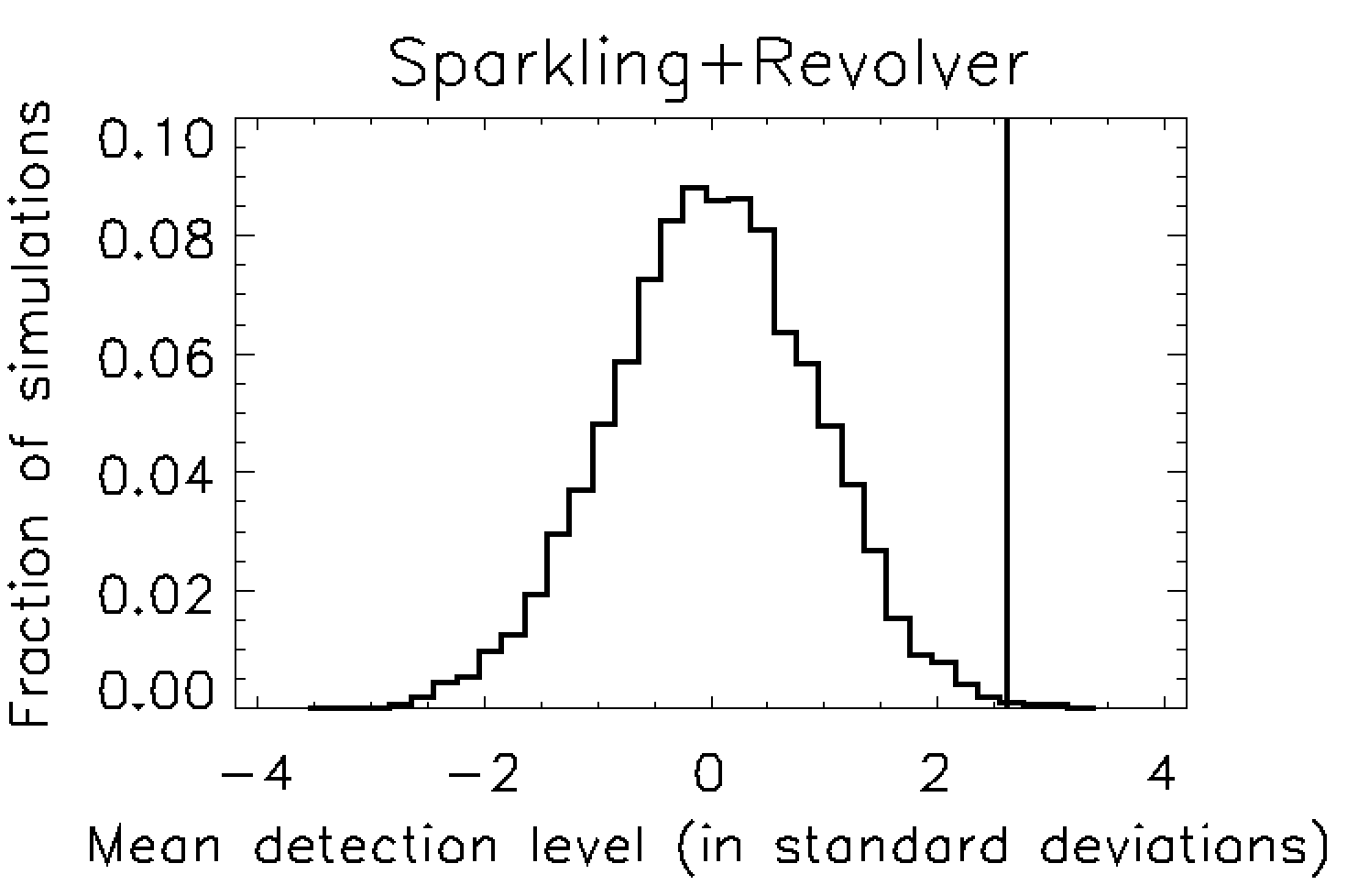}
  \includegraphics[width=0.495\linewidth]{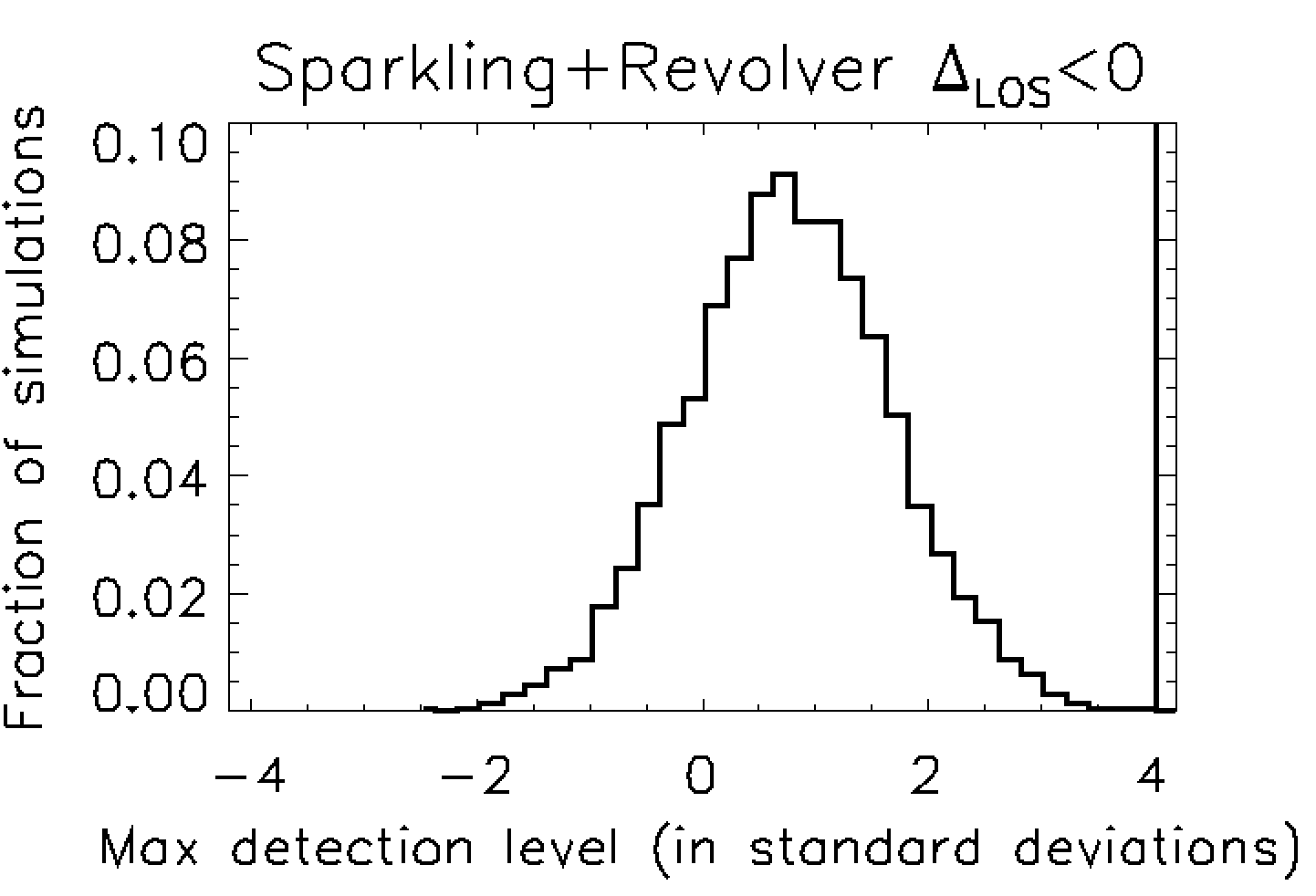}
  \includegraphics[width=0.495\linewidth]{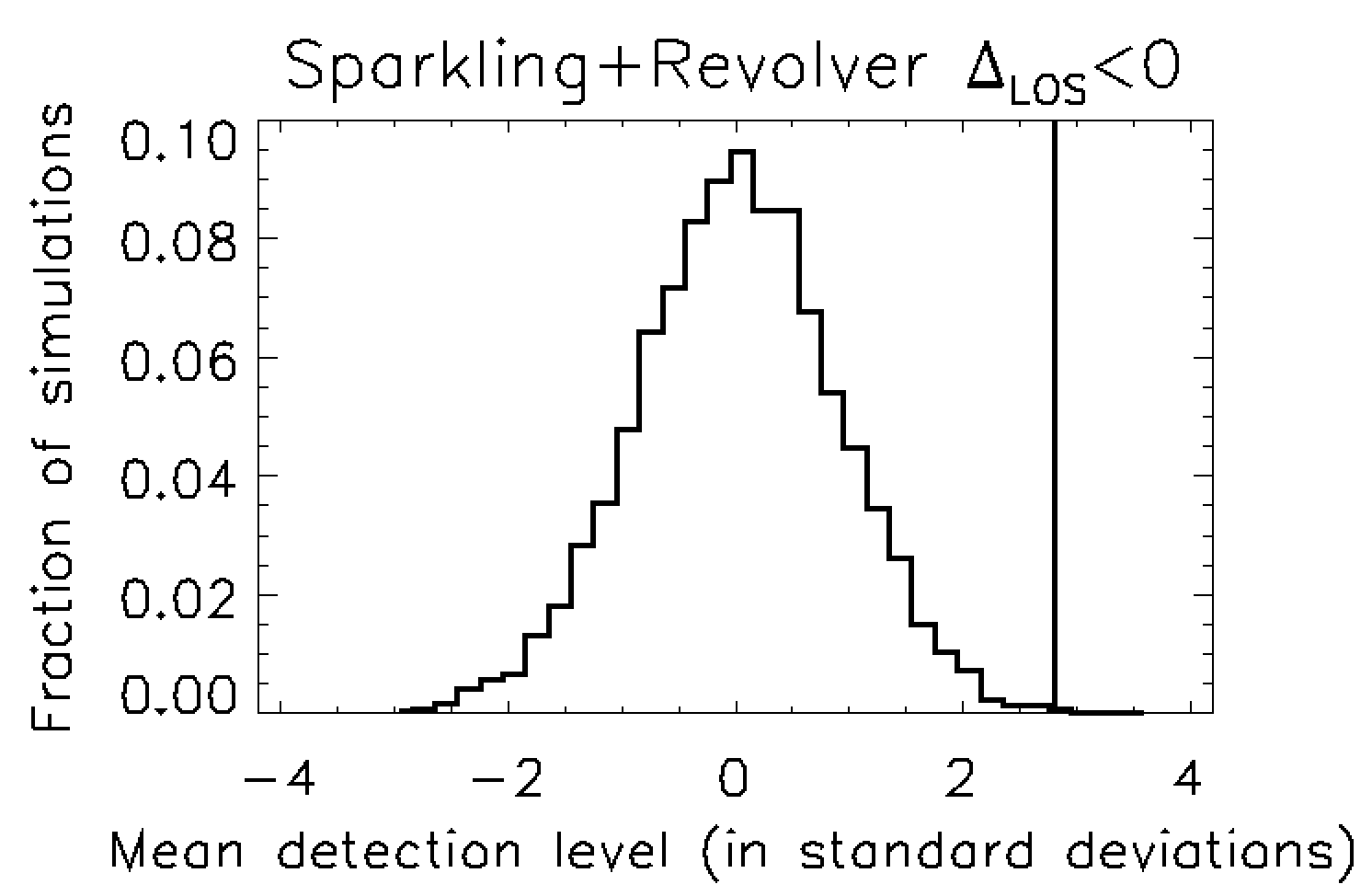}
\caption{The histograms show (upper left panel) the maximum detection level for void temperatures ($0.2\%$ of the simulations have higher maximum detection level), (upper right panel) the mean detection level for void temperatures ( $0.2\%$ of the simulations have higher mean detection level), (lower left panel) the maximum detection level for void temperatures using only voids with $\Delta_{LOS}<0$ ( $0.03\%$ of the simulations have higher maximum detection level) and (lower right panel) the mean detection level for void temperatures using only voids with $\Delta_{LOS}<0$ ($0.09\%$ of the simulations have higher mean detection level). All cases are based on 9000 simulations and the free parameters (void size, wavelet scale, ratio of void radius) are either varied to optimize the detection for each simulation (for maximum detection) or the significances are averaged over the parameter values in Table \ref{tab:significances_comb} (mean detection). The vertical lines show the same number for the real data. \label{fig:maxmean_detlevel}} 
\end{figure}

In Table \ref{tab:significances_comb}, we can see the significances of our detection for all the choices which have been made. We calculate the significance for all of the same choices in all simulations, giving a similar table for each of the 9000 simulated CMB maps (1000 simulations are used for obtaining standard deviations). The problem of the look-elsewhere-effect is the fact that the significance we find varies depending on the which of the numbers in the table we choose. We bypass this problem in two different ways
\begin{enumerate}
\item {By taking the maximum value over the significances in the table for each single simulation}. For each simulation individually, we find the optimal combination of these parameters which maximizes the detection level for that given simulation. This detection level (measured in number of standard deviations) is then recorded. In Figure \ref{fig:maxmean_detlevel} we show a histogram of this maximum detection level for each simulation. The vertical line shows the same number for the actual data. Only $0.2\%$ of 9000 simulations have a higher maximum detection level.
\item {By taking the mean value over the significances in the table for each single simulation}. Clearly statistical fluctuations may show a high detection level in some simulated maps, given a sufficient number of simulations. However, for a real stable detection, we would expect most of the numbers in the table to be high since the expected size of the voids varies within the given wavelet scales tested and that a real effect would show some significance for all the different size samples and not just one sample. We therefore also take the mean over all significances in the table for each simulation. In Figure \ref{fig:maxmean_detlevel} we show this mean value for each simulation. We see again that only $0.2\%$ of the simulations show a similar or higher mean detection level (about half of these simulations are among the $0.2\%$ also having a higher maximum detection). This shows that it is highly uncommon in a simulation to find both such a high maximum detection level as we find in the data and that the high detection level is equally stable as in the data. We see in the same figure that when imposing $\Delta_\mathrm{LOS}<0$, the number of simulations with similarly high maximum and mean detection level is even smaller.
\end{enumerate}

\subsection{Void temperatures in mock catalogues}

\begin{figure}[htbp]
  \begin{center}
  \includegraphics[width=0.8\linewidth]{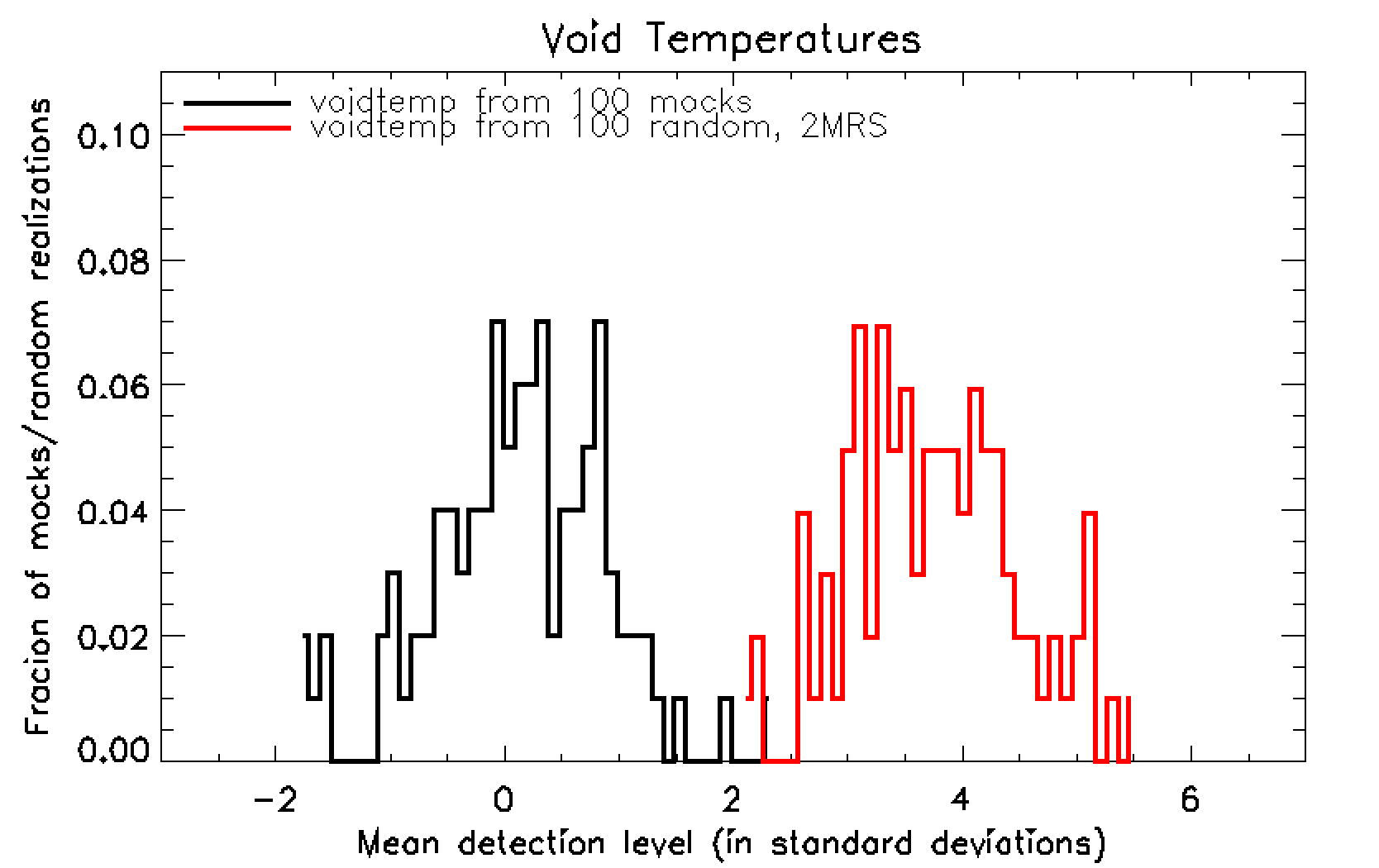}
  \includegraphics[width=0.8\linewidth]{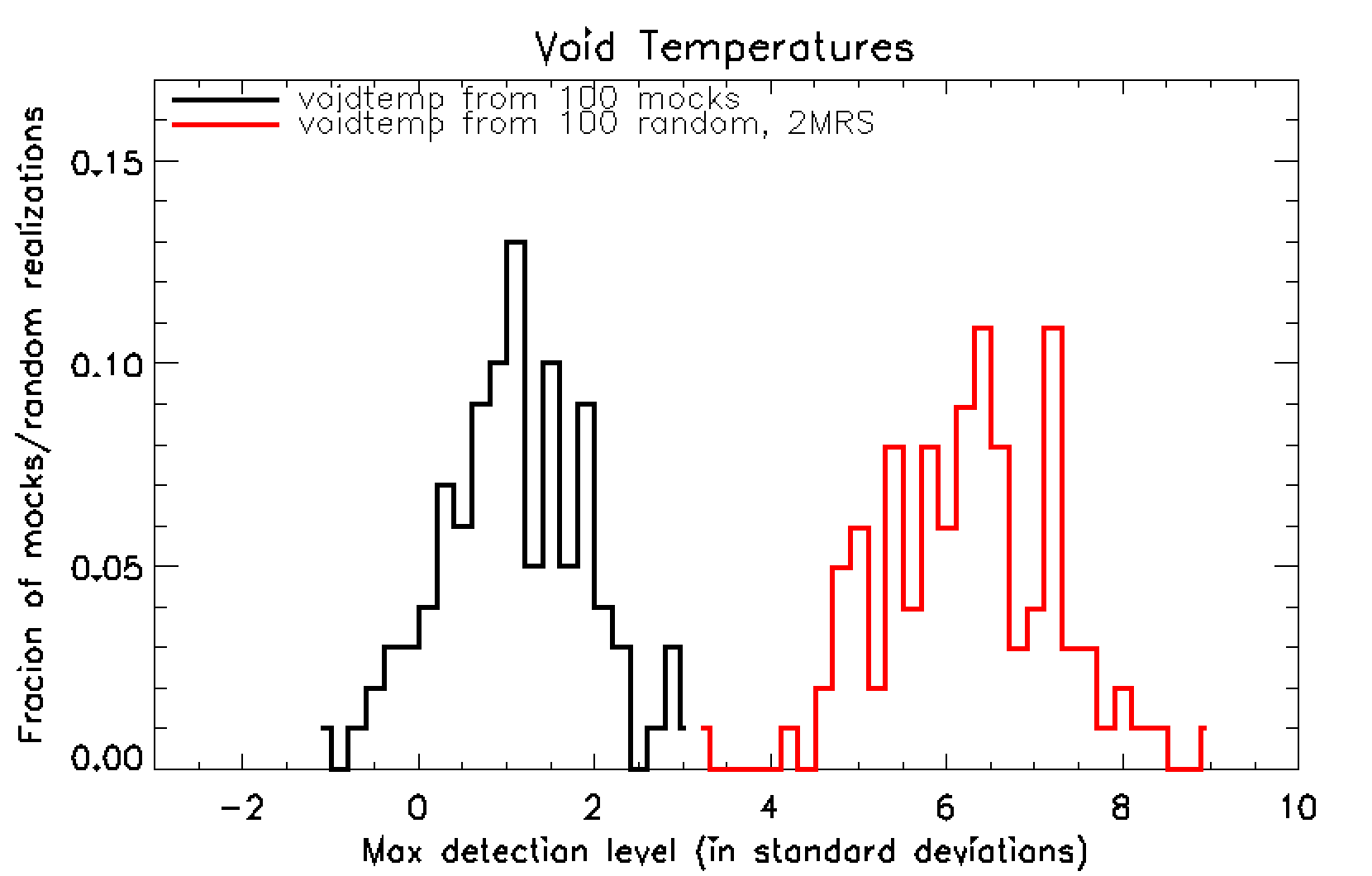}
  \end{center}
  \caption{ \label{fig:mocks} Mean (upper panel) and maximum (lower panel) detection levels in mocks (black histogram) and in 100 realizations of 2MRS voids. In the same way as in Fig. \ref{fig:maxmean_detlevel}, we have calculated the mean and maximum detection levels taken over the different parameter variations shown in Table \ref{tab:significances_comb} for the void sets of each individual mock catalogue and for 100 \sparkling void realizations based on 2MRS data. All temperatures are calculated on the actual Planck SMICA CMB map and significances calibrated on the corresponding CMB simulations.} 
\end{figure}

We have repeated the above described procedure for 100 mock void maps measuring the void temperature for Planck data for each mock with significances calibrated on void temperatures from CMB simulations for each individual mock void weight map. For each mock, we then created a table similar to Table \ref{tab:significances_comb} including the significance values for the same combinations of void sizes, fractions of radius and wavelet scales and from this table, we created the maximum and mean detection level, similar to Fig. \ref{fig:maxmean_detlevel}, for each mock. In Figure \ref{fig:mocks}, we compare these to the numbers obtained in the same manner for 100 realization of the \sparkling voids in 2MRS data. We can see that the maximum significances are higher in all 100 realizations of the observed voids than in any of the 100 mocks void sets, confirming our previous results.

\subsection{Dependence on CMB frequency map, cleaning process and Planck data release }

\begin{figure}[htbp]
  \includegraphics[width=0.9\linewidth]{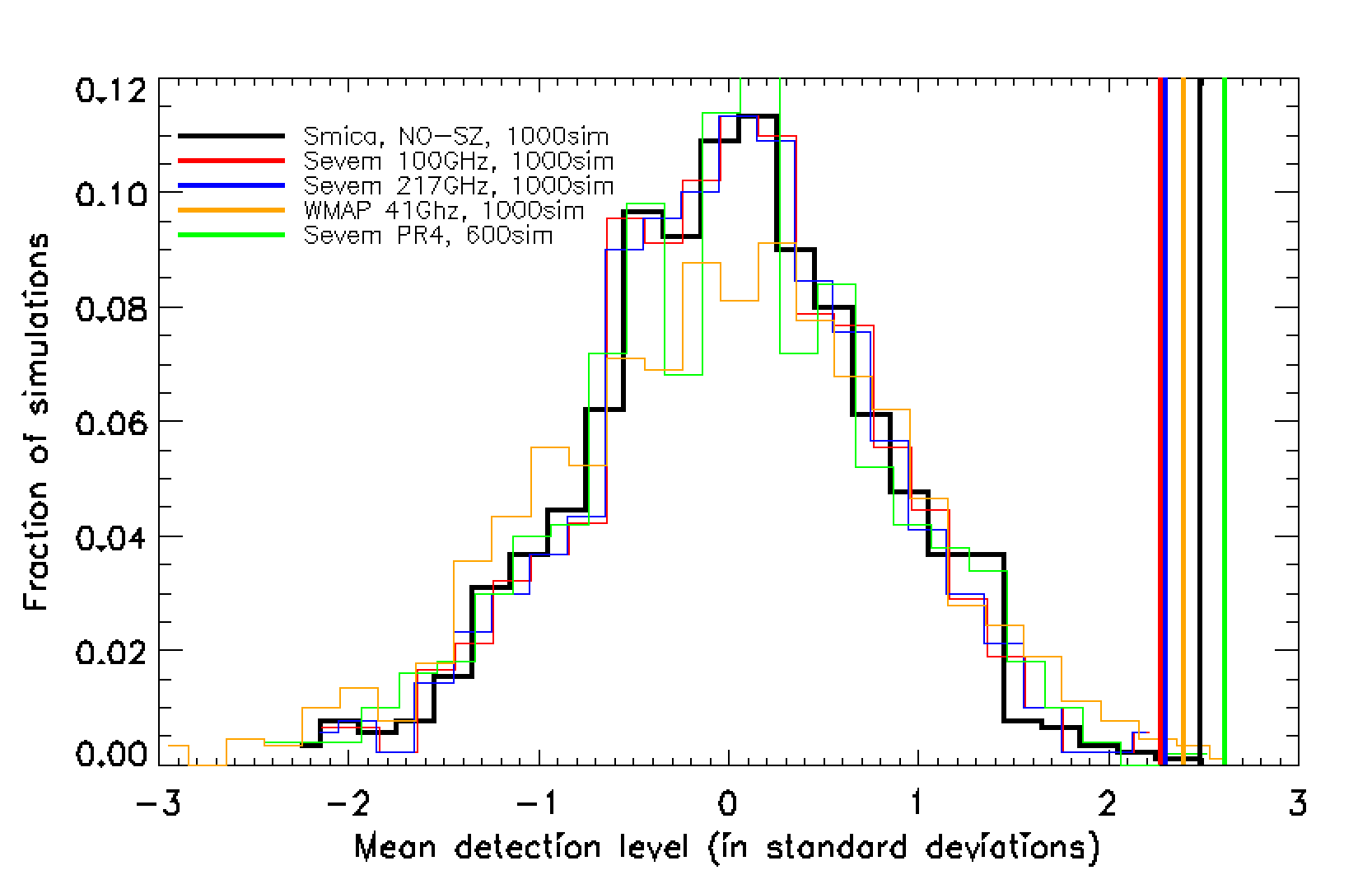}
  \caption{The histogram shows the mean detection level for void temperatures over the free parameters (void size, wavelet scale, ratio of void radius) for each of our 1000 simulations using only voids with $\Delta_{LOS}<0$. The vertical line shows the same number for the real data. Results are shown for the Smica NO-SZ map where contamination from the SZ-effect has been removed, the SEVEM cleaned frequency bands at 100GHz and 217GHz, the WMAP cleaned Q-band at 41Ghz and the SEVEM cleaned map from the last PR4 Planck release. \label{fig:mean_detlevel}} 
\end{figure}

In Figure \ref{fig:mean_detlevel}, we show the mean detection level, estimated in the same manner as in Figure \ref{fig:maxmean_detlevel}, but using different CMB maps. We have chosen to focus on the sample of voids giving the strongest detection, the voids with $\Delta_\mathrm{LOS}<0$ and test the robustness of the significance for this case. The black histogram shows again the SMICA component separation algorithm results from the Planck PR3 data release already shown in the lower right panel of Figure \ref{fig:maxmean_detlevel}, but now using the CMB map based on the SMICA algorithm optimized on the removal of the Sunyaev-Zeldovich (SZ) effect from the data.

The green histogram shows the corresponding results for the CMB map obtained from the SEVEM component separation algorithm for the newer Planck PR4 data release. Note however the much higher uncertainty for this result as only 600 simulated maps were produced for PR4 as opposed to the 10 000 SMICA maps in PR3. The remaining three histograms show the maps obtained from the individual $100$GHz and $217$Ghz Planck channels using the SEVEM component separation method as well as the WMAP $41$GHz channel. Note that the beam size, noise level and foreground residuals level are different in these different maps giving rise to different histograms. In particular, the WMAP 41GHz channel has a much larger beam and noise level than the Planck maps, giving rise to a broader histogram.

We have also analysed the void temperatures for the CMB maps created from the NILC and Commander component separation algorithms, using 1000 simulations of each. We found for both sets of maps that none of the 1000 simulated maps have a similarly high maximum or mean detection level as the data, consistent with the above results for SMICA and SEVEM. We see that for all these different cases, the significance of the results remains similar and robust.

\subsection{Possible temperature anticorrelations between areas of filaments and voids}

In the Introduction, we already mentioned our worry of a possible expected anticorrelation between galaxy temperatures and void temperatures. Due to the large scale fluctuations of the CMB, a simulation which by coincidence (or by a physical process) exhibits cold galaxy temperatures could have a higher probability of having hot void temperatures. The reason for this is more easily seen in Figure \ref{fig:tempmap}. There is little overlap between the parts of the sky occupied by the nearby galactic filaments (upper panel) and the parts of the sky occupied by the nearby voids (lower panel). In order to keep the mean temperature over the full sky zero while the CMB around galaxies is cold, the void areas need to be hot, creating a quadrupolar structure which would be evident if one combines the two panels of the figure. We show that this quadrupole indeed introduced a strong anticorrelation and that by removing the quadrupole from simulated CMB maps (and from the actual data), this anticorrelation is strongly reduced.

In order to test the presence of such an anticorrelation, we look at the outliers of the simulations: among our 10 000 simulated CMB maps, we choose the 100 simulations with the coldest galaxy temperature. If an anticorrelation is present, we would expect the void temperatures for these simulations to be warm and vice versa. For the main galaxy sample of H2025, there is no CMB simulation which exhibits a galaxy temperature as cold as in the real data. We therefore choose to look at the sample of all large spiral galaxies and not only those in the massive galactic filaments which are the ones giving the strongest detection. In this case, there are 15 of the 10 000 simulated CMB maps with a galaxy temperature colder than in the data. Using this sample, it is easier to compare with actual data: do simulated CMB maps with galaxy temperatures which are even colder than in the data also show void temperatures as hot as, or hotter, than in the data?

\begin{figure}[htbp]
  \begin{center}
  \includegraphics[width=0.9\linewidth]{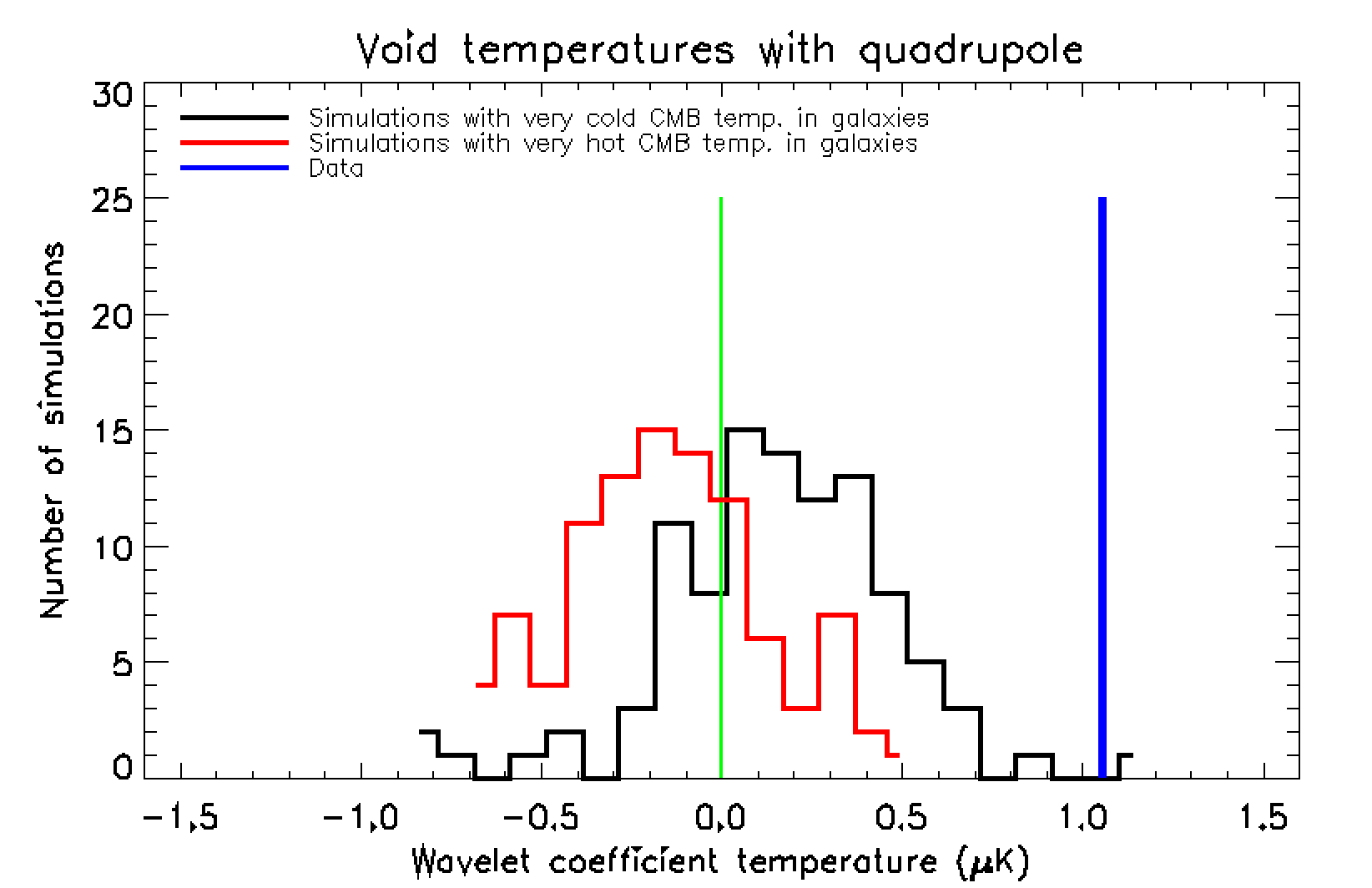}
  \includegraphics[width=0.9\linewidth]{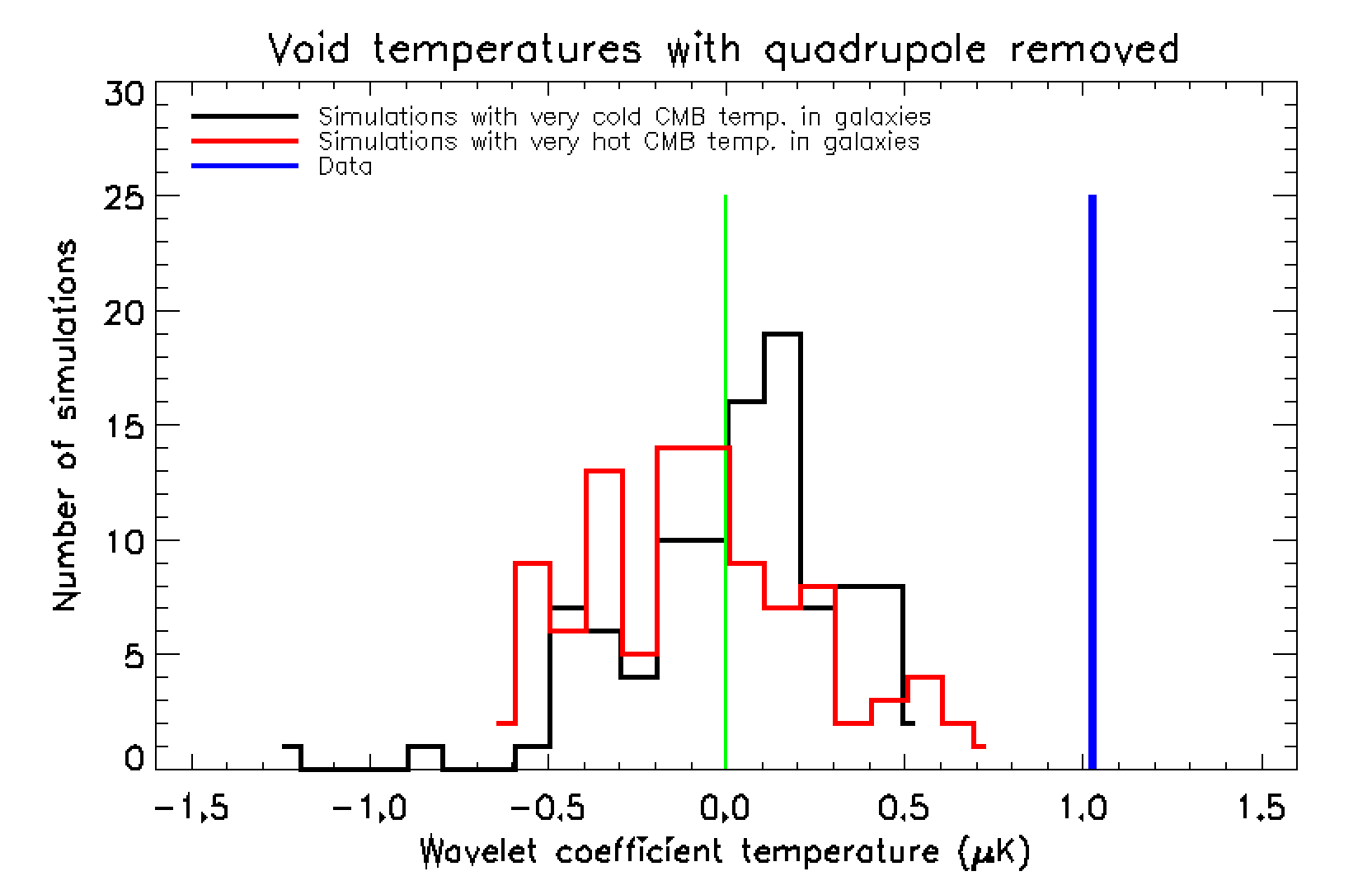}
  \end{center}
  \caption{The void wavelet temperatures of voids with $R_\mathrm{void}$ within the largest quartile in 100 simulated CMB maps without (upper panel) and with (lower panel) the quadrupole removed. Black histograms show the void temperatures of the 100 simulations (of 10 000) with coldest mean CMB temperature around galaxies and red histograms show the 100 simulations with the hottest temperature around galaxies. Blue line show the void temperature in Planck data. The histogram is shown with very small bins such that individual simulations are visible.\label{fig:correlations}} 
\end{figure}

In Figure \ref{fig:correlations} we show the void temperatures of the 100 CMB simulations which exhibit the coldest and the 100 CMB simulations which exhibit the hottest galaxy temperatures. Upper panel shows results without removing the quadrupole and the lower panel shows the results with quadrupole removed. We can clearly see that in the case where the quadrupole is maintained, the simulations with cold galaxy temperatures show hotter void temperatures and vice versa. There is even one of the cold galaxy simulations with a hotter void temperature that in the data. When the quadrupole is removed, as expected, this trend is almost completely removed. The histograms are superposed and one can even see that the largest tail of the simulations with cold galaxy temperatures also show cold, not hot, void temperatures.

Measuring the correlation coefficient between galaxy and void temperatures for the largest quartile of voids while keeping the quadrupole, we find an anticorrelation of $10\%$ to $30\%$ depending on the fraction of void radius used. After removing the quadrupole, this anticorrelation is reduced to between $2\%$ to $5\%$. These tests show the importance of removing the quadrupole from the CMB simulations before testing galaxy and void temperatures. We see from Figure \ref{fig:correlations} that the small anticorrelation remaining after subtracting the quadrupole, is unable to account for the high void temperatures found in the data.

\section{Conclusions}

We have found that CMB photons passing through nearby voids are significantly hotter than expected whereas CMB photons passing through the overdensities in the nearby galactic halos are correspondingly colder than in simulations (see Figure \ref{fig:tempmap} for an illustration). Our results show the opposite of the predicted ISW effect in the local Universe \citep{iswfull} due to a cosmological constant, but consistent with what one would observe if an unknown mechanism makes the gravitational potentials in the recent Universe ($z\lesssim0.1$) increase, causing a negative ISW effect. An independent confirmation of our results was reported in \cite{tunnels} where they confirm this sign change of the ISW effect at low redshift using low density cylinders elongated along the line of sight, cosmic tunnels, instead of voids.

Recent results from DESI \citep{desi1,desi2} show evidence for dynamical dark energy and possible phantom crossing at higher $z$. In \cite{nonaccelerating}, supernova data were corrected for progenitor age bias, giving strong evidence for dynamical dark energy and a non-accelerating Universe at low redshift, and thereby again opening the possibility for a gravitational potential evolution and ISW effect in the nearby Universe which is not consistent with the standard $\Lambda$CDM prediction.

At high redshifts ($z\gtrsim0.1$) the positive ISW effect consistent with a cosmological constant has been detected at the $3-4\sigma$ level, but with several reports of a significant excess in some data sets and redshift ranges \citep{iswpuzzle}. In \cite{iswlocal2} the correlation between the integrated projected mass density from 2MASS in the redshift range $z=[0,0.1]$ using photometric redshifts and WMAP CMB data was found to be within the expectation for the ISW effect at the largest scales. The main difference from our work is the use of photometric redshifts instead of the spectroscopic redshifts, the focus on much larger angular scales and that they average over a much larger redshift range than in this work. Note further \cite{bossquasars} where a negative ISW effect is claimed also for $z>1.5$.

The DESI results seem to indicate that the properties and evolution of the dark energy may be poorly understood. Could unknown properties of dark energy give rise to strongly increasing gravitational potentials at small redshifts? An example of a theory which predicts such a behaviour is given in \cite{ula}. In that work, ultra light axions play the role of dark energy at early times and behave like dark matter in recent times thereby increasing the potentials and giving rise to the ISW effect changing sign as shown in their Figure 5. Also in theories of Galilean gravity such a sign change can occur \citep{galilean1,galilean2}. Further work is necessary to test whether these theories or similar mechanism are fully consistent with the observed signal and other cosmological observations.

We have shown that many aspects of our findings are consistent with the expected behaviour of a (inverse) ISW effect, (1) that the signal is frequency independent, (2) that the effect increases with increasing void sizes, (3) the rise in temperature is higher in the central parts of the voids and falls to zero at about one void radius and (4) that the effect is stronger in voids with an extended underdense region along the line of sight. The results have been shown robust to changes in versions of the CMB Planck map, the component separation algorithm and the Planck frequency used for creating the CMB map. In the Appendix, we also show that the results persist when splitting the data in two separate galactic hemispheres, in two different redshift ranges and when using only \sparkling or only \revolver voids separately. For the voids having an underdensity along the line of sight, if we assume the standard ISW scenario with a cosmological constant, we would expect a lower temperature in these voids and thereby a smaller detection. We find, on the contrary, a larger temperature and a larger detection in these voids compared to other voids. This would be more consistent with an anomalous dark energy and ISW effect where an underdensity increases the CMB temperature.

The amplitude of the ISW effect at larger redshift is normally calculated by cross-correlating the maps of large scale structure, often represented as maps of galaxy number density, with the CMB fluctuations, assuming linear evolution of density perturbations. Due to the proximity of structures at $z<0.03$, such a cross-correlation would require detailed knowledge of the density field at non-linear scales since individual galaxies and voids here occupy a substantial number of pixels, often spanning several degrees on the sky. A more accurate comparison with the expected ISW effect is therefore deferred to future study as it requires calculating the ISW effect using a detailed model of the nearby density field.

\begin{acknowledgements}
Thanks to Dennis Fremstad, Hans Winther, Farbod Hassani, Abdolali Banihashemi, Prajwal Puttasiddappa and Sigurd K. Næss for discussions and suggestions. Results in this paper are based on observations obtained with Planck (http://www.esa.int/Planck), an ESA science mission with instruments and contributions directly funded by ESA Member States, NASA, and Canada. We acknowledge the use of NASA’s WMAP data from the Legacy Archive for Microwave Background Data Analysis (LAMBDA), part of the High Energy Astrophysics Science Archive Center (HEASARC). The simulations were performed on resources provided by UNINETT Sigma2 - the National Infrastructure for High Performance Computing and  Data Storage in Norway". Some of the results in this paper have been derived using the HEALPix package \citep{healpix}
\end{acknowledgements}

\begin{appendix}

\section{Testing validity of the results on data splits}

Here we show detailed results of several data splits. In this way, we can test consistency in a new way: If the observed effect originates form a physical process and not from coincidence, one would expect the observed heating of voids to be present throughout all parts of the data. Here we look at the void temperature results using each void finder algorithm separately, measuring the void temperature in the northern and the southern galactic hemisphere separately and finally measuring the void temperatures in two different redshift shells. Dividing the data in two, there is less data in each part and therefore larger uncertainties, impeding the division of the data in even smaller parts. With larger uncertainties, the distributions from simulations become broader and more simulations spuriously have high values, similar to the data. We therefore expect lower significance values in each independent part of the data. The high significances in the main part of the paper therefore appear as a result of combining data where all parts of the data show similar trends and the combination results in lower uncertainties and higher significances.

We first look at results using \sparkling and \revolver voids separately in Table \ref{tab:significances_sparkling} and Table \ref{tab:significances_revolver}. We see that in both cases, although the voids positions and sizes are quite different, we find $2-3\sigma$ detections. As can be seen in Figure \ref{fig:voidmaps}, the uncertainty for \revolver results are larger given the small area on the sky occupied by the voids. For \revolver the significances for the smaller voids are less than $2\sigma$ while the larges quartile of voids shows larger significances than the corresponding void sizes for \sparkling. For \sparkling, we also see in Table \ref{tab:significances_sparkling_los} the general trend that voids with $\Delta_\mathrm{LOS}<0$ show higher CMB temperatures. For \revolver, most voids satisfy this condition and no significant difference is therefore seen when limiting $\Delta_\mathrm{LOS}<0$ for \revolver voids (results not shown).

\begin{table}[htbp]
\caption{Significances of \sparkling void temperatures on wavelet transformed SMICA maps (with quadrupole removed). }
  \begin{center}
    \label{tab:significances_sparkling}
    \begin{tabular}{c|c|c|c|c}
wav. sc. & ph. sc. & $\Delta_{23}<0$ & $R>R_\mathrm{median}$ & $R>R_\mathrm{quartile}$\\
\hline
$4^\circ$ & $10^\circ$ & $2.2\sigma$ & $2.3\sigma$ &  $2.9\sigma$\\
$4^\circ$ (0.75$R$)& $10^\circ$ & $2.0\sigma$  & $2.1\sigma$ & $3.0\sigma$\\
$4^\circ$ (0.25$R$) & $10^\circ$ & $2.2\sigma$  & $2.2\sigma$ & $2.5\sigma$\\
$5^\circ$ & $12.5^\circ$ & $2.4\sigma$ & $2.4\sigma$ & $3.0\sigma$ \\
$3^\circ$ & $7.5^\circ$ &  $2.0\sigma$ & $2.1\sigma$ & $2.8\sigma$ \\
\hline
    \end{tabular}
  \end{center}
\tablefoot{Standard deviations are calibrated using 1000 simulated SMICA CMB maps. Results are given for different samples, wavelet scales and fractions of void radii (results given for 1/2 void radius when not otherwise specified). }
\end{table}

\begin{table}[htbp]
\caption{Significances of \revolver void temperatures on wavelet transformed SMICA maps (with quadrupole removed). }
  \begin{center}
    \label{tab:significances_revolver}
    \begin{tabular}{c|c|c|c|c}
wav. sc. & ph. sc. & $\Delta_{23}<0$ & $R>R_\mathrm{median}$ & $R>R_\mathrm{quartile}$\\
\hline
$4^\circ$ & $10^\circ$ & $1.0\sigma$ & $1.7\sigma$ &  $3.2\sigma$\\
$4^\circ$ (0.75$R$)& $10^\circ$ & $0.7\sigma$  & $1.9\sigma$ & $3.4\sigma$\\
$4^\circ$ (0.25$R$) & $10^\circ$ & $1.4\sigma$  & $1.6\sigma$ & $2.7\sigma$\\
$5^\circ$ & $12.5^\circ$ & $1.1\sigma$ & $1.8\sigma$ & $3.3\sigma$ \\
$3^\circ$ & $7.5^\circ$ &  $0.9\sigma$ & $1.7\sigma$ & $3.1\sigma$ \\
\hline
    \end{tabular}
  \end{center}
\tablefoot{Standard deviations are calibrated using 1000 simulated SMICA CMB maps.  Results are given for different samples, wavelet scales and fractions of void radii (results given for 1/2 void radius when not otherwise specified). Note that \revolver voids occupy much less sky fraction than the \sparkling voids and therefore have a much larger uncertainty in CMB temperature and measured significance.}
\end{table}

\begin{table}[htbp]
\caption{Significances of \sparkling void temperatures on wavelet transformed SMICA maps (with quadrupole removed) using only voids with $\Delta_{LOS}<0$.  }
  \begin{center}
    \label{tab:significances_sparkling_los}
    \begin{tabular}{c|c|c|c|c}
wav. sc. & ph. sc. & $\Delta_{23}<0$ & $R>R_\mathrm{median}$ & $R>R_\mathrm{quartile}$\\
\hline
$4^\circ$ & $10^\circ$ & $2.4\sigma$ & $2.5\sigma$ &  $3.1\sigma$\\
$4^\circ$ (0.75$R$)& $10^\circ$ & $2.5\sigma$  & $2.6\sigma$ & $3.6\sigma$\\
$4^\circ$ (0.25$R$) & $10^\circ$ & $2.3\sigma$  & $2.3\sigma$ & $2.6\sigma$\\
$5^\circ$ & $12.5^\circ$ & $2.4\sigma$ & $2.4\sigma$ & $2.9\sigma$ \\
$3^\circ$ & $7.5^\circ$ &  $2.4\sigma$ & $2.5\sigma$ & $3.2\sigma$ \\
\hline
    \end{tabular}
  \end{center}
\tablefoot{Standard deviations are calibrated using 1000 simulated SMICA CMB maps. Results are given for different samples, wavelet scales and fractions of void radii (results given for 1/2 void radius when not otherwise specified).}
\end{table}

\begin{figure}[htbp]
  \includegraphics[width=0.9\linewidth]{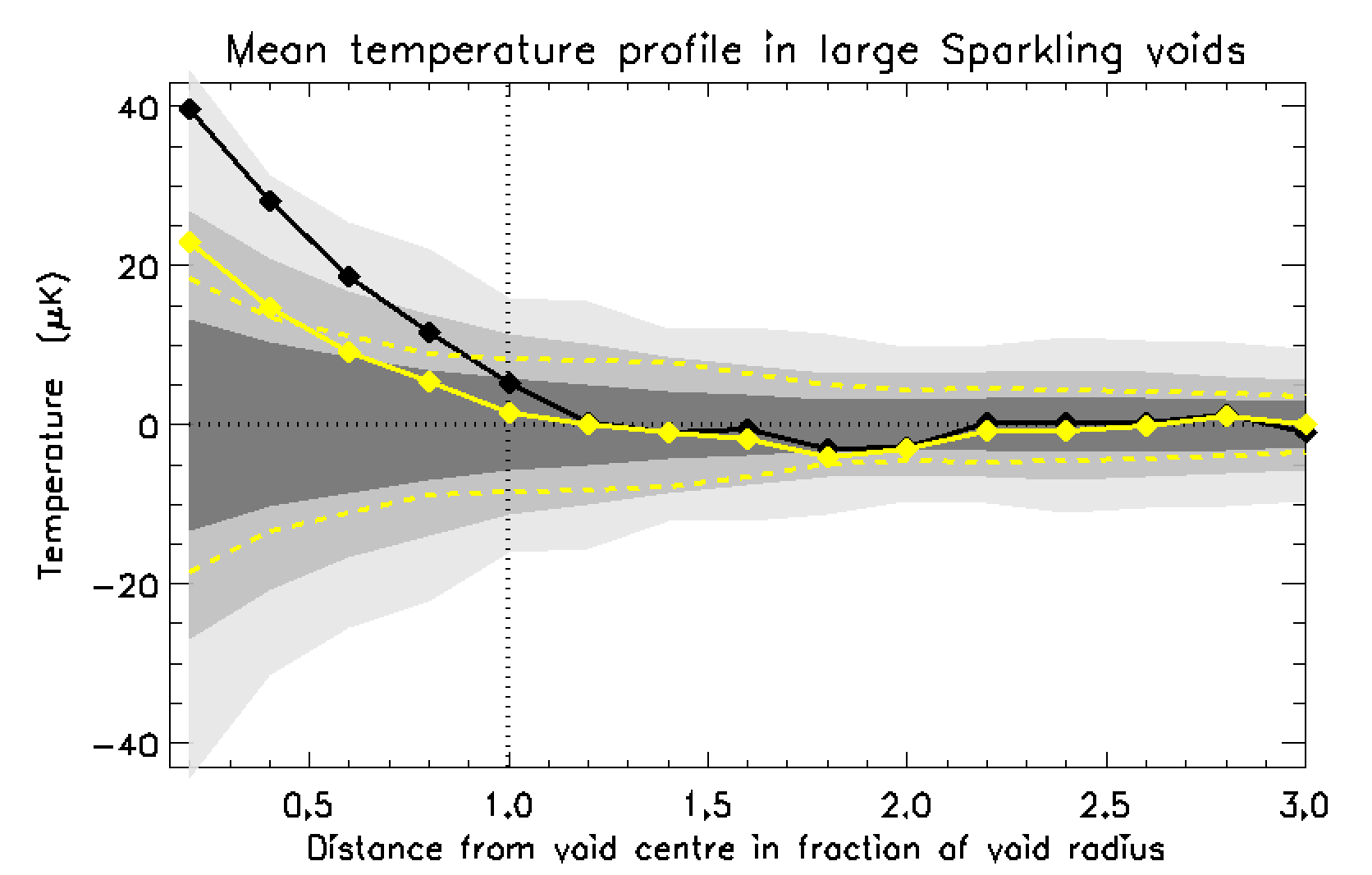}
  \caption{Same as Figure \ref{fig:voidprof} but only for \sparkling voids. \label{fig:voidprof_sparkling}} 
\end{figure}

\begin{figure}[htbp]
  \includegraphics[width=0.9\linewidth]{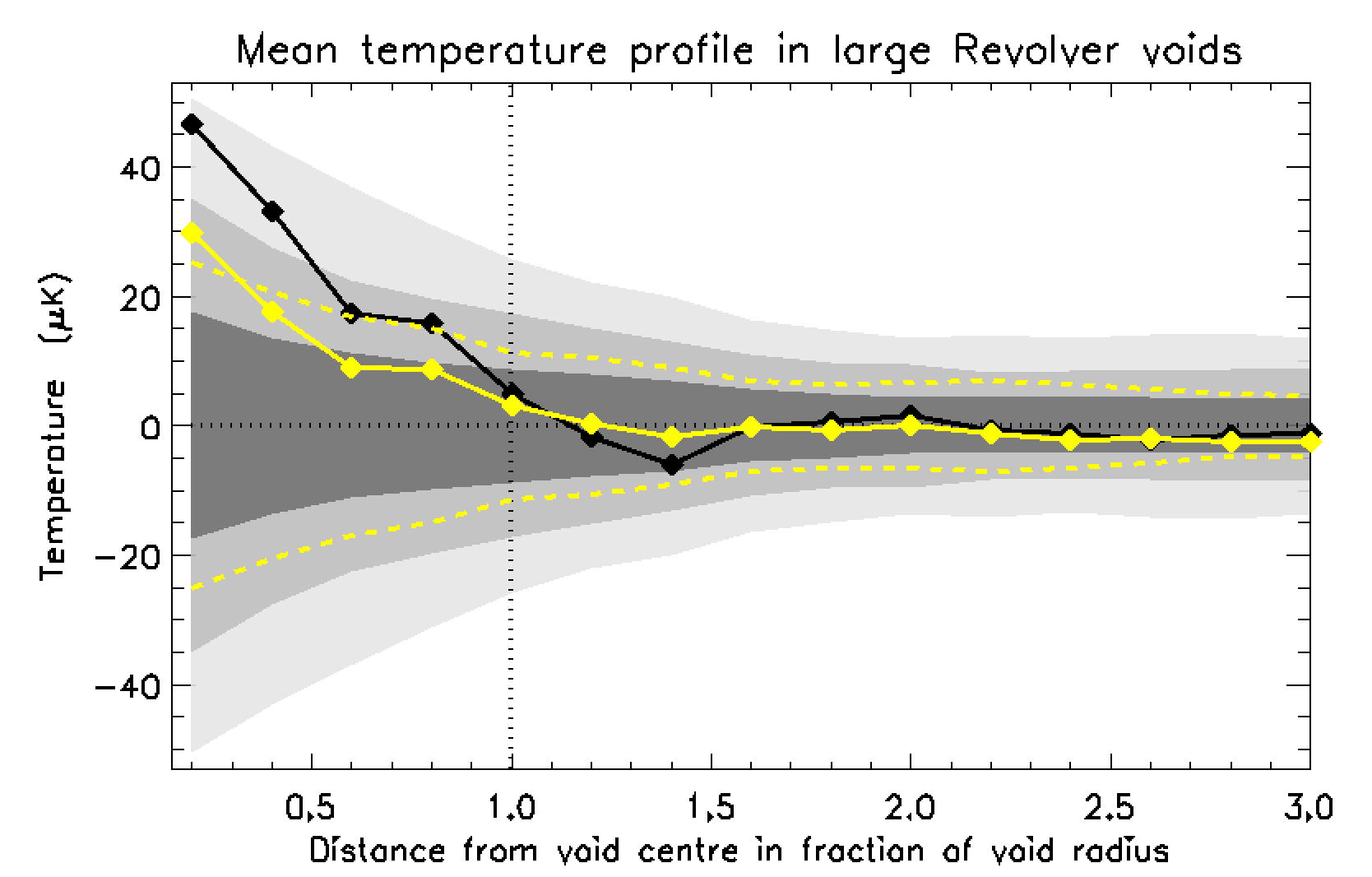}
  \caption{Same as Figure \ref{fig:voidprof} but only for \revolver voids. \label{fig:voidprof_revolver}} 
\end{figure}

In Figure \ref{fig:voidprof_sparkling} and Figure \ref{fig:voidprof_revolver} we show the temperature profiles for \sparkling and \revolver voids respectively. Both sets of voids separately show consistent profiles to the combined case in Figure \ref{fig:voidprof}.

\begin{figure}[htbp]
  \includegraphics[width=0.495\linewidth]{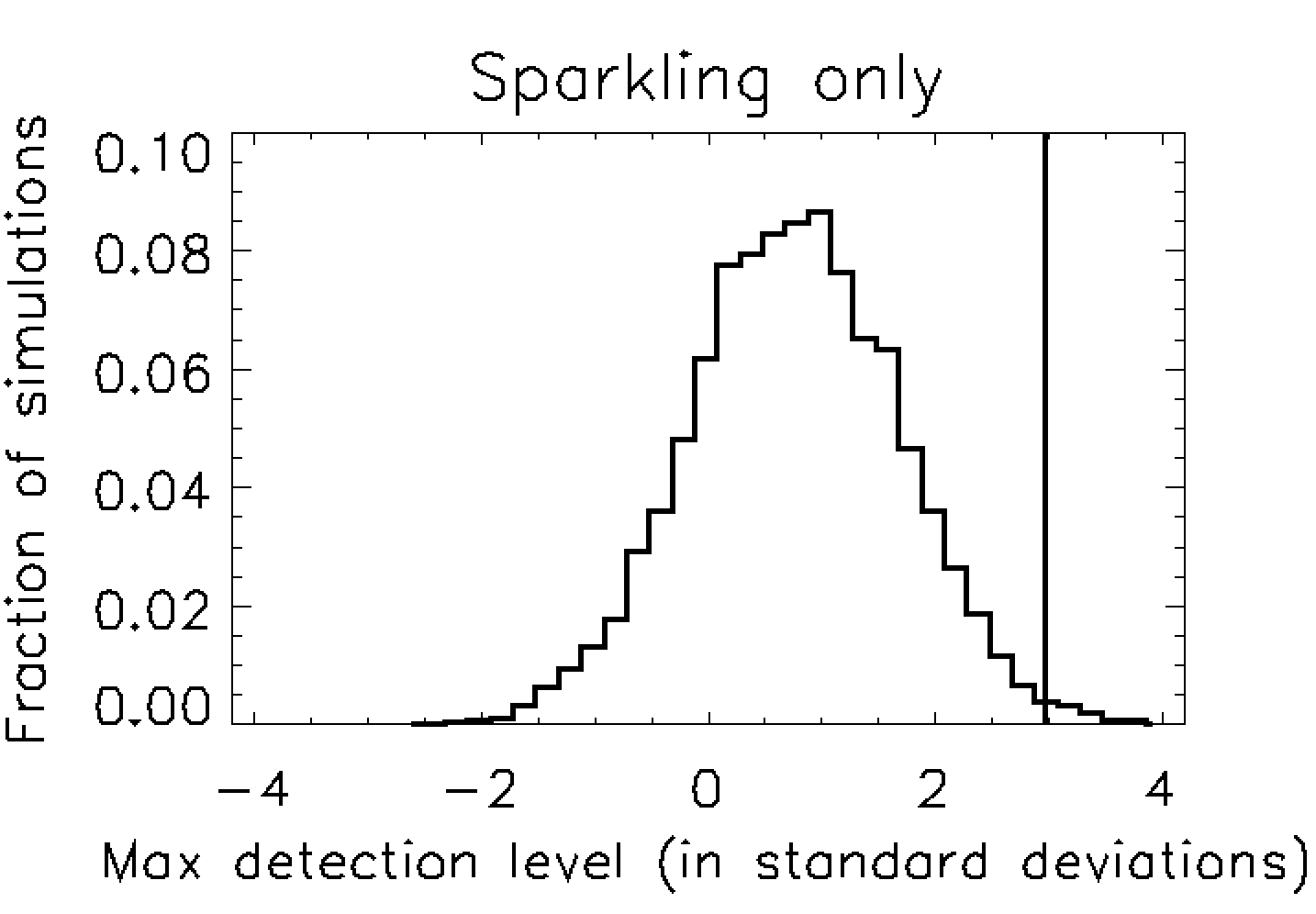}
  \includegraphics[width=0.495\linewidth]{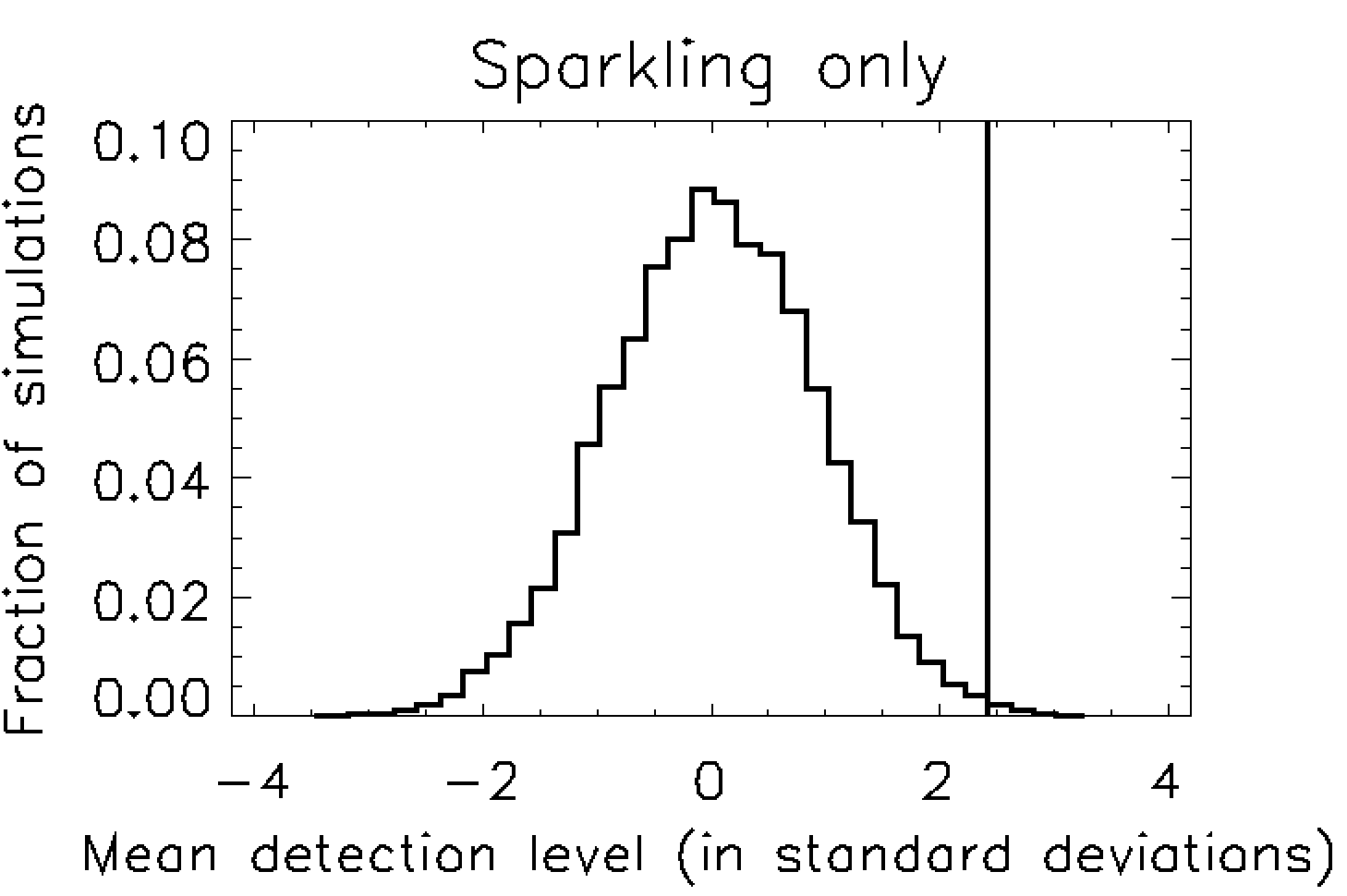}
  \includegraphics[width=0.495\linewidth]{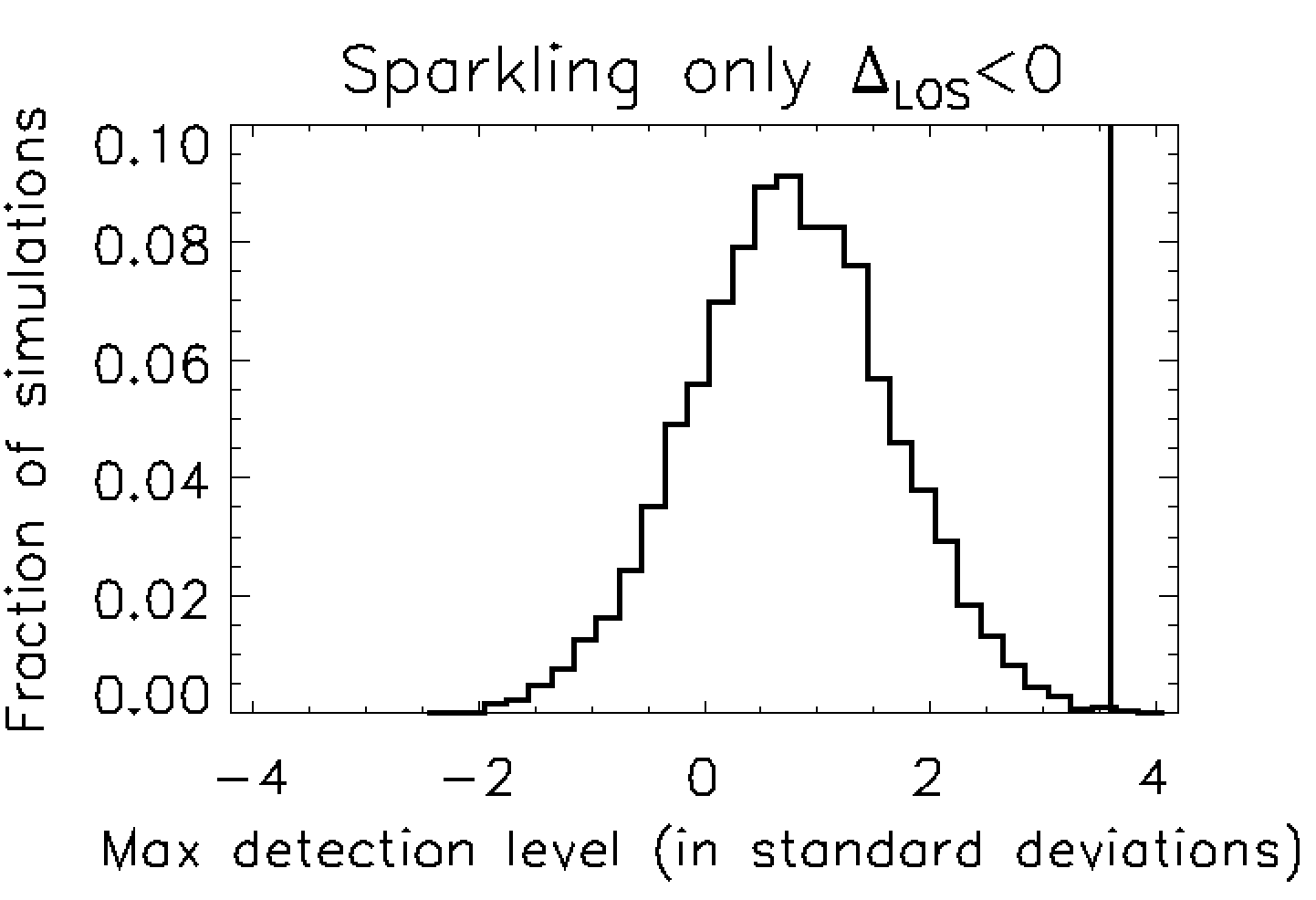}
  \includegraphics[width=0.495\linewidth]{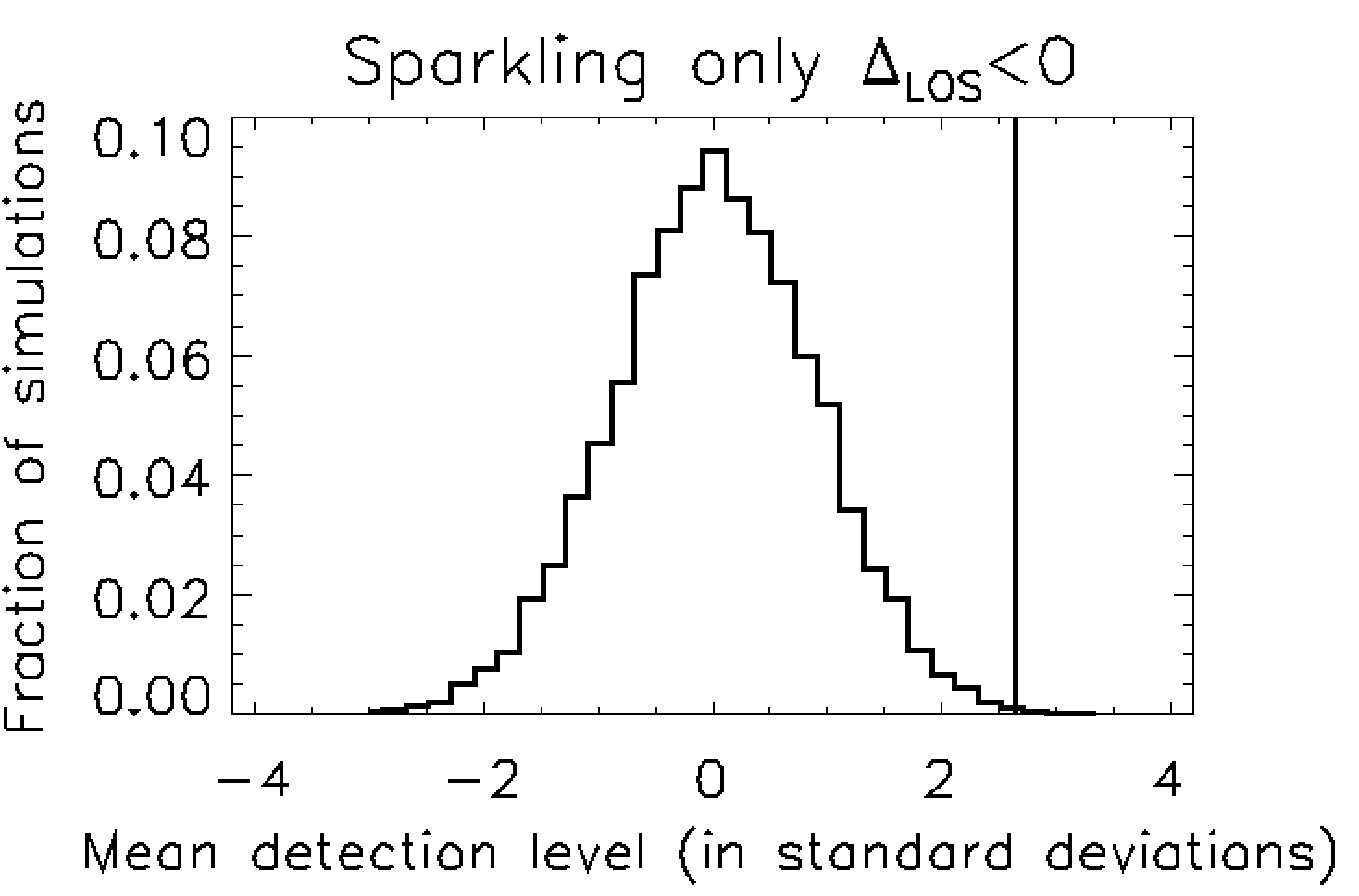}
  \includegraphics[width=0.495\linewidth]{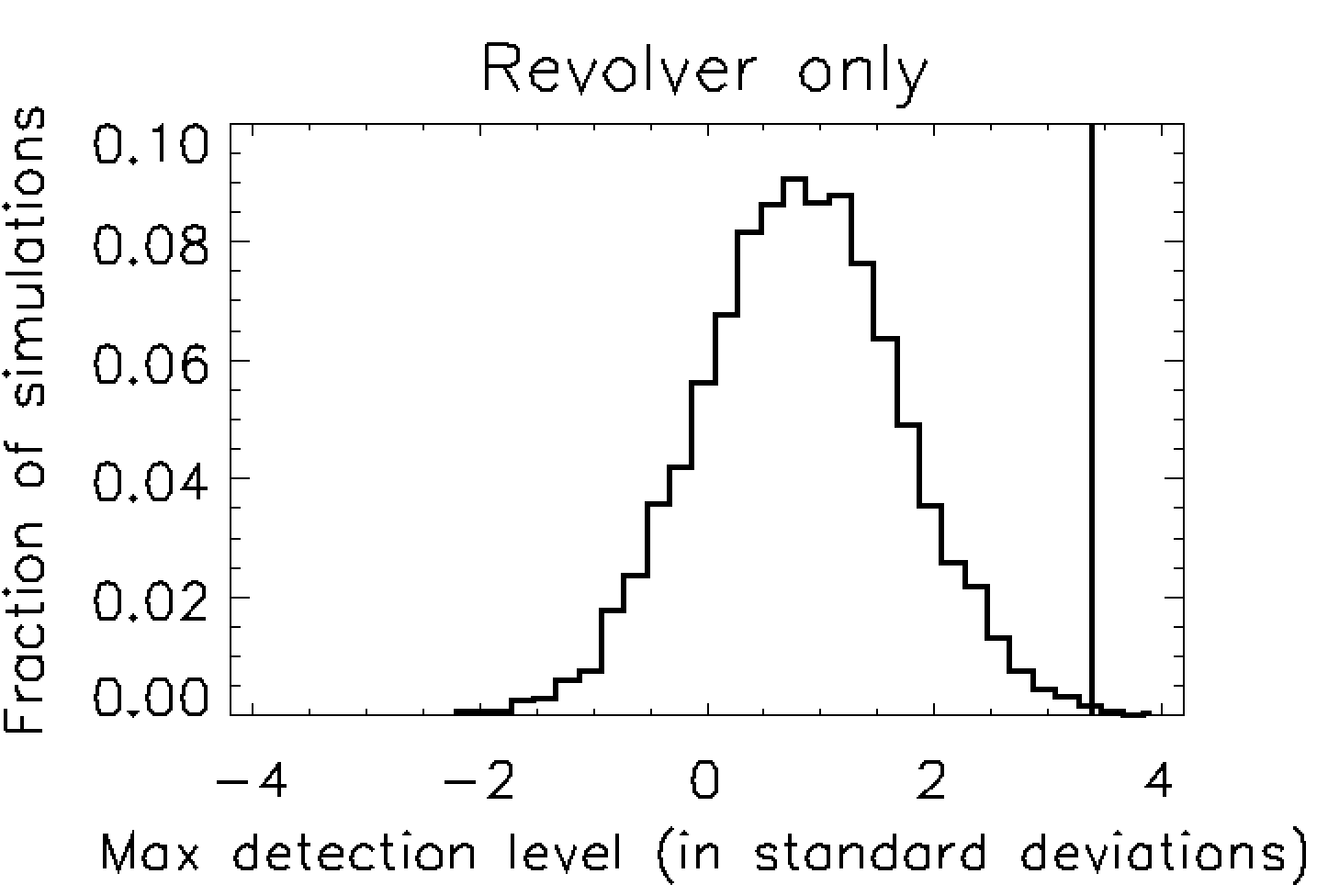}
  \includegraphics[width=0.495\linewidth]{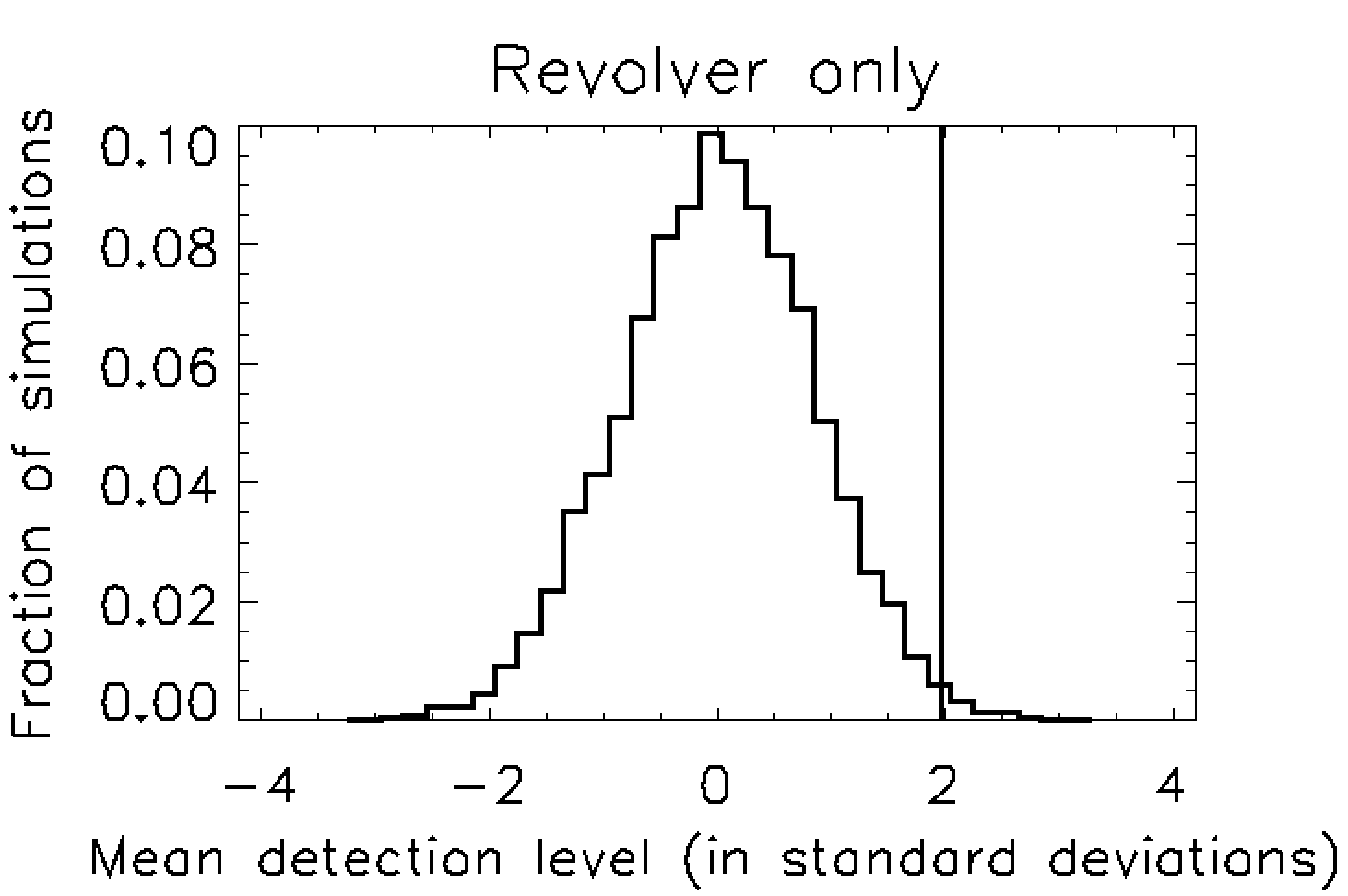}
  \caption{ \label{fig:maxmean_detlevel2} Same as Figure \ref{fig:maxmean_detlevel}, but for \sparkling and \revolver voids separately. Left column: max values, right column: mean values, upper row: \sparkling open voids, middle row: \sparkling voids with $\Delta_{LOS}<0$, lower row:\revolver voids.}
\end{figure}

In Figure \ref{fig:maxmean_detlevel2}, we show the result of correcting for the look-elsewhere-effect for \sparkling and \revolver voids separately. Again we see that both mean and maximum detection levels are in the far tail of the distribution obtained from simulations. For \sparkling alone, only $0.4\%$ of the simulations have similar og higher mean detection and $0.9\%$ a higher maximum detection. This decreases to $0.1\%$ when limiting to voids within undersensities along the line of sight. For \revolver with void temperatures which are not significantly high for  for smaller voids as can be seen in Table \ref{tab:significances_revolver}, the maximum and mean detection level when including all numbers in the table are still significant, only $0.2\%$ of the simulations have a similar maximum detection when optimizing parameters for each simulation, and only $0.9\%$ of the simulations have a similar or higher mean detection level.

\begin{figure}[htbp]
  \includegraphics[width=0.495\linewidth]{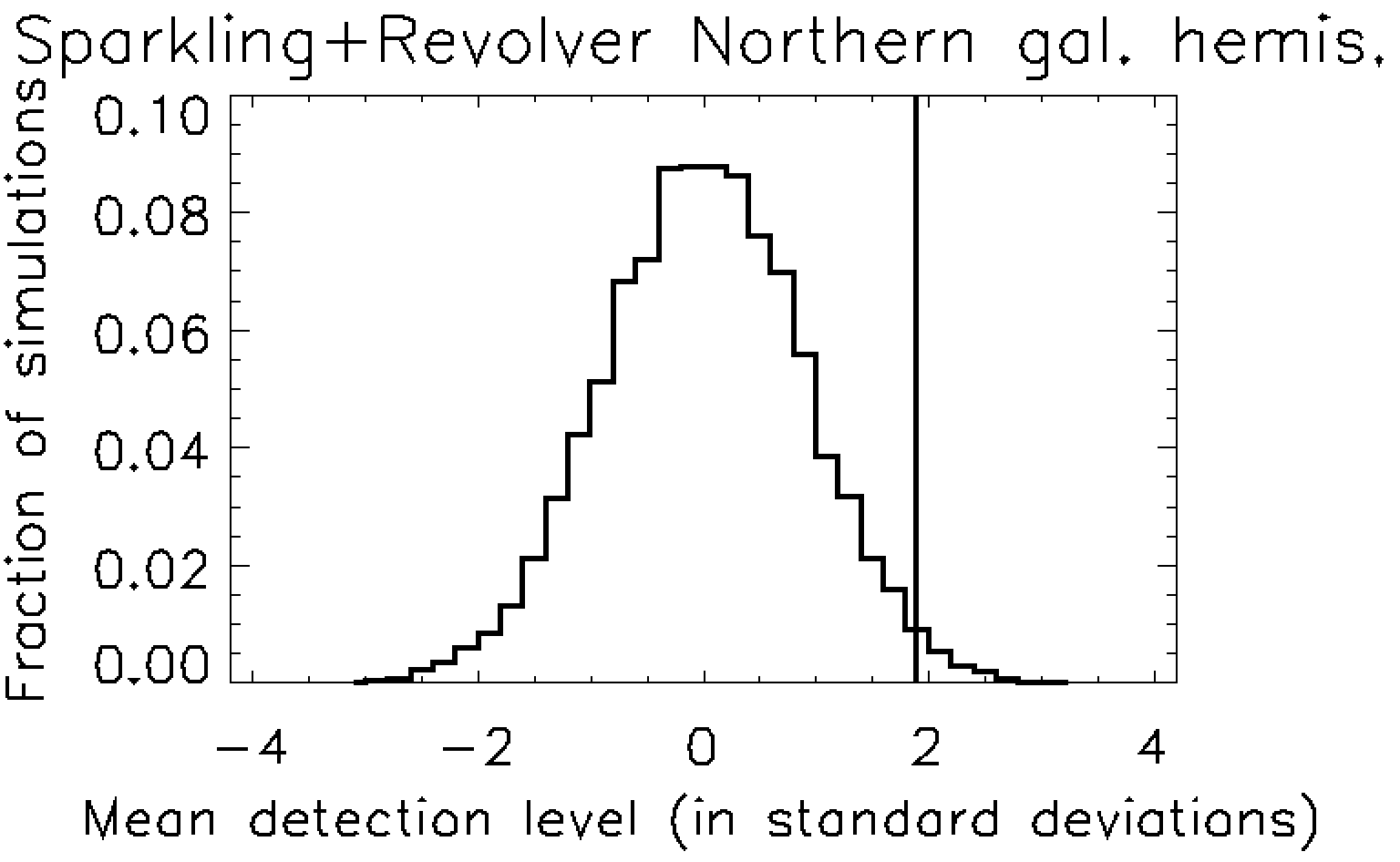}
  \includegraphics[width=0.495\linewidth]{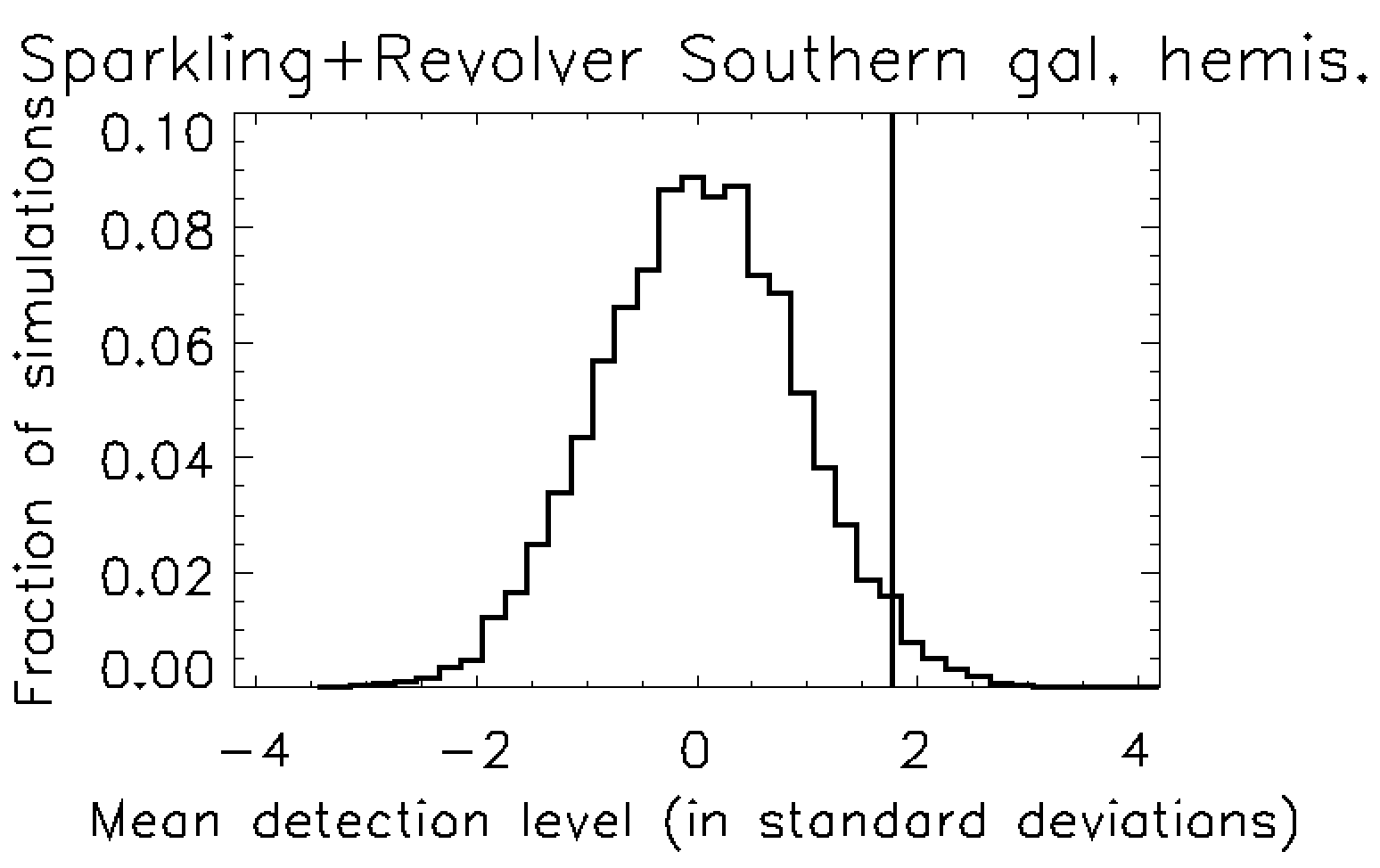}
  \includegraphics[width=0.495\linewidth]{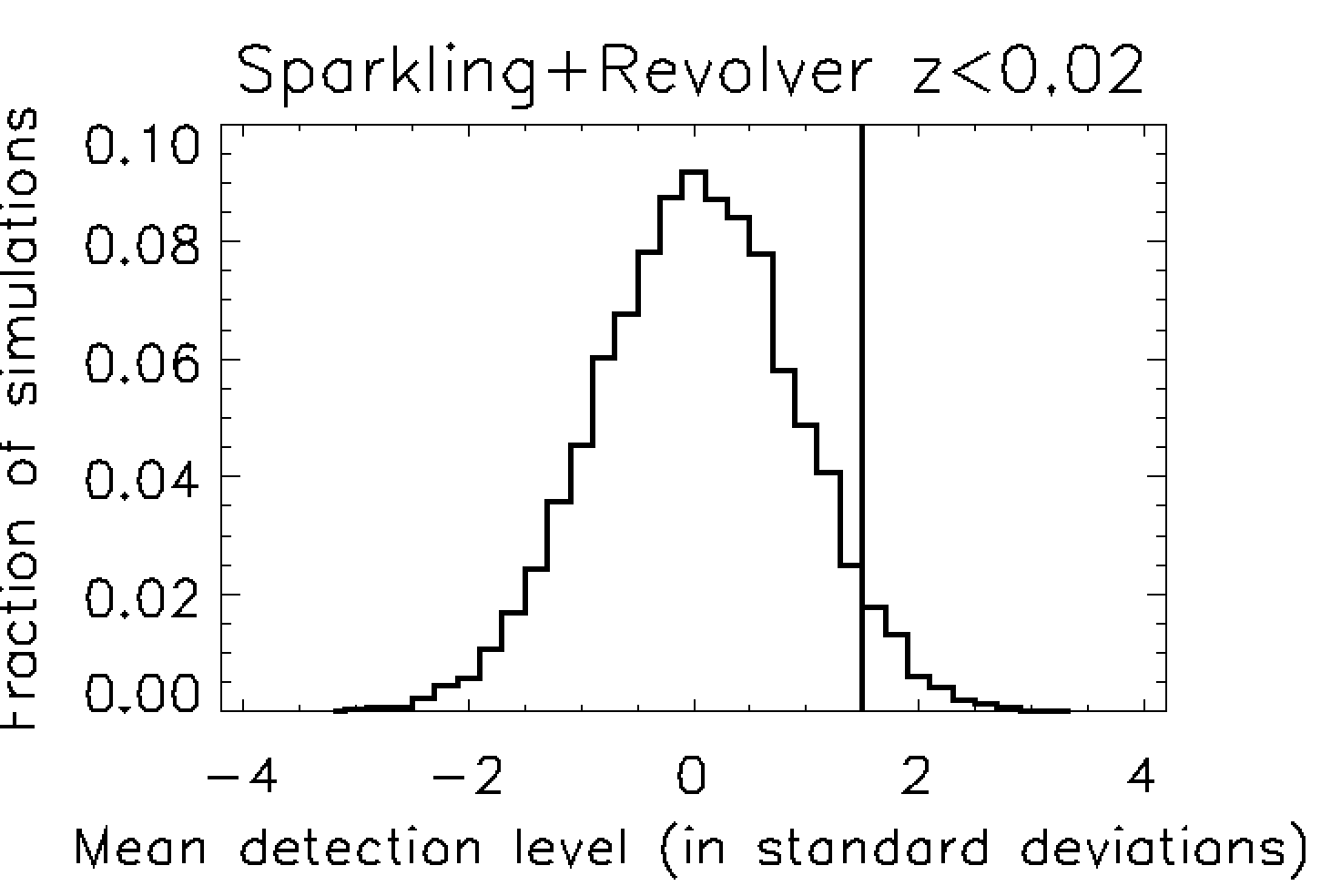}
  \includegraphics[width=0.495\linewidth]{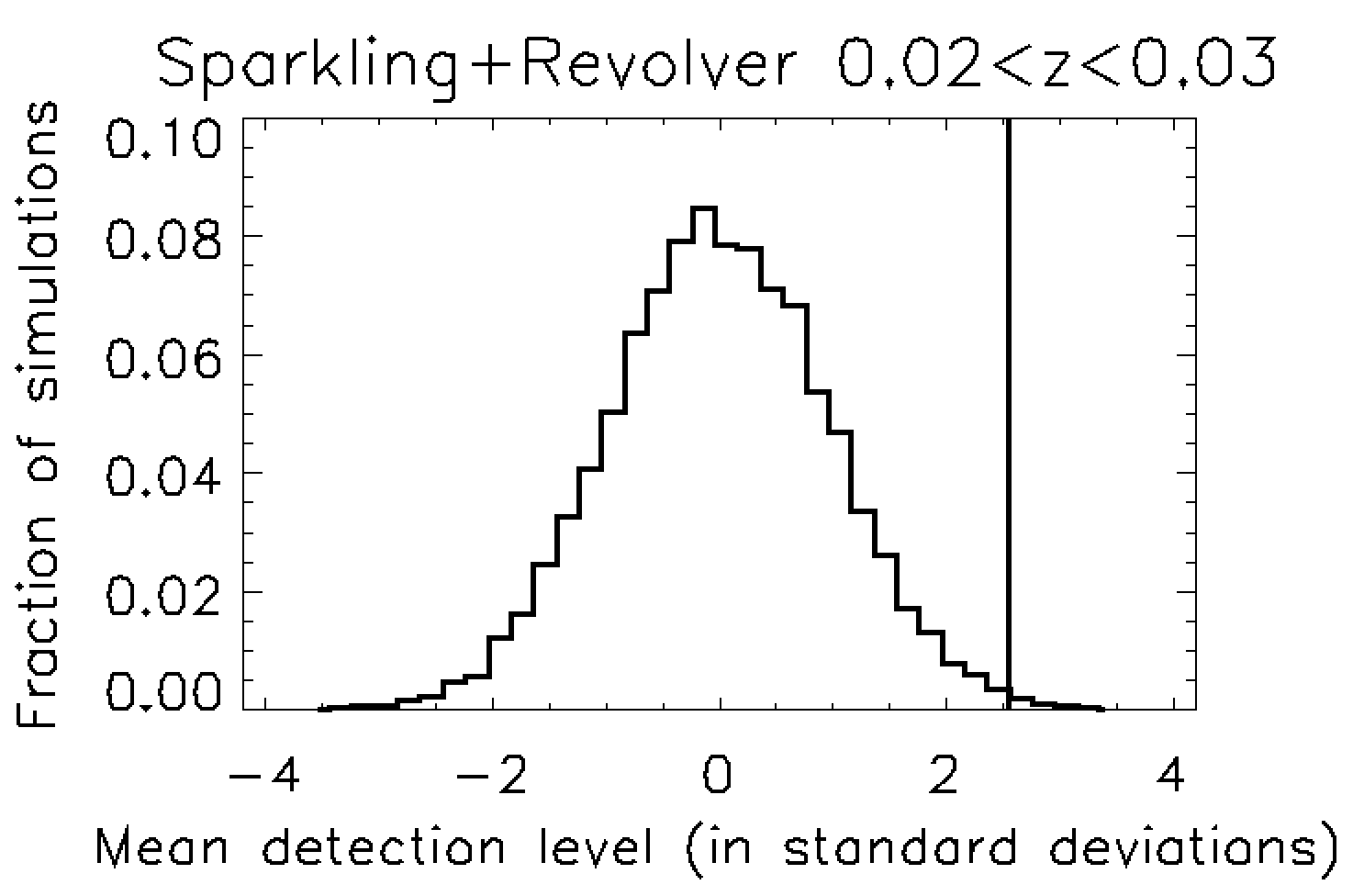}
    \caption{ \label{fig:mean_detlevel_split} Mean detection level (calculated as in Figure \ref{fig:maxmean_detlevel}) for void temperatures measured in norther galactic hemisphere only (upper left), southern galactic hemisphere only (upper right), lower redshift shell $z<0.02$ only (lower left) and higher redshift shell $0.02<z<0.03$ only (lower right).}
\end{figure}

\begin{figure}[htbp]
  \includegraphics[width=0.495\linewidth]{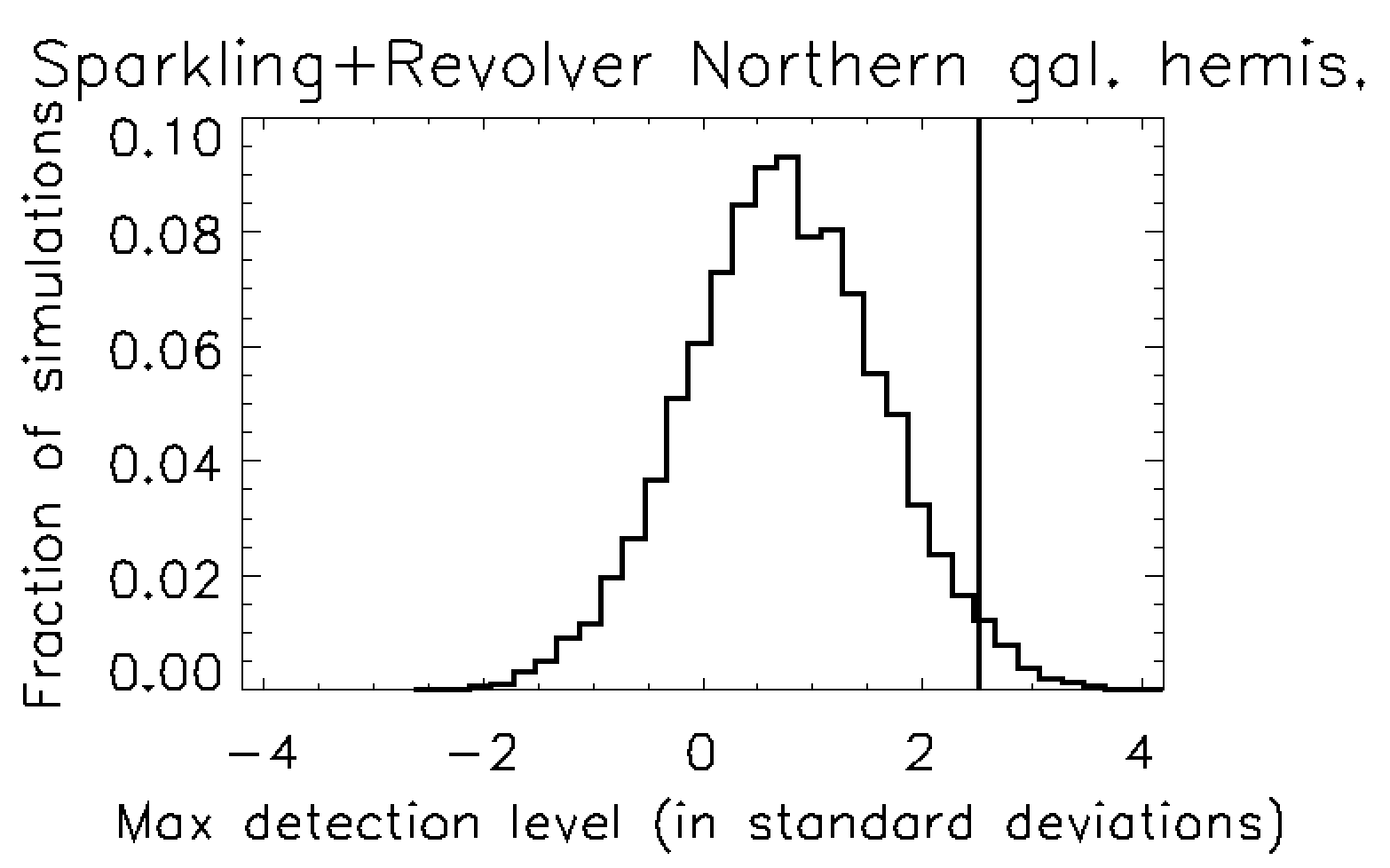}
  \includegraphics[width=0.495\linewidth]{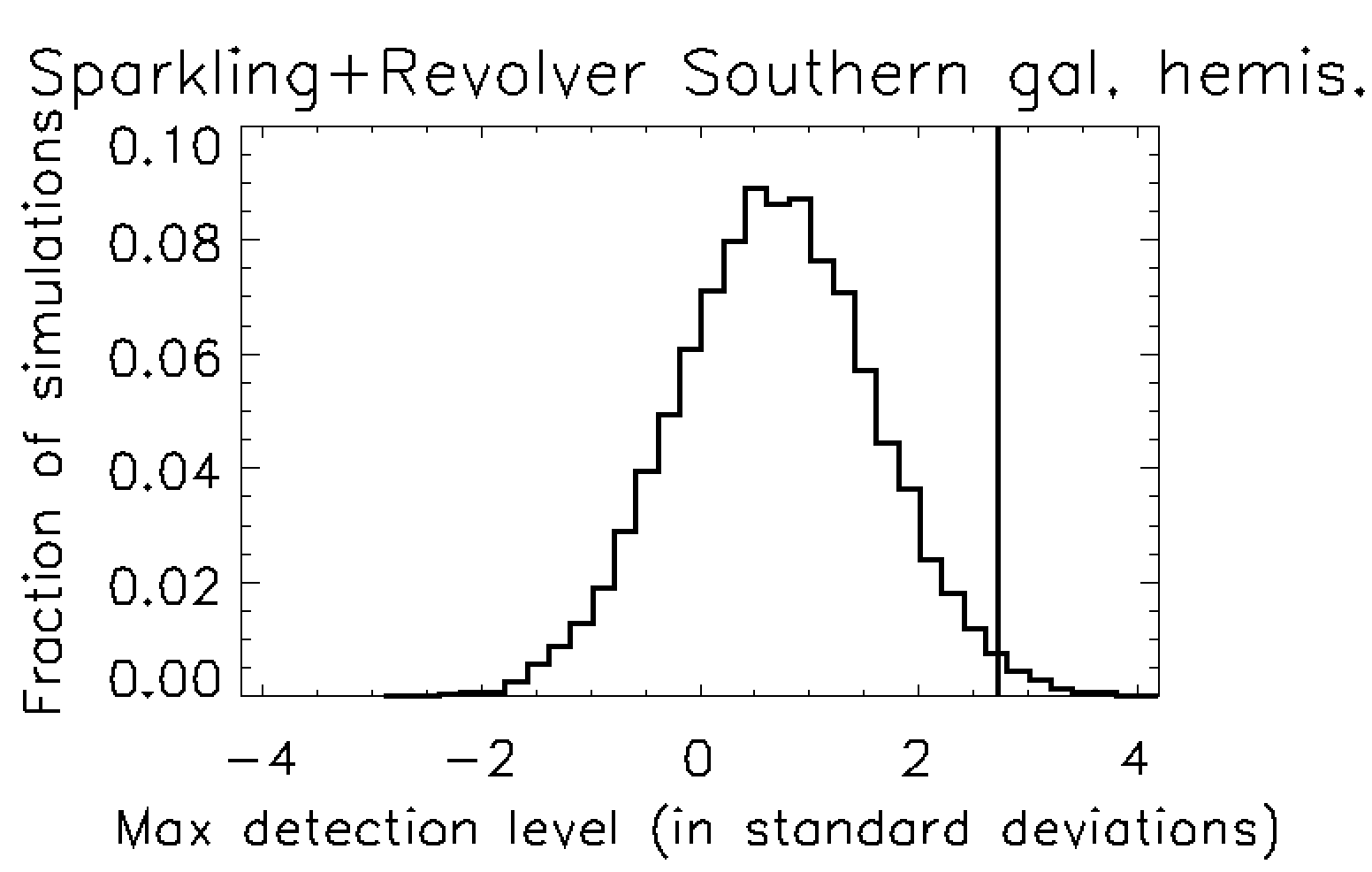}
  \includegraphics[width=0.495\linewidth]{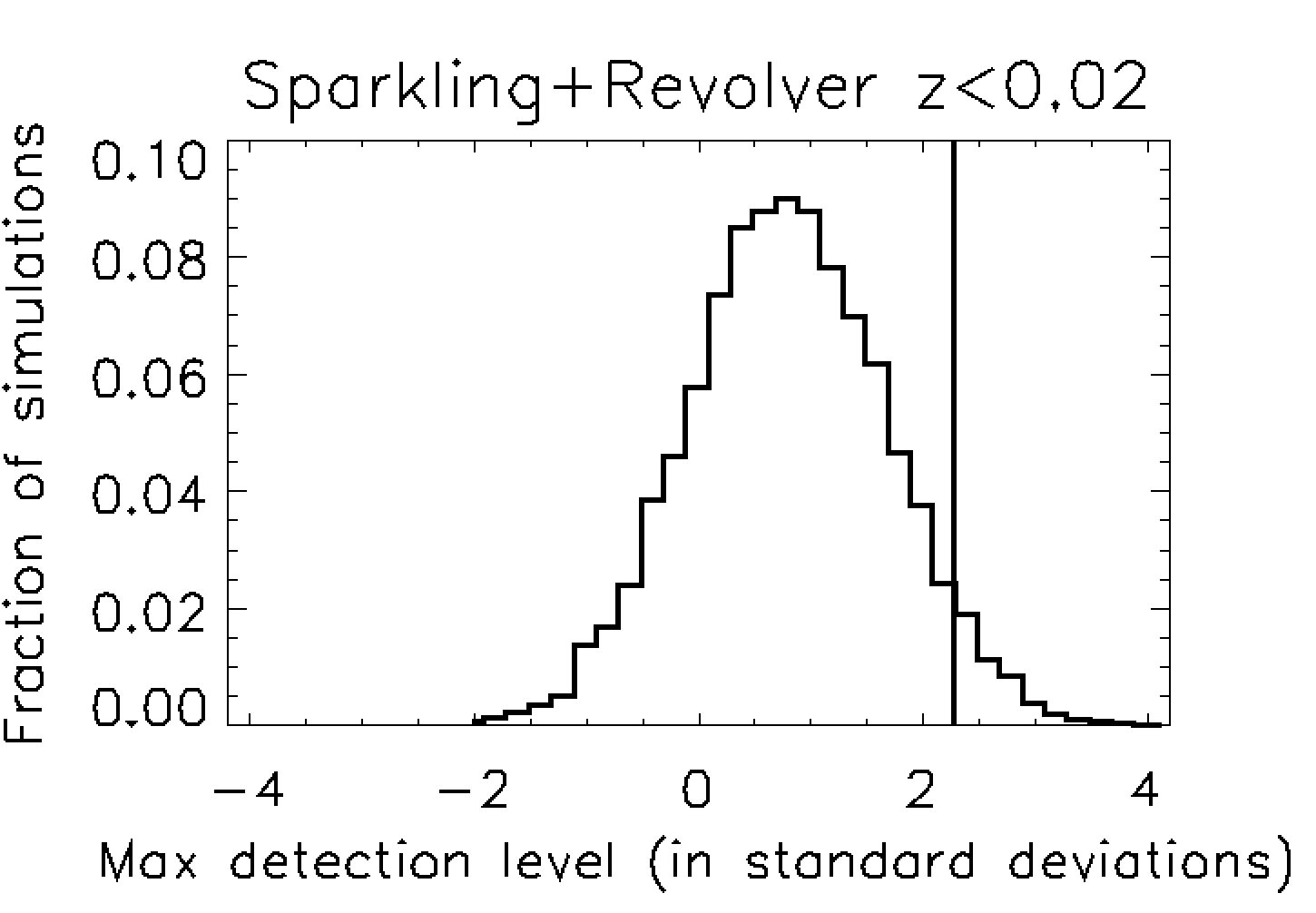}
  \includegraphics[width=0.495\linewidth]{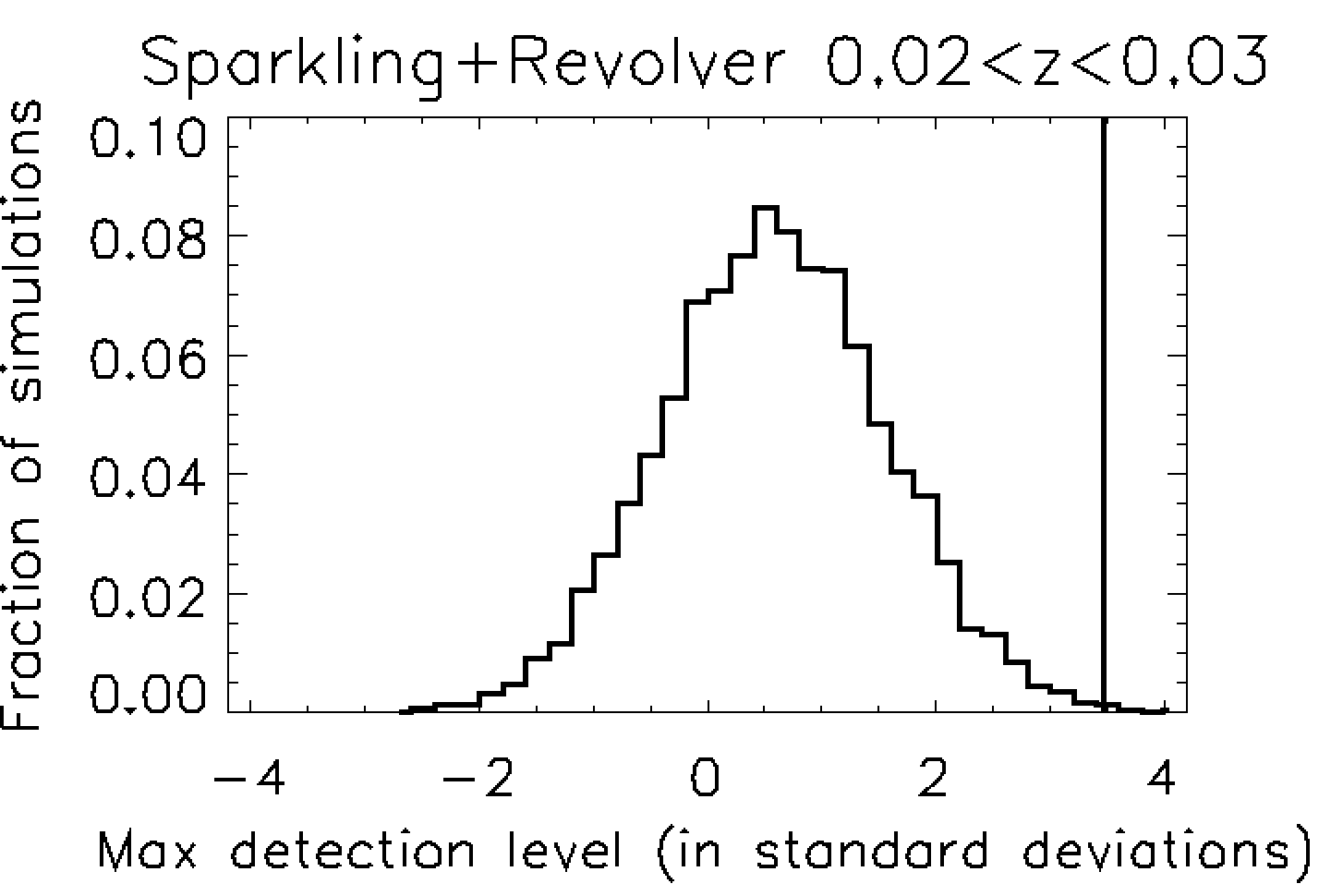}
    \caption{ \label{fig:max_detlevel_split} Maximum detection level (calculated as in Figure \ref{fig:maxmean_detlevel}) for void temperatures measured in norther galactic hemisphere only (upper left), southern galactic hemisphere only (upper right), lower redshift shell $z<0.02$ only (lower left) and higher redshift shell $0.02<z<0.03$ only (lower right).}
\end{figure}

\begin{figure}[htbp]
  \includegraphics[width=0.495\linewidth]{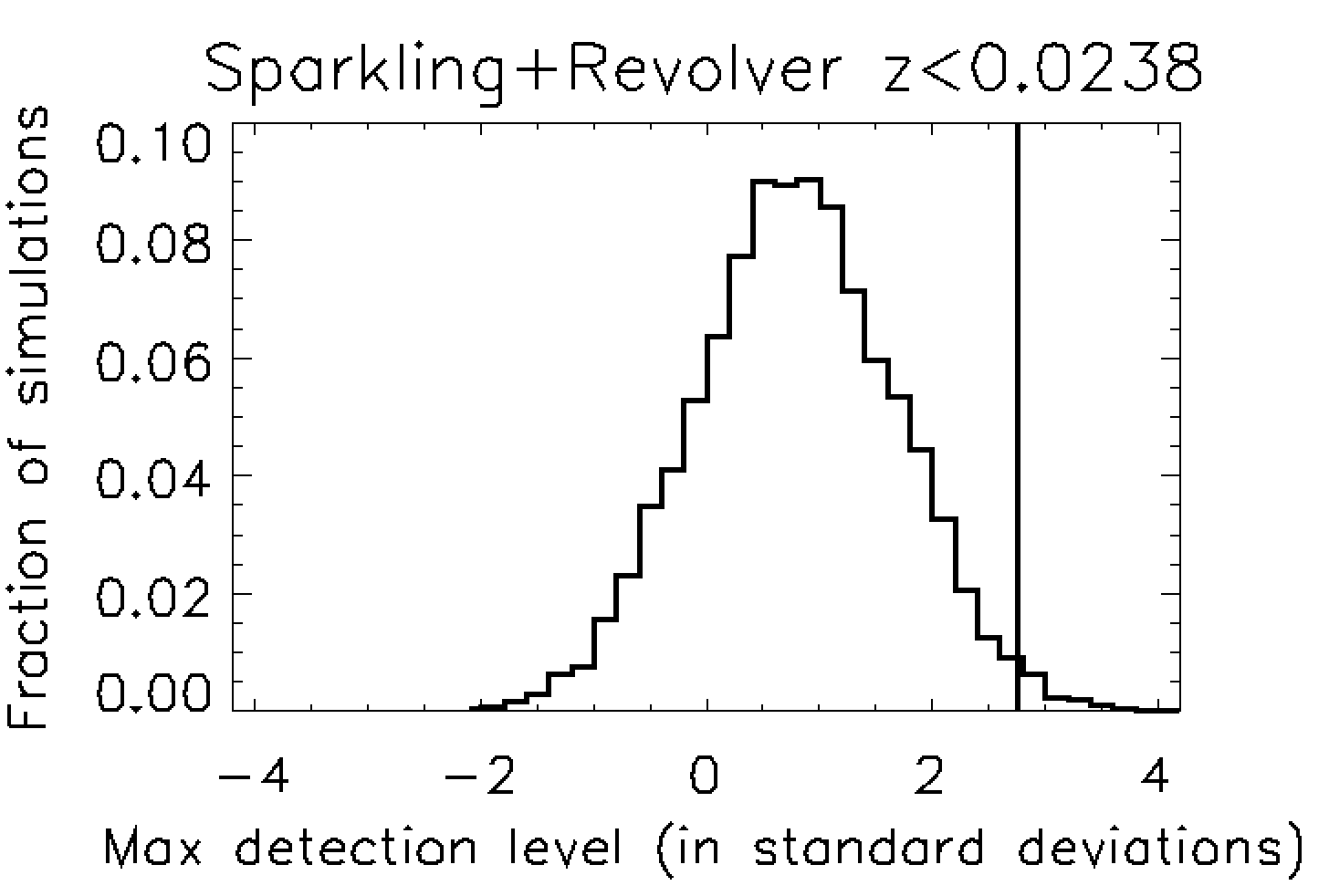}
  \includegraphics[width=0.495\linewidth]{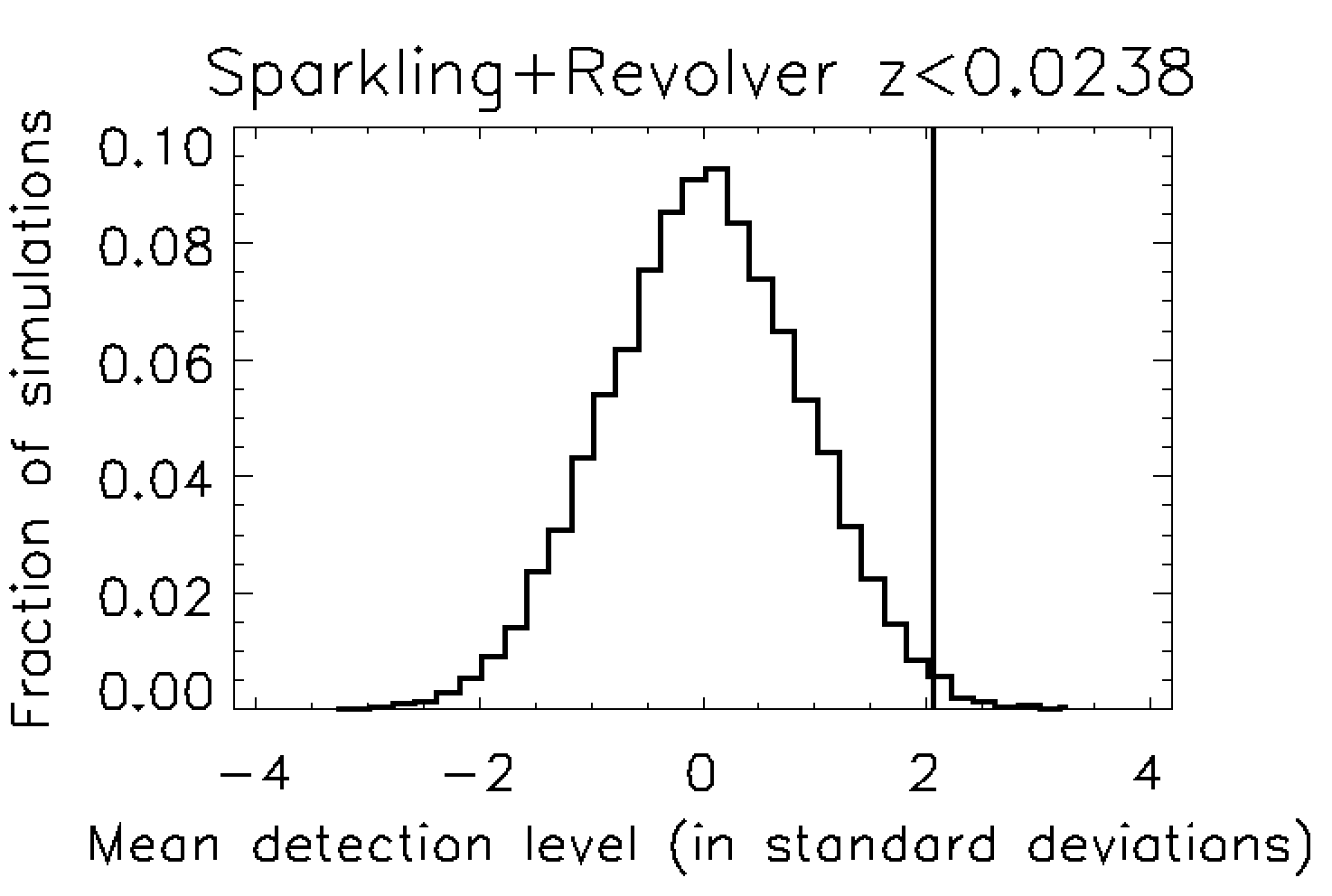}
  \includegraphics[width=0.495\linewidth]{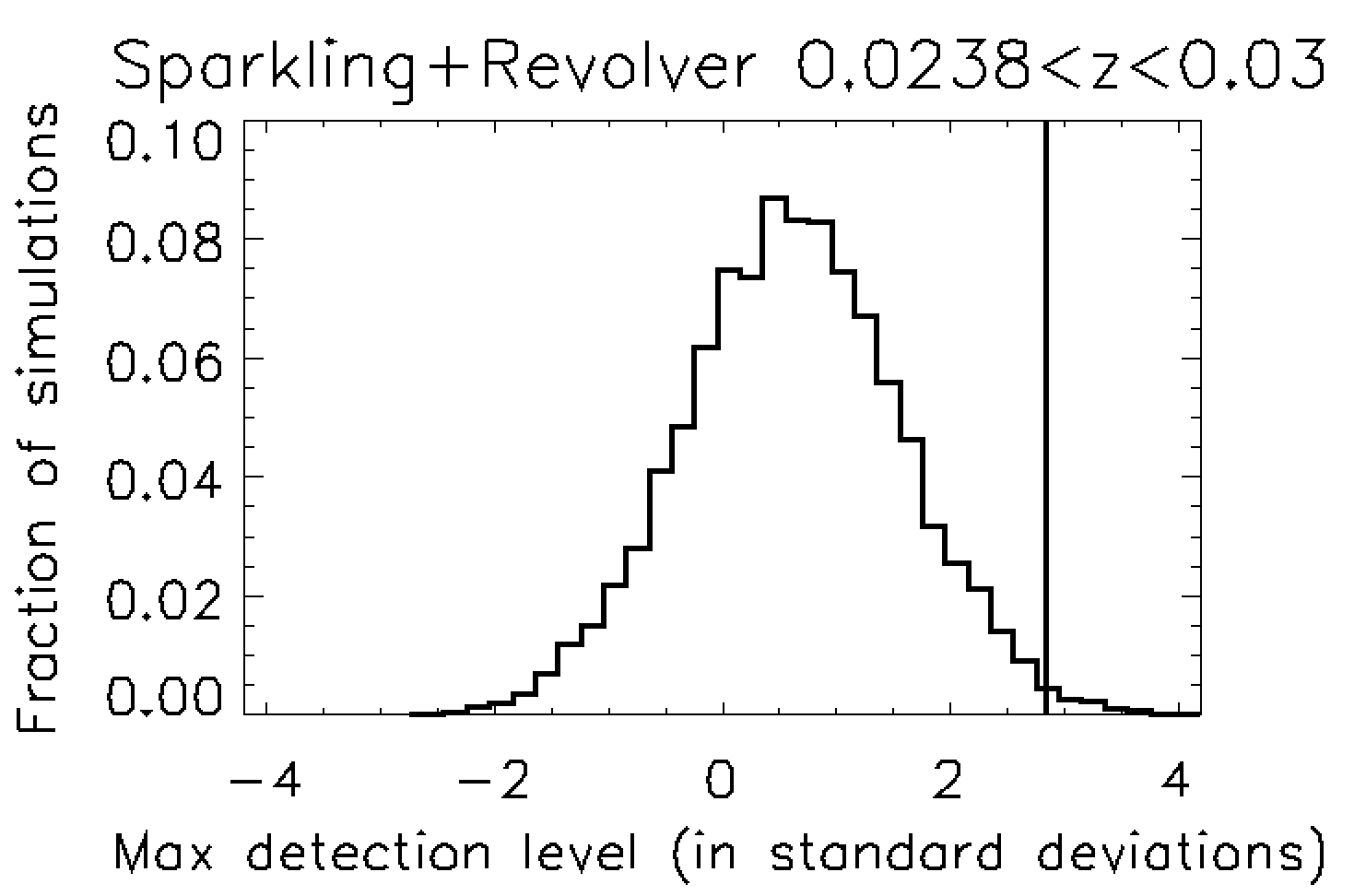}
  \includegraphics[width=0.495\linewidth]{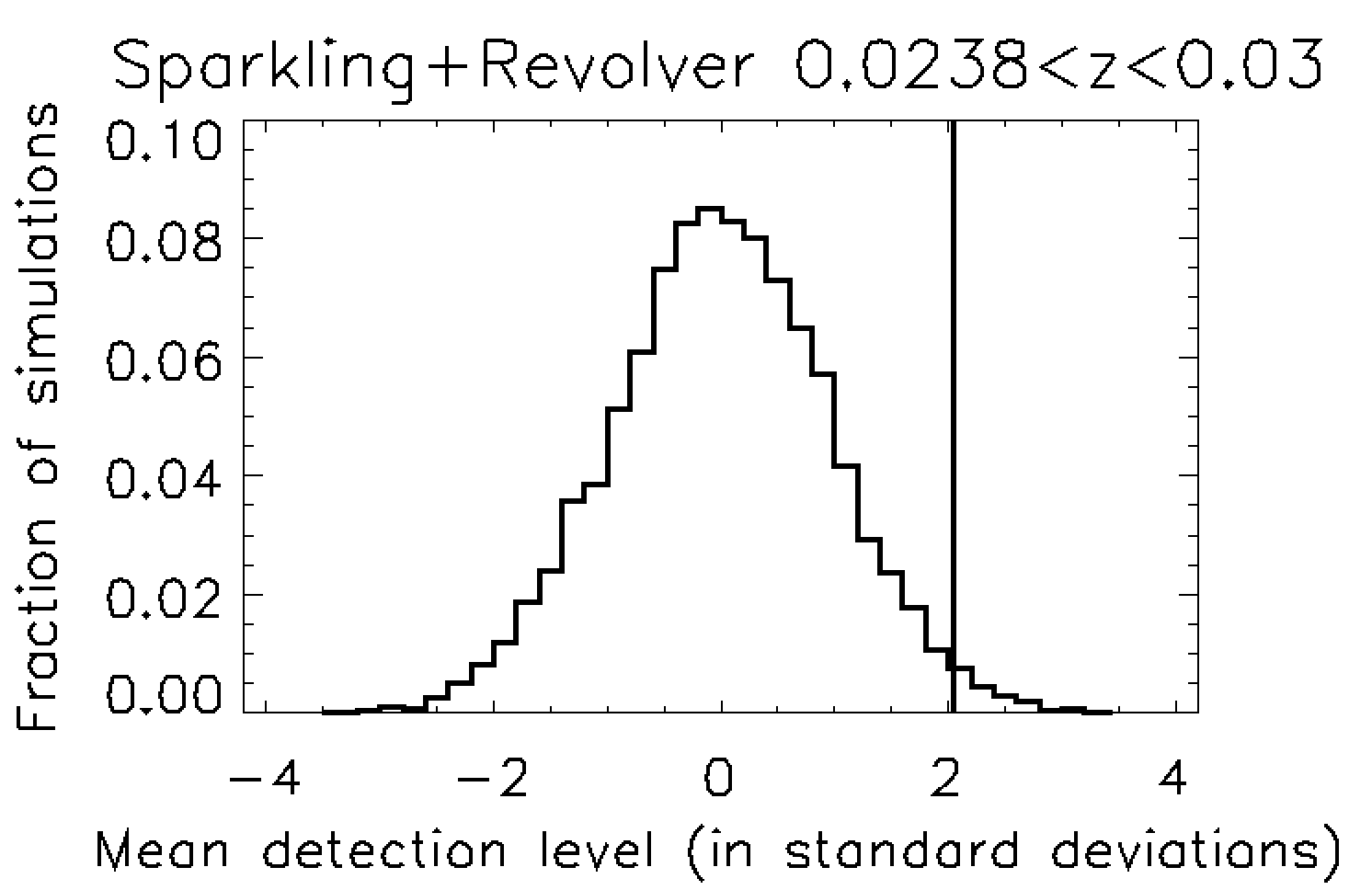}
    \caption{ \label{fig:max_zsplit2} Same as Figure \ref{fig:maxmean_detlevel}, but for the equal volume redshift split. Upper panels: Maximum (left) and mean (right) detection level using only voids in the most nearby shell $z<0.0238$. Lower panels:  Maximum (left) and mean (right) detection level using only voids in the second shell $0.0238<z<0.03$.}
\end{figure}

In Figure \ref{fig:mean_detlevel_split} and Figure \ref{fig:max_detlevel_split}, we have show results for two separate data splits:
\begin{itemize}
\item calculating the void temperature only in the northern or the southern galactic hemisphere
\item calculating the void temperature using the voids weight map obtained only for voids in the inner redshift shell $z<0.02$ or the outer redshift shell $0.02<z<0.03$.
\end{itemize}.
The figure shows again the maximum and minimum detection levels taken over all choices of void radius fraction, void size sample and wavelet scale. We see a very similar detection level for the northern and southern hemispheres, in both cases about $2\%$ of the simulations have larger detections. For the redshift shells, while both shells clearly show a strongly positive mean void temperature, the inner shell shows less detection than the outer shell, with about $5\%$ simulations having larger detections in the inner shell and $0.3\%$ in the outer shell. Note however that the volume of the inner shell $z<0.02$ is much smaller than the outer shell $0.02<z<0.03$. In order to further test consistency between redshift shells, we also split the observed volume inside $z<0.03$ in two equal volume shells, $z<0.0238$ and $0.0238<z<0.03$. In Figure \ref{fig:max_zsplit2} we show the results. Now the significance level is very similar for both redshift shells with about $1\%$ simulations with higher detections in both shells. In summary, the results show a remarkable consistency between all these very different data splits.

\end{appendix}

\end{document}